\documentclass{aa}
\pdfoutput=1

\usepackage{graphicx}
\usepackage{txfonts}
\usepackage{pictex}
\usepackage{latexsym}
\usepackage{graphics}
\usepackage{afterpage}
\usepackage[]{hyperref}
\usepackage[section]{placeins}
\usepackage{float}
\usepackage{units}

\begin{document} 

   \title{A treasure hunt in the pool of spectra classified as high-redshift QSOs by the spectroscopic pipeline of the SDSS DR16}

   \author{Helmut Meusinger\inst{1}
          }

   \institute{Th\"uringer Landessternwarte, Sternwarte 5, D-07778 Tautenburg, Germany 
                   \email{meus@tls-tautenburg.de}
                 }

  \date{Received XXXXXX; accepted XXXXXX}

 \abstract{
The discovery of outsiders in the form of unusual, rare, or even unknown object types is important as they can provide useful information about otherwise hidden physical phenomena and processes. 
The present study takes advantage of the fact that the automated spectroscopic pipeline of the Sloan Digital Sky Survey (SDSS) occasionally assigns uncommon spectra to high-redshift QSOs.
This paper presents an analysis of $\sim 4000$ spectra that are QSOs with redshifts  $z>4.5$ according to the spectroscopic pipeline of the SDSS DR16.
It turns out that, after excluding non-classifiable spectra of low quality and those from three special plates, only 26 \% are high-$z$ QSOs,
50\,\% are QSOs at lower redshifts,  16\,\% are galaxies,  and 8\,\% are stars.  A significant proportion of the latter three categories prove to be unusual 
and are re-assigned here to a variety of rare types. The results of the re-evaluation are summarised in a catalogue. 
 }


  \titlerunning{Unusual spectral types among high-$z$ QSOs from SDSS DR16}
  \authorrunning{H. Meusinger}

   \maketitle

\section{Introduction}\label{Intro}

The Sloan Digital Sky Survey \citep[SDSS;][]{York_2000} has revolutionised the  database of Extragalactic astronomy. SDSS-I/II contains spectra for about 4\,$10^4$ QSOs with redshifts up to $z\approx 5.5$ \citep{Schneider_2010}.  The spectra were obtained by a pair of multi-object spectrographs with a total of 640 fibres, 
which produced an average resolving power of $\approx 2000$ in the wavelength interval 3800\,\AA\ to 9200\,\AA.
In the next steps,  SDSS-III and SDSS-IV used an upgraded pair of spectrographs for the Baryon Oscillation Spectroscopic Survey \citep[BOSS;][]{Dawson_2013}  and the Extended Baryon Oscillation Spectroscopic Survey  \citep[eBOSS;][]{Dawson_2016} that have a total of 1000 fibres and an extended wavelength coverage up to  
10\,400\,\AA. Multiple target selection algorithms were applied and large numbers of faint QSOs and galaxies were detected. The SDSS Data Release 16 \citep[DR16;][]{Ahumada_2020} contains more than four million spectra of QSOs, galaxies, and stars.  The final SDSS-IV QSO catalogue from the DR16  \citep[QCDR16,][]{Lyke_2020} lists 750\,414 QSOs. One of the opportunities and at the same time one of the challenges arising from the huge number of spectra is the possibility of discovering rare,  unusual or even previously unknown types of QSOs, galaxies, and stars. 

For the vast majority of the SDSS QSOs, the spectral energy distribution (SED) in the optical and UV is well represented  by composite spectra \citep[e.g.][]{VandenBerk_2001}.  However,  due to the multiple physical processes involved, QSOs exhibit a wide range of spectral properties that can differ greatly from the standard spectrum. It is widely agreed that objects whose properties deviate significantly from the norm may hold a key to important physical processes,  e.g. related to the evolution of active galactic nuclei (AGNs) and their impact on the host  galaxies \citep[see review by][]{Harrison_2024}. The SDSS has confirmed the existence of rare populations of QSOs with unusual spectral properties, such as a red continuum,  unusual broad absorption lines (BALs), weak or absent emission lines in combination with an unobscured continuum (WLQs),  or double-peaked broad emission lines 
\citep[e.g.][]{Fan_1999, Fan_2002, Hall_2002,  Strateva_2003, Plotkin_2008}. Another rare phenomenon is that of the
changing-look QSOs (CLQs), which show extreme temporal changes that correspond to a transition between different QSO types \citep[e.g.][]{LaMassa_2015,  Potts_2021, Komossa_2024}. For statistical analyses it is important to increase the number of known objects of such types. Larger, but still relatively small, samples have been discovered by using specialised methods for the detection of outsiders \citep[e.g.][]{Meusinger_2012, Fustes_2013, Wei_2013,  Meusinger_2014,  Meusinger_2016,  Baron_2017, Doorenbos_2021, Reis_2021, Tiwari_2025}.

The main aim of the present study is also to find rare and unusual spectra, but using a different method.  The automated spectroscopic pipeline of the SDSS tends to assign a high redshift to spectra of outliers, which can be those of low quality or of unusual types.   The result of an SQL query for high-$z$ QSOs should therefore be treated with caution if the aim is actually to obtain a sample of high-$z$ QSOs. On the other hand, it can also prove to be a kind of treasure trove for rare and unusual spectra.  This paper presents the results of an analysis of nearly 4000 spectra to which the spectroscopic pipeline of the SDSS DR16 has assigned redshifts $z>4.5$. The analysis and classification of the spectra are described in Sect.\,\ref{sect:Re_classification}, the resulting object sample is presented in Sect.\,\ref{sect:Sample},  some unusual types are briefly discussed in Sect.\,\ref{sect:Rare_types}.  Due to the variety of object types involved, it is not possible to discuss them in detail. Readers are encouraged to do further research.  The Appendices present the description of the catalogue and a number of illustrating spectra.  The objects are labelled here with their SDSS name, with the exception of the figures, where the designation ‘SDSS’ is omitted for reasons of space.  Spectra are labelled with plate-MJD-fiberID. 

The  Lambda Cold Dark Matter ($\Lambda$CDM) cosmology is assumed with $H_0 = 73$ km s$^{-1}$ Mpc$^ {-1}$,  $\Omega_{\rm \Lambda} =0.73$, and $\Omega_{\rm M}=0.27$.

\section{Re-classification of the spectra}\label{sect:Re_classification}

The list of spectra for the present study  was extracted via SQL query from the catalogue SpecObjAll on the Catalog Archive Server of the SDSS DR16.   The only selection criteria were the classification as QSOs with $z > 4.5$ 
and the requirement that the zWarning flag is either zero (i.e. no problems) or 64 (i.e. negative fluxes occur). 
The resulting list contains 3955 spectra, which are  intentionally not limited to those from plates of good quality.
The spectra were downloaded from the SDSS Science Archive Server  using the bulk load service.
The original idea was to use the software tool ASPECT \citep{inderAu_2012} for the construction of a Kohonen self-organising map \citep[SOM;][]{Kohonen_2001}.  The SOM technique is an artificial neural network algorithm that uses unsupervised learning to sort the spectra by similarity. In previous studies, this method has proven to be efficient in the selection of rare, unusual objects from large samples of  SDSS spectra \citep{Meusinger_2012,Meusinger_2014,Meusinger_2016, Meusinger_2017}.  However, initial tests with a SOM for a subsample of spectra from the present sample have shown that this method is less efficient here, as the sample is extremely inhomogeneous. This led to the decision to `manually' classify the spectra individually.

\begin{figure}[htbp]
\begin{center}
\includegraphics[viewport= 0 0 1170 540,width=7.1cm,angle=0]{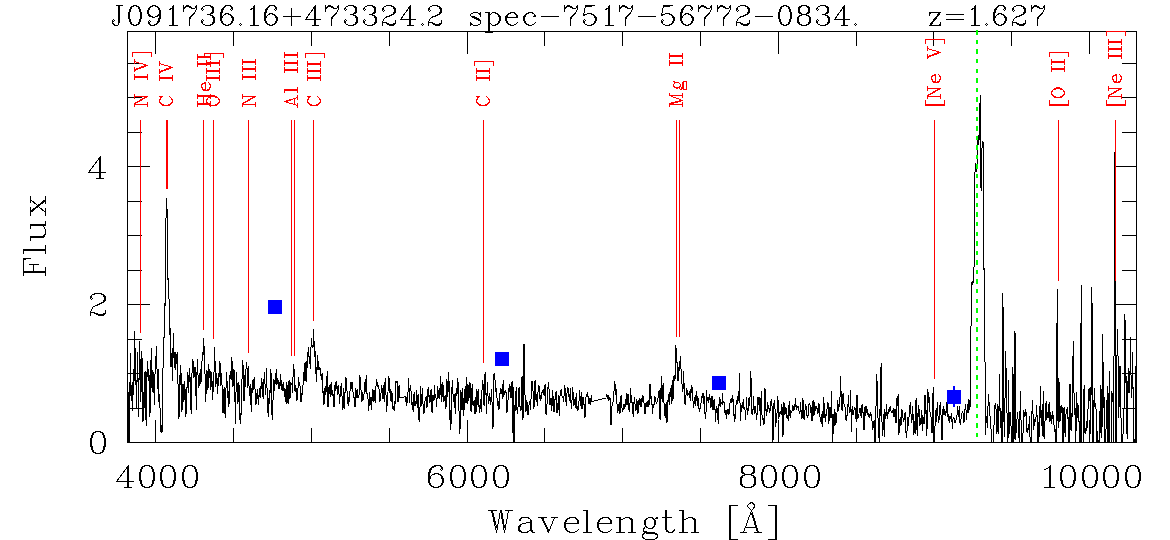} 
\includegraphics[viewport= 20 -150 256 256,width=1.6cm,angle=0]{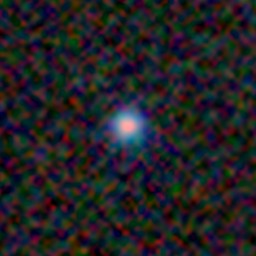} \\
\includegraphics[viewport= 0 0 1170 540,width=7.1cm,angle=0]{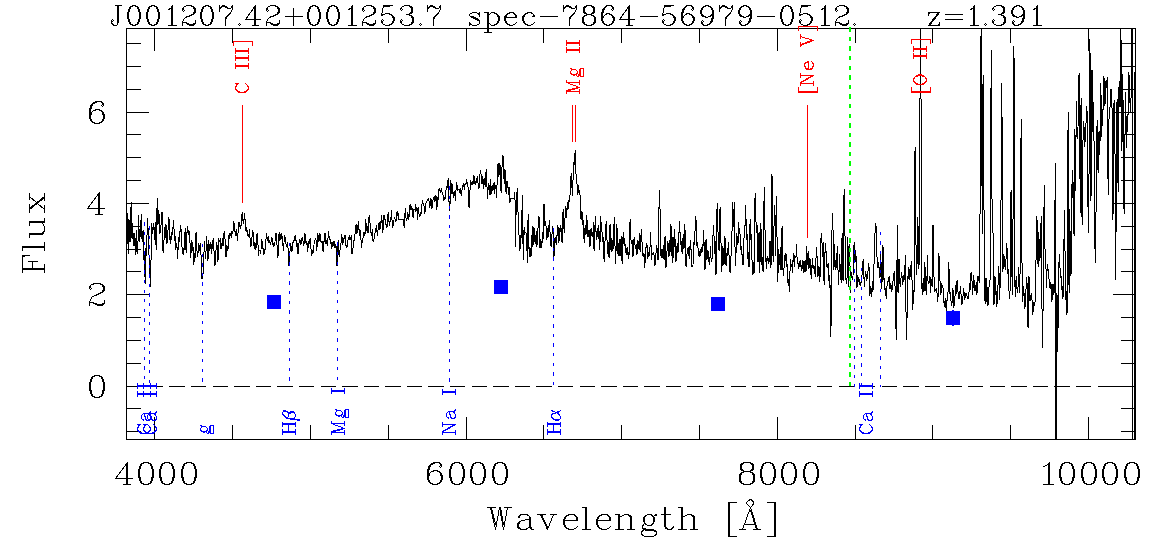} 
\includegraphics[viewport= 20 -150 256 256,width=1.6cm,angle=0]{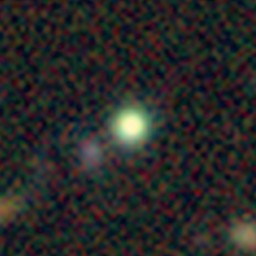} \\
\includegraphics[viewport= 0 0 1170 540,width=7.1cm,angle=0]{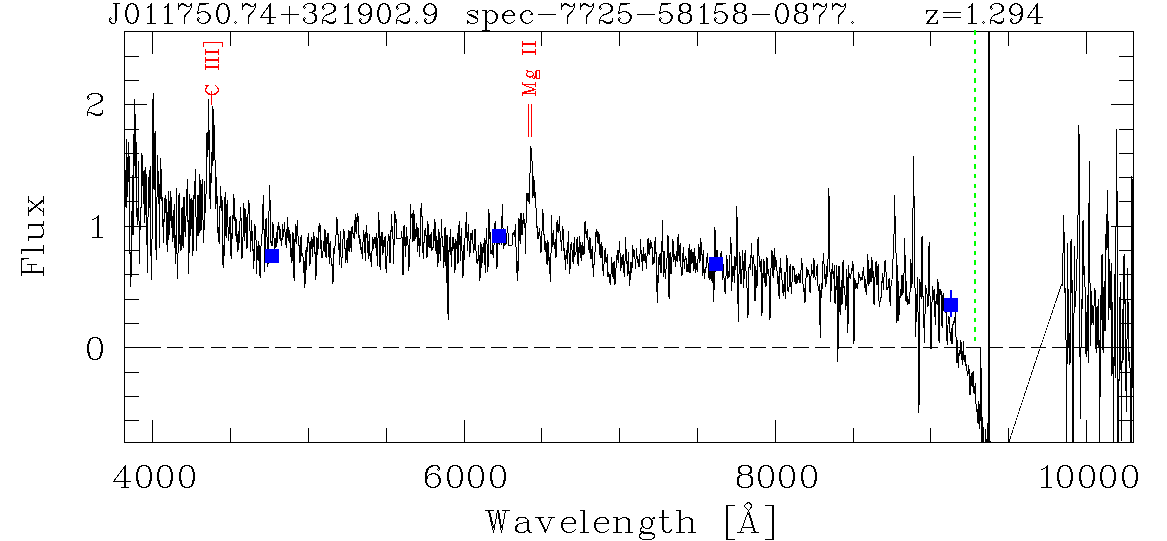} 
\includegraphics[viewport= 20 -150 256 256,width=1.6cm,angle=0]{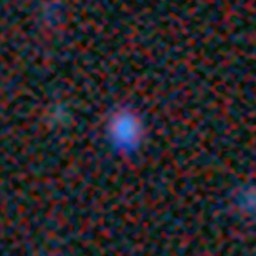} \\
\includegraphics[viewport= 0 0 1170 540,width=7.1cm,angle=0]{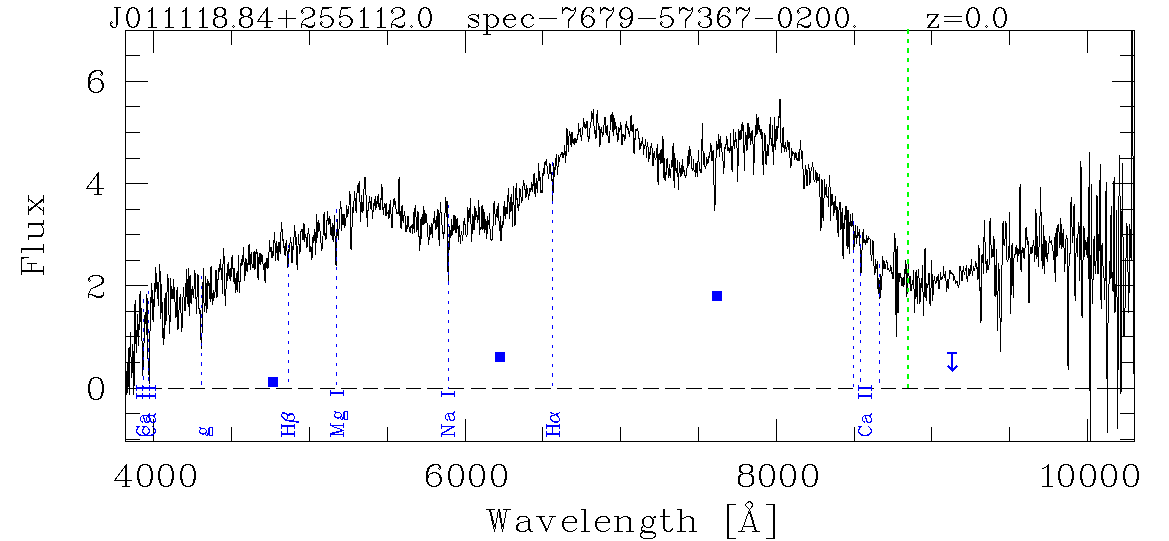} 
\includegraphics[viewport= 10 -90 156 156,width=1.7cm,angle=0]{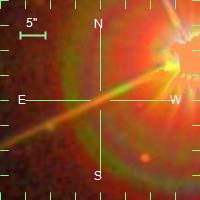} 
\caption{Four examples of problematic spectra with special features that can significantly influence automated classification (see text).
Emission lines are labelled at the top, absorption lines at the bottom. 
The vertical green lines in the spectra plots mark the positions of the Ly\,$\alpha$ line for the redshift from the SDSS.
Blue squares indicate fluxes derived from SDSS photometry.
The image of J011118 is from the SDSS, all others are from the LS (15 arcsec side length).
}
\label{fig:special_cases_1}
\end{center}
\end{figure}

\begin{figure}[htbp]
\begin{center}
\includegraphics[viewport= 0 0 1170 540,width=7.1cm,angle=0]{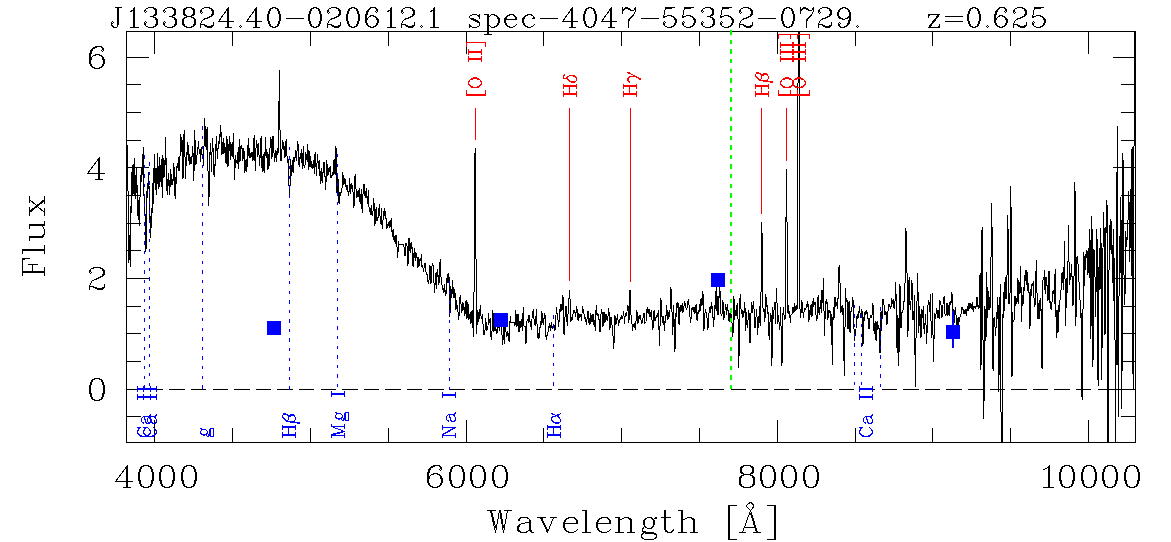} 
\includegraphics[viewport= 20 -150 256 256,width=1.6cm,angle=0]{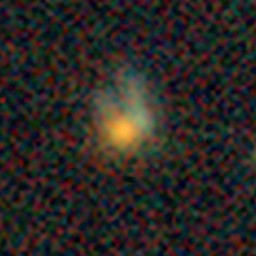} \\
\includegraphics[viewport= 0 0 1170 540,width=7.1cm,angle=0]{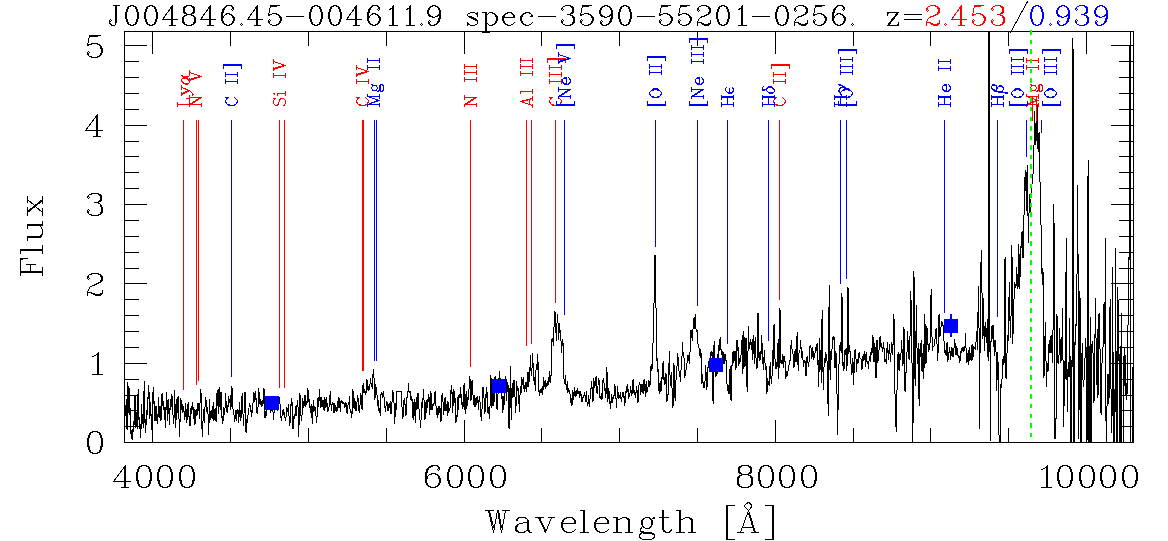} 
\includegraphics[viewport= 20 -150 256 256,width=1.6cm,angle=0]{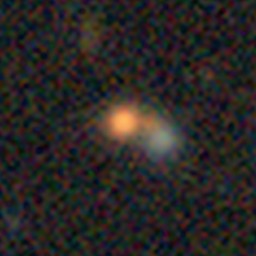} \\
\caption{As Fig.\,\ref{fig:special_cases_1}, but two cases of overlapping spectra.
}
\label{fig:special_cases_2}
\end{center}
\end{figure}

The spectra were inspected in three consecutive runs, i.e. each spectrum was evaluated three times. The tool {\tt zshift} \citep{Meusinger_2017} was used, which displays the  spectrum graphically and allows the user to manually change the value of $z$ and thus the markers for the positions of the typical spectral lines in order to find  the most probable redshift. That is, the re-classification is essentially based on  spectral lines.  In contrast to the spectroscopic pipeline of the SDSS,  no template spectra were used here, which is clearly an advantage for the discovery of unusual spectra.  Redshifts were preferably measured on narrow emission lines, if present.  Each inspection resulted in a re-determination  of $z$, the spectral type (QSO, galaxy, star, or unknown) and, if possible, a subtype (Table\,\ref{tab:subtypes}).   In a number of cases, additional information was used for classification.  In particular, the images from the DESY Legacy Surveys \citep[LS,][]{Dey_2019}  were helpful in interpreting overlaps between the spectra of different sources  (Fig.\,\ref{fig:special_cases_2}) or supported the classification as a galaxy.  In some cases, the proper motion data from Gaia DR3  \citep{Gaia_2023} and the MIR colours from the Wide-Field Infrared Survey Explorer  \citep[WISE;][]{Wright_2010} helped to distinguish between stars and QSOs.  Suspicious problematic spectra were checked by comparison with SEDs based on the  SDSS griz magnitudes 
(Figs.\,\ref{fig:special_cases_1}, \ref{fig:reddest_NonBALQSOs}, \ref{fig:polars}).  In the case of degraded quality of the spectrum, a  bad-quality flag was set, which can be  `noisy', `corrupt', or  `spike'.  An example for the latter is shown in the bottom panel of Fig\,\ref{fig:special_cases_1}. Noisy and/or distorted spectra can still be classified, but not those in the category `spike'. The reliability of the estimated $z$ was coded with the flag $r_z$, which can have the values 3 (no doubt; at least two lines securely identified), 2 (classification secure, but $z$ slightly uncertain), 1 (classification very likely, but $z$ only tentative), and 0 (undefined, type = unknown).  The latter category consists of 862 spectra (22\,\%) in which no significant spectral features could be recognised.

The distinction between galaxies and QSOs is a bit blurred.  A source was considered a galaxy if only typical stellar absorption lines and/or narrow emission lines at low $z$ were identified. In general, a source was classified as QSO if broad emission or broad absorption lines were recognised (exceptions are two supernovae of type Ia)  or if,  at low $z$,  the line ratio [\ion{O}{iii}]/H$\beta$ appears very high and the [\ion{O}{iii}]  line relatively broad.   A more correct distinction  based on diagnostic line ratios is beyond the scope of this study.   A small number of spectra that show the typical QSO emission lines  at $z \ga 0.8$ without broad components were tentatively classified as QSOs of type 2.  For a considerable number of spectra, only one narrow double emission line is seen, which could be identified with [\ion{O}{ii}]\,3727,3729 of a galaxy,  whereby the [\ion{O}{iii}]\,4959,5007 lines are out of the spectral window or not recognisable due to a disturbance of the spectrum. 

Reliable redshifts (i.e. $r_z \ge 2$) were estimated for altogether 2976 spectra  (75\,\%),  one third of which have
the bad-quality flag set. The low quality is certainly the main reason for incorrect determinations by the SDSS pipeline in these cases. For another part, which is smaller but not insignificant here, special spectral characteristics that are not due to intrinsic properties of the targets probably play a decisive role.  \citet{Flesch_2021} has demonstrated that incomplete de-blending and other effects in the SDSS pipeline lead to problematic or wrong classification results.
Four examples of problematic spectra from the present sample are shown in Fig.\,\ref{fig:special_cases_1}, from top to bottom:\footnote{Here and in all the following plots of spectra, flux means $F_\lambda$ in units of $10^{-17}$\,erg\,s$^{-1}$\,cm$^{-2}$\,\AA$^{-1}$.}
(a) a false peak at $\sim 9300$\,\AA\ that can be misinterpreted as the Ly\,$\alpha$ line, 
(b) an offset between the blue and the red part of the spectrum \citep[see][Fig.\,15]{Bolton_2012}, 
(c) a gap between $\sim 9300$ and 9900\,\AA, 
(d) a glare from a diffraction spike of a bright nearby star (bad-quality flag =  `spike')\footnote{No object is visible at this position in the LS image.}.
Another interesting case is the overlapping of spectra from different objects close to the line of sight. 
Figure\,\ref{fig:special_cases_2} shows two examples:
a galaxy with a star in the foreground (top) and a QSO with an emission-line galaxy in the foreground (bottom). 
These examples illustrate that the spectroscopic pipeline  from the SDSS DR16 tends to classify unusual spectra as high-$z$ QSOs. Conversely, this means that there is a high probability of finding rare, unusual objects among the spectra classified as high-$z$ QSOs.  This is the main focus of the current work (Sect.\,\ref{sect:Rare_types}).

The spectra originate from a total of 2205 spectroscopic plates from the SDSS, BOSS, and eBOSS surveys.
The plates 7260, 7261, and 7262  stand out due to their exceptionally high average detection rate of 198 high-$z$ QSOs per plate, compared with an average of 1.5 for the remaining plates.   These three plates belong to the special  program `Orion-Taurus'.  The fields are located outside the footprint area of the SDSS main survey at low Galactic latitudes ($b < 20\degr$). The associated strong reddening due to dust in our Galaxy is most likely the main reason for a high rate of misidentifications in these special fields.  The re-classification did not confirm  a single high-$z$ QSO on all three plates. Nevertheless, these spectra were included in the catalogue of the present work for the sake of completeness. On plate 7260, 97\%  of the spectra were re-classified as stars.  On plate 7261, 52\% are galaxies and 42\% are stars. On plate 7262, the by far largest proportion (70\%) was identified as galaxies, among them many with very red spectra (Sect.\,\ref{subsect:Red_QSOs}).

\section{Description of the sample}\label{sect:Sample}

\subsection{Composition of the entire sample}\label{sect:Composition}

The results from the re-classification of the 3955 spectra from a total of 3351 sources are summarised in the catalogue (see Appendix\,\ref{sect:Catalogue}).  For the 482 sources with more than one spectrum in the catalogue, a 'best' spectrum was selected, which in most cases, but not always, is the one with the SDSS flag  SciencePriority = 1. The agreement of the revised $z$-values from the double and multiple spectra is generally very good.  A relative deviation $\Delta z/(1+z) > 0.05$ is found for only one object.\footnote{For \object{SDSS\,J043836.52+261941.9} the SQL query results in the two spectra spec-7262-56683-0635 and spec-7262-56659-0629.  The former is an emission line galaxy at $z = 0.062$ whereas the latter indicates a star and is very likely to be assigned to 
\object{SDSS\,J043836.21+261938.3}.}

\begin{figure}[htbp]
\begin{center}
\includegraphics[viewport= 10 0 932 482,width=8.7cm,angle=0]{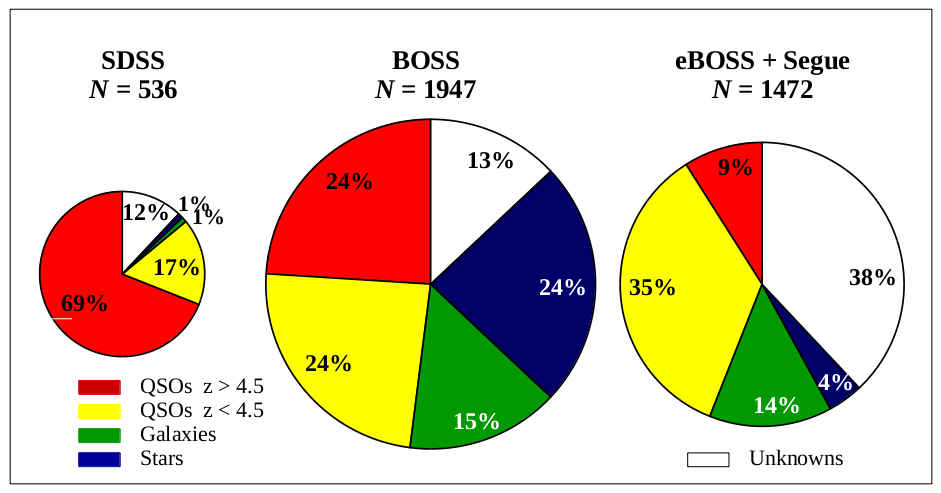} \\
\includegraphics[viewport= 0 0 932 482,width=8.8cm,angle=0]{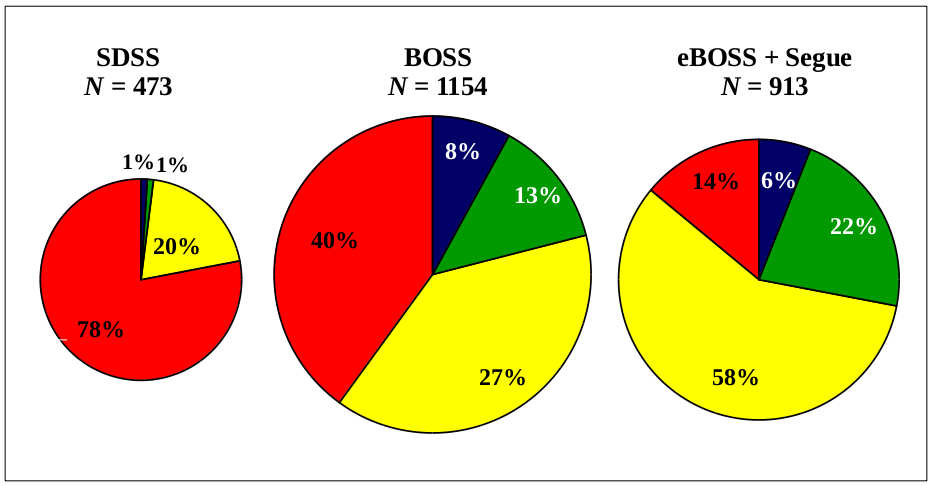} 
\caption{
Top: Revised spectral types per survey for all 3955 spectra. 
Bottom: As above, but only for the 2540 classified spectra of known type (QSO, galaxy or star)  after exclusion of the BOSS plates 7260, 7261, and 7262 (Orion-Taurus fields). 
}
\label{fig:diagram_type_per _survey}
\end{center}
\end{figure}

About half of the spectra was confirmed to belong to QSOs, but only 24\,\% proved to have $z > 4.5$, 28\,\% are QSOs at $z \le 4.5$  (with a mean $\bar{z} = 1.72$; hereafter labelled as `low-$z$ QSOs'),  13\,\% are galaxies, 13\,\%  stars,  22\,\% cannot be classified.  If the non-classifiable spectra (unknowns, $r_z = 0$) and those from the special plates are excluded, 26\,\% of the objects turn out to be high-$z$ QSOs,  50\,\% are low-$z$ QSOs, 16\,\%  galaxies, and 8\,\% stars.  There are clear differences between SDSS, BOSS, and eBOSS (Fig.\,\ref{fig:diagram_type_per _survey}).  The `contamination'  by low-$z$ QSOs and galaxies is highest with eBOSS and lowest with SDSS.

\begin{figure}[htbp]
\begin{center}
\includegraphics[viewport= 20 0 740 740,width=4.4cm,angle=0]{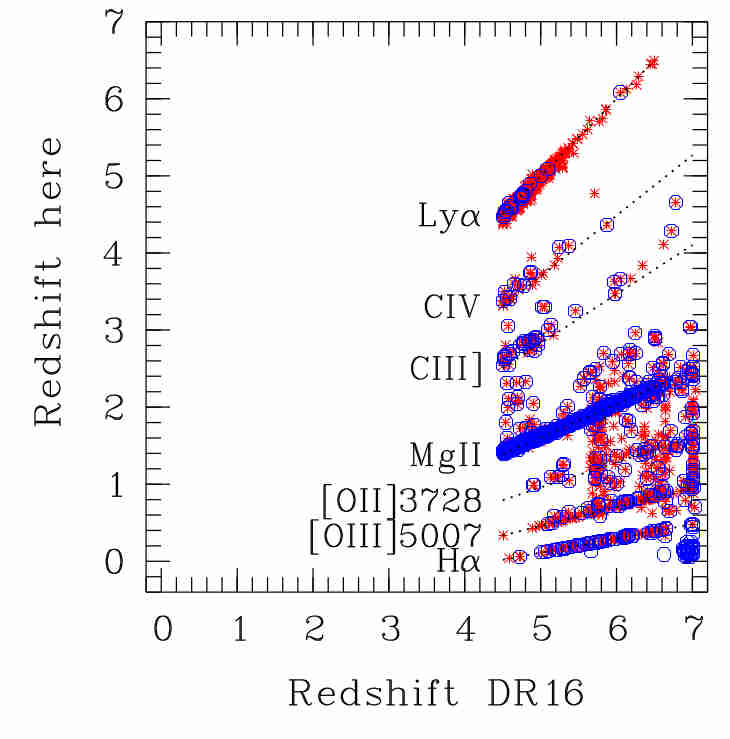} \ 
\includegraphics[viewport= 20 0 740 740,width=4.4cm,angle=0]{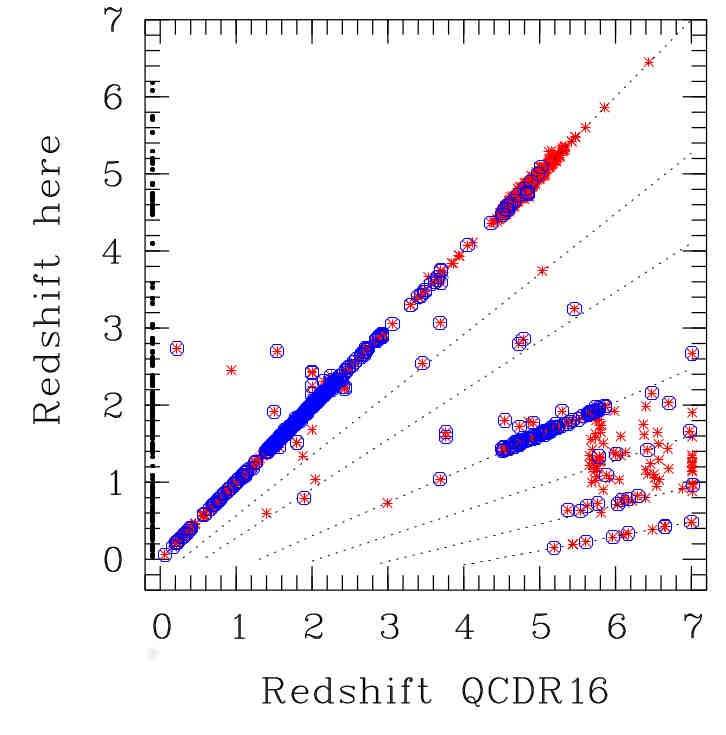} 
\caption{
Revised redshift from the present work versus redshift from the SDSS DR16  pipeline (left) and versus redshift from the QCDR16 \citep{Lyke_2020} (right)  for the QSOs with $r_z\ge 2$.  QSOs with unusual spectra (Sect.\,\ref{subsect:low_z_QSOs}) are marked by blue circles.  Black dotted lines: relations that result when one of the emission lines labelled at the the left-hand side was interpreted as Ly$\alpha$ in the comparison sample. The dots at $z({\rm QCDR16}) <0$ indicate sources that are not included in the QCDR16.
}
\label{fig:zHere_zSDSS}
\end{center}
\end{figure}

In the left panel of Fig.\,\ref{fig:zHere_zSDSS}, the revised $z$ values are compared with the SDSS DR16 redshifts for QSOs with $r_z \ge 2$.  A large proportion of the incorrect $z$-determinations by the pipeline is obviously due to a confusion of  lines. This applies in particular to unusual BAL QSOs where the continuum break due to  \ion{Mg}{ii} and \ion{Fe}{ii} BALs on the blue side of the \ion{Mg}{ii} line shows a certain similarity with the break on the blue side of the Ly\,$\alpha$ line due to the Ly\,$\alpha$ forest  (Sect.\,\ref{sect:FeLoBALs}).  However, it is also worth noting that a remarkable proportion of data points lie between the dotted lines, which indicates more complex spectra. The comparison with the QCDR16 results in a much better agreement (right panel of Fig.\,\ref{fig:zHere_zSDSS}). 
78\,\% of the low-$z$ QSOs  with $r_z \ge 2$ from the present sample have entries in the QCDR16,   where the relative deviation of the $z$-values is less than 0.1  for 79\,\% of them. The present sample contains 216 QSOs that are not listed in the QCDR16.

\subsection{Galaxies and stars}\label{subsect:Rare_QSOs}

\begin{figure}[htbp]
\begin{center}
\includegraphics[viewport= 20 20 590 740,width=6.2cm,angle=270]{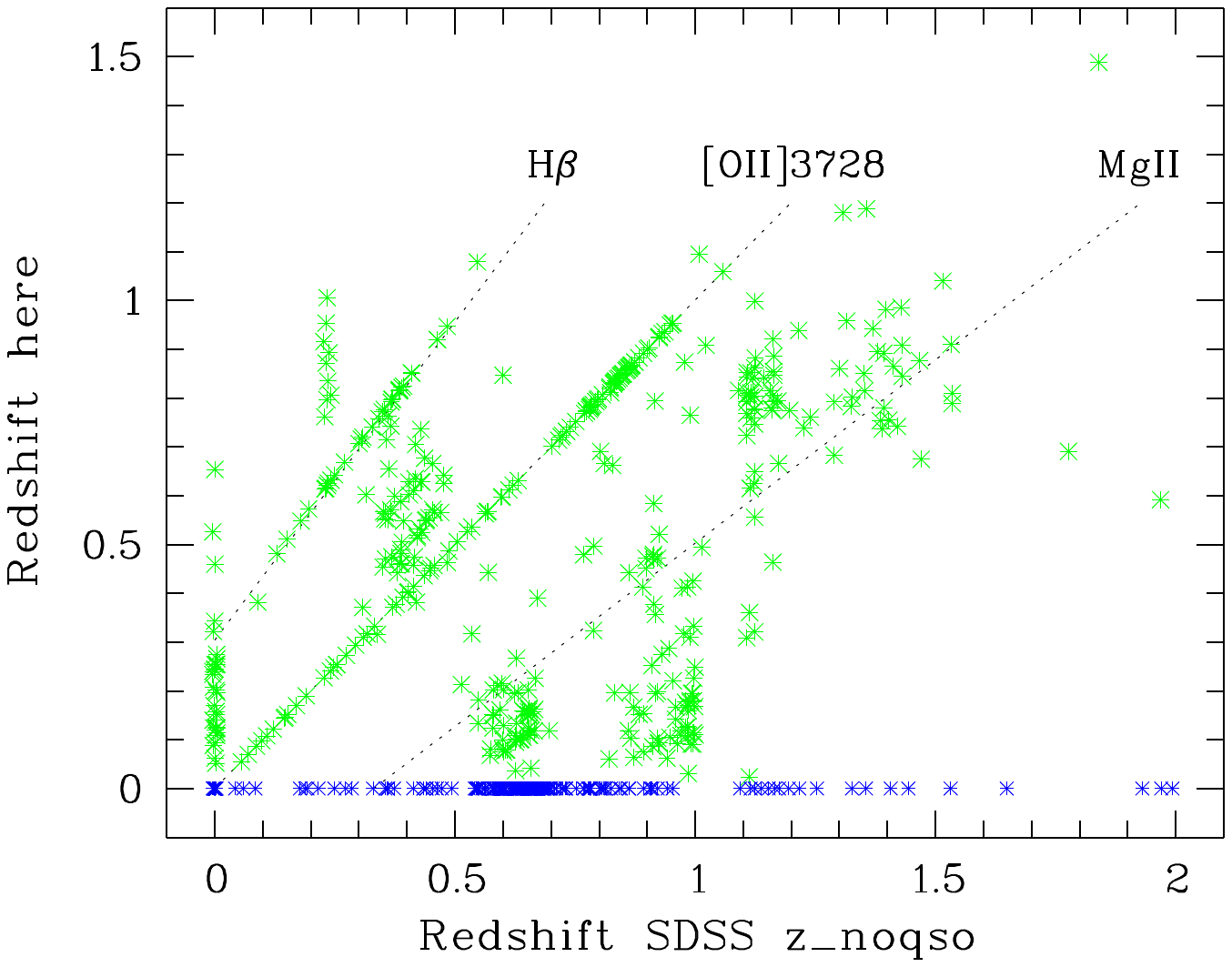} 
\caption{Revised redshift versus  $z_{\rm noqso}$ for the sources with $r_z\ge 2$, which were classified here as galaxies (green)  or stars  (blue).   Black dotted lines: relations that result if the line identified here with  [\ion{O}{ii}]\,3728  was identified by the SDSS with one of the lines labelled at the top.
}
\label{fig:zHere_znoqso}
\end{center}
\end{figure}

The redshifts of the galaxies cover the range from $z = 0.024$ to 1.488, in most cases (75\,\%) measured on emission lines. In addition to the $z$ value based on QSO template spectra, BOSS and eBOSS provide the alternative redshift estimation $z_{\rm noqso}$ from no QSO templates.  This parameter is available for 98\,\% of the stars and for more than 99\,\% of the galaxies with $r_z \ge 2$. Among the galaxies, 97\,\% have $z_{\rm noqso} > 0$.  The agreement with $z$ from the present study is moderate at best (Fig.\,\ref{fig:zHere_znoqso}). For less than half,  the relative deviation is $|z - z_{\rm noqso}|/(1+z) < 0.2$.

In most cases, the classification as a star is based on the detection of typical stellar absorption lines at $z = 0$. 
A total of 539 spectra were assigned to stars, of which 518 have $r_z \ge 2$.  Among those with available $z_{\rm noqso}$, 90\,\%  have  $z_{\rm noqso} = 0$. A subtype was specified for about one third of them.  A remarkably high proportion of 18\,\% are emission line stars,  13\,\% were classified as late-type stars. 15 stars are very likely  white dwarfs (WDs), some of them in a binary system with a late-type star.  Four stars are classified as Extragalactic supernovae (Sect.\,\ref{subsect:SNe}), three of which have $r_z \ge 2$.  A particular surprise was the discovery of possible cyclotron humps in the spectra of five stars (Sect.\,\ref{subsect:Polars}).

\subsection{MIR colours}\label{subsect:MIR}

\begin{figure}[htbp]
\begin{center}
\includegraphics[viewport= 00 00 600 420,width=8.8cm,angle=0]{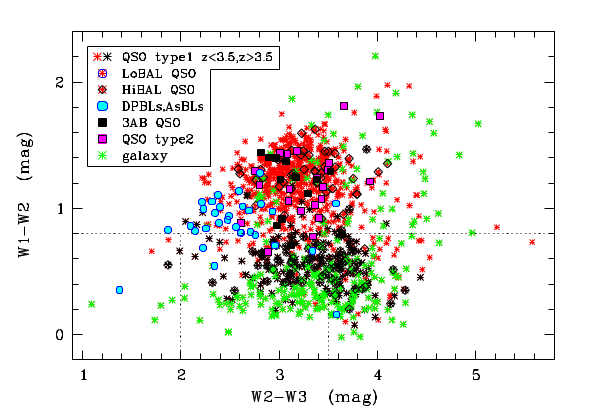} 
\caption{WISE colour-colour  diagram for 1160 QSOs and galaxies. 
The horizontal line marks the AGN threshold from \citet{Stern_2012},
the dotted vertical lines separate the regions of passive (left), intermediate (middle) and  star-forming galaxies (right) \citep{Jarrett_2017}.  The subtypes mentioned in the insert are explained in Sect.\,\ref{sect:Rare_types}.
}
\label{fig:WISE_CCD}
\end{center}
\end{figure}

The WISE MIR colours  provide a useful tool, with some cautions, for selecting AGNs and distinguishing AGNs from 
star-formation activity \citep[e.g.][]{Stern_2012, Jarrett_2017}.  Figure\,\ref{fig:WISE_CCD} shows the two-colour diagram based on the magnitudes from the AllWISE data release \citep{Cutri_2021}  in the first three WISE bands at 3.4, 4.6, and 12 $\mu$m. Only sources with errors $< 0.5$\,mag for $W3$ and $< 0.2$\ mag for $W1$ and $W2$ are plotted. About three quarters of the galaxies populate the galaxy locus at $W1-W2 < 0.8$,  about 15\%  are located in the upper right corner at $W1-W2 > 0.8$ and $W2-W3 > 3.5$ and could be LINERs or ULIRGs \citep[see][]{Jarrett_2017}.  A part of the galaxies that lie above the demarcation line host possibly AGNs of type 2.  QSOs with $z \la 3$ are concentrated in the `AGN region'  centred at  $W1-W2 \approx 1.3$ and $W2-W3 \approx 3.5$,  whereas QSOs at higher redshift are found at  $W1-W2 < 0.8$ (as can be seen also in Fig.\,\ref{fig:colour_z_QSOs}, bottom).

\subsection{FIRST radio counterparts}\label{subsect:FIRST}

A cross-match with the catalogue from the Faint Images of the Radio Sky at Twenty-centimeters (FIRST) survey \citep{Helfand_2015}  results in a positional coincidence of less than $3\arcsec$ for 13\,\% of the QSOs in the present sample.  For comparison, \citet{Lyke_2020} found matches for 3\,\% among the QSOs in the QCDR16.   Only 7\,\% (33/498) of the QSOs with revised redshifts $z > 4.5$ (and $r_z \ge 2$)  were identified with FIRST sources.  The highest redshift for a QSO detected by FIRST is $z = 6.291$ for \object{SDSS\,J161425.13+464028.9}, but this value is probably due to a misinterpretation of a false peak at 8850\,\AA\  with the Ly\,$\alpha$ line (as in the top panel of Fig.\,\ref{fig:special_cases_1}).  The highest reliable redshift of a QSO with FIRST detection is $z = 5.308$ for \object{SDSS\,J161425.13+464028.9}.

\subsection{General remarks on the low-$z$ QSO sample}\label{subsect:low_z_QSOs}

In the following, the sample of objects re-classified as QSOs at $z\le4.5$, i.e. below the original selection threshold for high-$z$ QSOs, is referred to as the `low-$z$ QSO sample'.   It is the main source of the unusual spectra presented in the next Section.  Table\,\ref{tab:peculiars} provides an overview of the frequency of different rare spectral  types among the low-$z$ QSOs. Remarkably, the 10 peculiarity types contain 94\,\% of this sample.

\begin{table}[htbp]
\caption{Numbers of QSOs  of ten different unusual types (see Sect.\,\ref{sect:Rare_types}) among the low-$z$ QSOs with $r_z \ge 2$.}
\begin{tabular}{lr} 
\hline\hline 
\noalign{\smallskip}
Type          &   Number   \\
\hline 
\noalign{\smallskip}
All  &  948  \\ 
\hline 
\noalign{\smallskip}
Noticeable red continuum (without BAL QSOs) & 247 \\
FeLoBALs & 140 \\
LoBALs & 104 \\
HiBALs & 85 \\
Strong associated absorption lines & 184 \\
3000\,\AA\ break & 12 \\
Type 2  &  28 \\
WLQs  & 29 \\
Double-peaked or asymmetric broad Balmer lines & 34 \\
Strong Fe emission & 25 \\
\hline                      
\end{tabular}
\label{tab:peculiars}                    
\end{table}

\begin{figure}[htbp]
\begin{center}
\includegraphics[viewport= 00 00 650 450,width=8.8cm,angle=0]{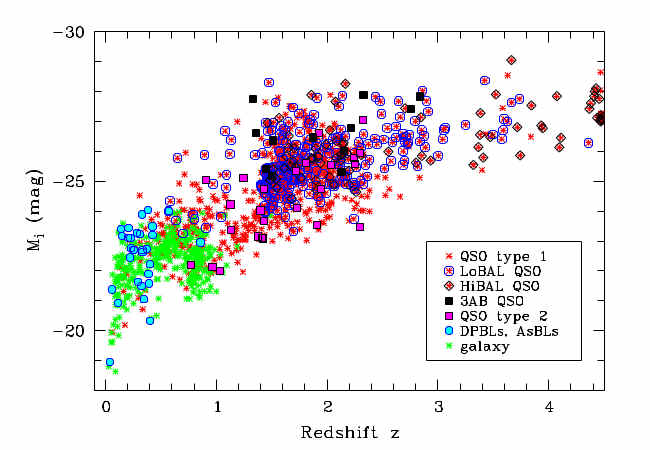}
\caption{Absolute i-band magnitude versus redshift for the 1245 sources with reliable ($r_z \ge 2$) redshifts $z \le 4.5$ and available i band magnitude.  
}
\label{fig:Mi_z}
\end{center}
\end{figure}

Sometimes a brightness threshold is used to distinguish QSOs from AGNs with lower luminosity,  e.g. absolute  i band magnitude $M_i < -22$ \citep{Schneider_2010, Shen_2011}.   Figure\,\ref{fig:Mi_z} shows the $M_i - z$ relation for the present sample.  $M_i$ was computed following \citet{Kennefick_2008},  the i band magnitudes were corrected for Galactic extinction and a standard continuum $K$ correction was applied.  As expected, the sources categorised as QSOs have bright magnitudes and follow the relation for a sample combined from surveys of different flux limits. 

17\,\% of the low-$z$ QSOs  are positionally coincident with FIRST radio sources within 3 arcsec.  Following \citet[][Eq. 5]{Ivezic_2002}, the ratio of the radio to optical flux density in the i band  is used here to compute the  (logarithmic) estimate of the radio loudness $R_i$.  The condition $R_i > 1$, i.e. radio-loud (RL), is fulfilled by 14\,\% of the low-$z$ QSOs. The relatively high radio detection fraction of the misclassified QSOs  is unlikely due to the fact that SDSS targeted radio sources for spectroscopy even when they were not selected as QSO candidates because of their optical colours (Sect.\,\ref{sect:Discussion}).

\section{Rare and unusual spectral types}\label{sect:Rare_types}

For most of the unusual types presented in this Section, a  selection of spectra is presented in Sect.\,\ref{sect:spectra}, where each panel shows the flux $F_{\lambda}$ as a function of the wavelength $\lambda$ in the observer's frame. 
The names of the object and of  the spectrum as well as the redshift are  given at the top of each panel.  Vertical lines indicate the positions of the emission lines (red) and absorption lines (blue) that typically occur in QSO spectra.

\subsection{Low-$z$ QSOs}\label{subsect:rare_lowz_QSOs}

\subsubsection{Inverted or flat UV continuum}\label{subsect:Red_QSOs}

The  optical and UV continuum of the SDSS QSOs can be approximated by a power law $F_{\lambda} \propto \lambda^{\alpha}$ with $\rm \alpha = -1.54$  \citep{VandenBerk_2001}. In addition to broad emission lines and high luminosity,  this blue continuum is one of the defining characteristics of type 1 QSOs.   A small population of type 1 QSOs however, shows red continua.  It has been suggested that the red continuum is due to dust extinction in the host galaxy and  that red QSOs represent an intermediate evolution stage between merger-driven star-forming galaxies and unobscured QSOs of type 1  \citep[e.g.][]{Hopkins_2005, Kim_2018, CalistroRivera_2024, Fawcett_2021, Fawcett_2023, Fawcett_2025}. 

\begin{figure}[htbp]
\begin{center}
\includegraphics[viewport= 00 00 602 417,width=8.7cm,angle=0]{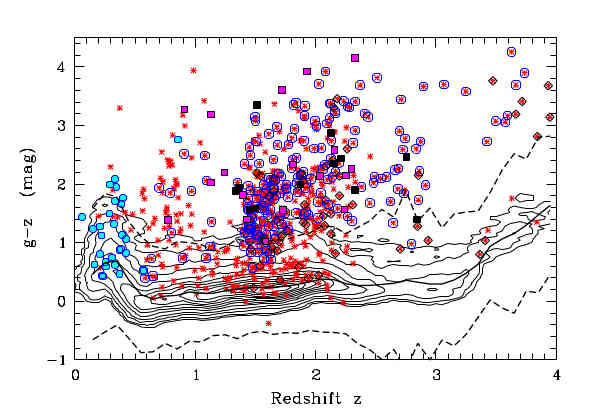} \\
\includegraphics[viewport= 00 00 602 417,width=8.7cm,angle=0]{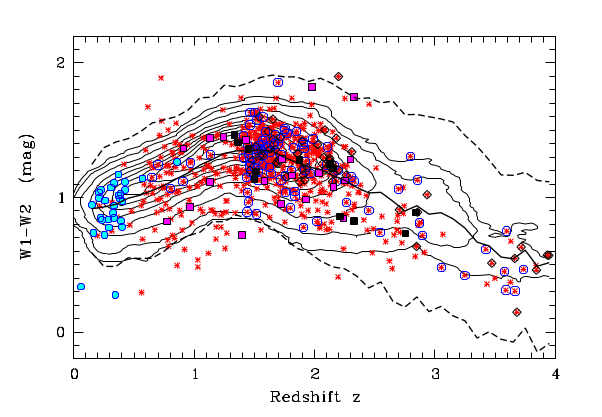} 
\caption{
Colour-redshift diagrams of the QSOs with $z < 4$ for photometric data from SDSS (top) and WISE (bottom). 
The symbols have the same meaning as in Fig.\,\ref{fig:Mi_z}. The SDSS colours were corrected for Galactic foreground extinction.  The MIR magnitudes $W1$ and $W2$ are from unWISE \citep{Schlafly_2019}.
The thin black contour lines show the population density of the QSOs from the SDSS DR7 QSO catalogue \citep{Shen_2011},  the thick black lines mark the median relation (solid) and the 3$\sigma$ deviation from the median (dashed). }
\label{fig:colour_z_QSOs}
\end{center}
\end{figure}

One of the striking features of the present sample is the high proportion of QSOs with flat or inverted spectra. About a quarter of the low-$z$ QSOs were categorised as `red'.  Including the BAL QSOs, which generally have redder colours than normal QSOs, it can be said that around 60\,\% have red spectra (Table\,\ref{tab:peculiars}).  In the top panel of Fig.\,\ref{fig:colour_z_QSOs}, the colour index $g-z$ is used to objectify the subjective assessment from the inspection of the spectra.  For comparison, the median $\mu(g-z)$  and its standard deviation $\sigma_{g-z}$ of the QSOs from the SDSS DR7 QSO catalogue \citep{Shen_2011} are plotted.  It turns out that about 68\,\% of the low-$z$ QSOs from the present sample are more than $3\sigma_{g-z}$ above the median and are therefore categorised as significantly red, of which  about half were classified as BAL QSOs.  In the MIR, on the other hand, the vast majority of the QSOs are close to the median relation  (Fig.\,\ref{fig:colour_z_QSOs}, bottom). The difference between the colour excess in the MIR and the optical  fits in with the interpretation that the emission  from the accretion disk is extinguished in the rest-frame UV.

\begin{figure}[htbp]
\begin{center}
\includegraphics[viewport= 00 00 670 502,width=8.8cm,angle=0]{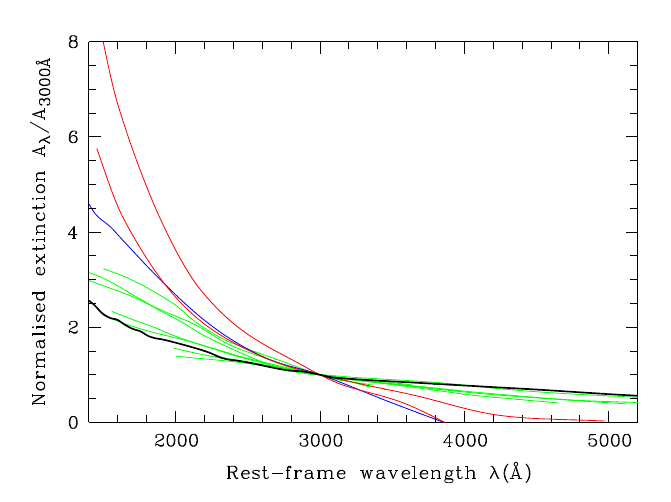} 
\caption{Normalised extinction $A_{\lambda}/A_{\rm 3000\,\AA}$ for nine red QSOs from Fig.\,\ref{fig:reddest_NonBALQSOs}  (green), two LoBAL QSOs (red), and one HiBAL QSO (blue). For comparison, the SMC EC is plotted in black.}
\label{fig:ECs}
\end{center}
\end{figure}

Figure\,\ref{fig:reddest_NonBALQSOs} shows the multi-band SEDs (MBSEDs)  for the ten non-BAL QSOs
with the highest relative colour excess $[(g-z) - \mu(g-z)]/\sigma_{g-z} \ga 10$. The highest value is measured for the type 2 QSO \object{SDSS\,J110511.15+530806.5}.  For comparison, the template spectrum of type 1 QSOs from \citet{Polletta_2007} is plotted, individually adjusted in the rest-frame near-infrared.  If it is assumed that the adjusted template corresponds to the intrinsic, unreddened QSO spectrum, the normalised extinction curve (EC) can be determined.  Figure\,\ref{fig:ECs} suggests that these ECs are steep in the UV, similar to that of the Small Magellanic Cloud (SMC) or even steeper. 

The evolution scenarios mentioned above also suggest  a causal link between  dust reddening and radio emission.
The present sample clearly confirms the earlier  finding \citep[e.g.][]{Ivezic_2002} that RL QSOs have redder median optical colours  and a larger proportion of objects with extremely red colours than normal QSOs.  The sub-sample of RL QSOs has an even higher proportion of red QSOs (84\,\%) than the parent sample (68\,\%). The mean distance from $\mu(g-z)$ is $1.64 \pm 0.09$\,mag for the RL QSOs, compared with  $1.29 \pm 0.03$\,mag for the QSOs without FIRST detection.   Conversely, the proportion of RL sources among the red QSOs is higher (18\,\%) then in the parent sample (14\,\%).

\subsubsection{LoBALs and FeLoBALs}\label{sect:FeLoBALs}

The physical interpretation of blue-shifted BALs in QSO spectra is still poorly understood, although the phenomenon 
has been known for a long time \citep{Lynds_1967}.  There is broad consensus that  BALs are  caused by powerful, sub-relativistic outflows \citep{Weymann_1991},  possibly  related to radiation-driven winds from the accretion disk. 
BAL QSOs  are usually divided into those that only show BALs of highly ionised atoms such as \ion{C}{iv}, \ion{Si}{iv}, \ion{N}{v}, \ion{O}{vi} (HiBALs)  and those that additionally show BALs of lower-ionisation ions such as \ion{Mg}{ii}, \ion{Al}{ii}, \ion{Al}{iii} (LoBALs) in the UV. A small number of QSOs additionally show  low-ionisation BALs from meta-stable exited stages of \ion{Fe}{ii} and/or \ion{Fe}{iii}.   These FeLoBAL QSOs are rare in optically selected QSO samples, it is assumed that only about 0.3\% of QSOs are of this type \citep{Trump_2006},
but the intrinsic fraction can be much higher \citep{Dai_2012}. BAL QSOs are probably best explained by a combination of orientation and evolution \citep{Choi_2022, Nair_2022} where  it is suggested that LoBALs and FeLoBALs represent an evolutionary stage when a dusty cocoon is being expelled by a powerful wind 
\citep[e.g.][]{Voit_1993, Yi_2022, Leighly_2024}.

It has long been known  that BALs can mimic high-$z$ QSOs \citep[e.g.][]{Appenzeller_2005}, which is nicely illustrated by  the high percentage of BAL QSOs in the `low-$z$ QSO sample' of the present study (Table\,\ref{tab:peculiars}): 35\,\% are classified as BALs, 15\,\% as FeLoBALs.    Overall, more than 50\,\% of the low-$z$ QSOs show strong absorption lines,  which were classified either as BALs if they are broad and blueshifted relative to the emission lines or as associated absorption lines (AALs)  when they are narrow and not significantly blueshifted. 
In the literature, a ‘BALnicity’ index  \citep[BI,][]{Weymann_1991} or an intrinsic absorption index \citep[AI,][]{Hall_2002} is often used to formally distinguish between BALs and AALs. However, the computation of theses indices is beyond the scope of the present paper.  It must also be mentioned again here that many spectra have a low signal-to-noise ratio, which makes their calculation difficult and very uncertain. 

LoBAL and FeLoBAL QSOs are known to be significantly redder in the UV than ordinary QSOs  \citep{Sprayberry_1992, Reichard_2003, Meusinger_2012, Peng_2024}.   SMC-like reddening by BAL dust was reported by \citet{Gaskell_2024}.  Reddening curves steeper than in the SMC were suggested for some samples of individual QSOs \citep[e.g.][]{Hall_2002, Jiang_2013, Leighly_2014, Leighly_2024}.  Figure\,\ref{fig:ECs} shows the normalised ECs of two LoBAL QSOs  (\object{SDSS\,J151131.03+431730.5} and \object{SDSS\,J134927.72+305251.0})  and one HiBAL QSO  (SDSS\,J125316.41+521042.8), whose continuum in the UV appears to be strongly inverted. 

\begin{figure}[htbp]
\begin{center}
\includegraphics[viewport= 20 20 577 422,width=5.5cm,angle=0]{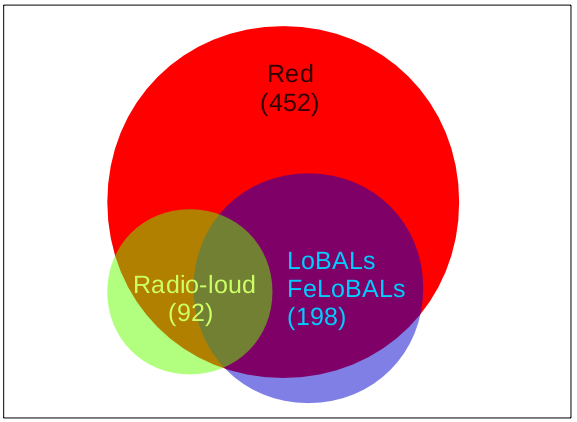} 
\vspace{0.5cm}
\caption{Venn diagram of the three subsamples of red  (red), LoBALs or FeLoBALs (blue), and RL QSOs (green).}
\label{fig:Venn}
\end{center}
\end{figure}

A remarkably large number of 53 LoBAL or FeLoBAL QSOs from the present sample with $r_z \ge 2$ are positionally coincident with FIRST radio sources, corresponding to a high radio detection fraction of 22\,\%.  
Among them are 44 RL QSOs ($R_i > 1$).  The Venn diagram in Fig.\,\ref{fig:Venn} illustrates the relative sizes and the overlap of the three subsamples of  UV-red QSOs, LoBAL and FeLoBAL QSOs, and RL QSOs.\footnote{The Venn diagram was created with BioVenn under http://www.biovenn.nl/.}  The subsamples were selected from the parent sample of the QSOs with  $z<4$ and photometric errors in the g and z bands less than 0.2 mag.

A selection of unusual FeLoBAL QSO spectra of good quality is shown in Fig.\,\ref{fig:LoBALs}. Some spectra have very broad overlapping troughs \citep{Hall_2002}.  In extreme cases, e.g. \object{SDSS\,J094317.61+541704.9}  \citep{Urrutia_2009, Meusinger_2012},  the continuum emission shortward of the \ion{Mg}{ii} line is nearly completely absorbed.  Such objects are expected to have particularly large column densities in their outflows. 
Others show many narrow troughs. In some cases, both the redshift of the emission lines and that of the absorption lines could be measured.  In other cases, only the  absorption redshift is available, where the red edge of the BAL trough was usually identified with the position of the corresponding line.  FeLoBAL spectra can be very complex,  for a more detailed discussion of similar spectra, see e.g. \citet{Hall_2002, Wang_2017, Choi_2020}. A thorough analysis of BAL spectra requires sophisticated tools such as SimBAL \citep{Leighly_2018}.

\subsubsection{3000\,\AA\ break}\label{sect:3ABQs}

The prototypes of this enigmatic and extreme rare class are two SDSS QSOs at $z=0.5$ and 1.353 referred to as  `mysterious objects'  by \citet{Hall_2002}.  The spectra are characterised by  a lack of typical UV emission lines, except broad \ion{Fe}{ii} emission and the [\ion{O}{ii}]\,3727,3729 doublet in some cases, 
in combination with a blue continuum at rest-frame wavelengths $\ga 3000$\,\AA\ and an inverted continuum at shorter wavelengths. The drop-off at $\sim 3000$\,\AA\ appears to be too steep to be explained by dust reddening with typical reddening laws  and does not clearly show the typical appearance of strong BALs.   The physical causes of these properties remain unclear.  Strong line absorption from gas clouds with broad smooth distributions of outflow velocities,  perhaps in combination with moderate dust reddening, seems to be the most convincing explanation
\citep[][]{Hall_2002, Green_2006, Meusinger_2016, Yi_2022}.

Using SOMs of about one million SDSS spectra, \citet{Meusinger_2016} selected 23 such ‘3000\,\AA\ Break QSOs’ (3ABQs, their sample A)  and a further 15 QSOs with spectra that resemble these to a certain extent, but not in all aspects (their sample B).  They concluded that 3ABQs are most likely extreme versions of FeLoBAL QSOs.  In the present study, an unexpectedly large number of 12 QSOs (15 spectra)  were classified as 3ABQs or similar to 3ABQs (Fig.\,\ref{fig:3ABQs}),  among them one  of the two prototypes from \citet{Hall_2002}, \object{SDSS\,J220445.26+003141.9},  four sources from \citet{Meusinger_2016}, and  one source that was classified as unusual BAL QSO by \citet{Meusinger_2012}. The other six 3ABQs were not yet mentioned explicitly in the literature.

When searching for 3ABQs, there is a risk of contamination by two other types of spectra.  On the one hand, a similar spectrum arises for a faint source that lies on a diffraction strike of a nearby bright star (bottom panel of Fig.\,\ref{fig:special_cases_1}).  Such a contamination could be easily ruled out by the inspection of the SDSS images. 
On the other hand, the spectra of peculiar DQ WDs show an abrupt drop at $\sim 6000$\,\AA\, which makes them similar to 3ABQs at $z \sim1$ in some respects \citep[see][]{Meusinger_2016}.  This possibility can be ruled out here. With only one exception, all 3ABQs are listed in the Gaia DR3 catalogue with a proper motion which is essentially zero (less than 3$\sigma$). The exception is \object{SDSS\,J081659.11+024715.5},  for which the redshift $z=1.500$ is certainly known from the [\ion{O}{ii}]\,3727,3729 doublet and the narrow \ion{Mg}{ii} absorption line.  The WISE colours of all 3ABQs match their classification as QSOs.

None of the 12 3ABOs is listed in the 5th edition of the Roma-BZCAT Multifrequency Catalogue of Blazars \citep{Massaro_2015}.  Four have FIRST radio counterparts, all four are point sources.  Two of them, \object{SDSS\,J081659.11+024715.5}  and \object{SDSS\,J145045.55+461504.2}, are RL.   For the latter, the flux density in the band  120 to 168\,MHz  is given in the  LoTSS catalogue  \citep{Shimwell_2019}. The combination with the flux density from FIRST   results in a steep radio spectrum with  $\alpha = 0.55$ ($S_\nu \propto \nu^{-\alpha}$).

\subsubsection{Weak emission lines}

QSOs with abnormally weak emission lines in the UV form the rare and puzzling subclass of WLQs.
A number of different scenarios were proposed,  with the idea of a shielding gas component between the central X-ray source and the broad emission line region currently  appearing particularly attractive  \citep[e.g.][and references therein]{Paul_2022}.   Based on optical variability data, \citet{Kumar_2025} suggest that the clumpiness of the torus material flowing into the central engine may play a key role.  Although the first WLQs were discovered three decades ago \citep{McDowell_1995},  it was not until the SDSS that statistical analyses based on larger samples became possible \citep[e.g.][]{Fan_1999, Diamond_2009, Plotkin_2010, Meusinger_2012}.   A sample of 365 WLQ candidates has been selected from Kohonen SOMs of SDSS DR7 spectra  \citep{Meusinger_2014},  including 46 WLQs that fulfil the strict criterion of equivalent widths for \ion{Mg}{ii}  and  \ion{C}{iv} smaller than $3\sigma$ below the median of SDSS QSOs.  

In the present study, 42 QSOs were postulated as WLQ candidates (Fig.\,\ref{fig:WLQs}). None of them  is listed as a blazar by \citet{Massaro_2015},  only four have a radio counterpart in FIRST, and only two are RL. The overlap with the WLQ sample from \citet{Meusinger_2014} contains only four objects.   Of the 13 WLQ candidates with $z>4.5$, nine belong to  the high-$z$ SDSS DR5 QSOs selected by \citet{Diamond_2009}, but only three of them were classified there as WLQs.  As with the other unusual QSOs in this paper, the classification should be considered provisional as the method is purely subjective,  but also because many spectra are noisy and BALs sometimes cannot be completely ruled out.  It must again be pointed out that a more thorough analysis is beyond the scope of this paper.

\subsubsection{Double-peaked broad Balmer lines}\label{sect:DPBLs}

28 spectra of 26 QSOs show the unusual property of double shoulders in the broad Balmer lines (Fig.\,\ref{fig:dp_broad_Balmer}).  A literature search reveals that 65\,\% of them were not previously recognised as such.
Such objects, which are rare among the SDSS QSOs, are usually referred to as   `double-peaked' AGNs \citep{Eracleous_1994, Strateva_2003}.  Among the 26 QSOs of this type, 22 are at $z < 0.5$, where the H$\alpha$ line falls within the spectral window of the SDSS.  This number corresponds to an astonishingly high share of 35\% of the QSOs with $ z < 0.5$  in the present sample.  A further 8 QSOs stood out due to a probably related phenomenon of strongly asymmetrical profiles of the broad Balmer lines.

\begin{figure}[htbp]
\begin{center}
\includegraphics[viewport= 0 0 530 605,width=4.3cm,angle=0]{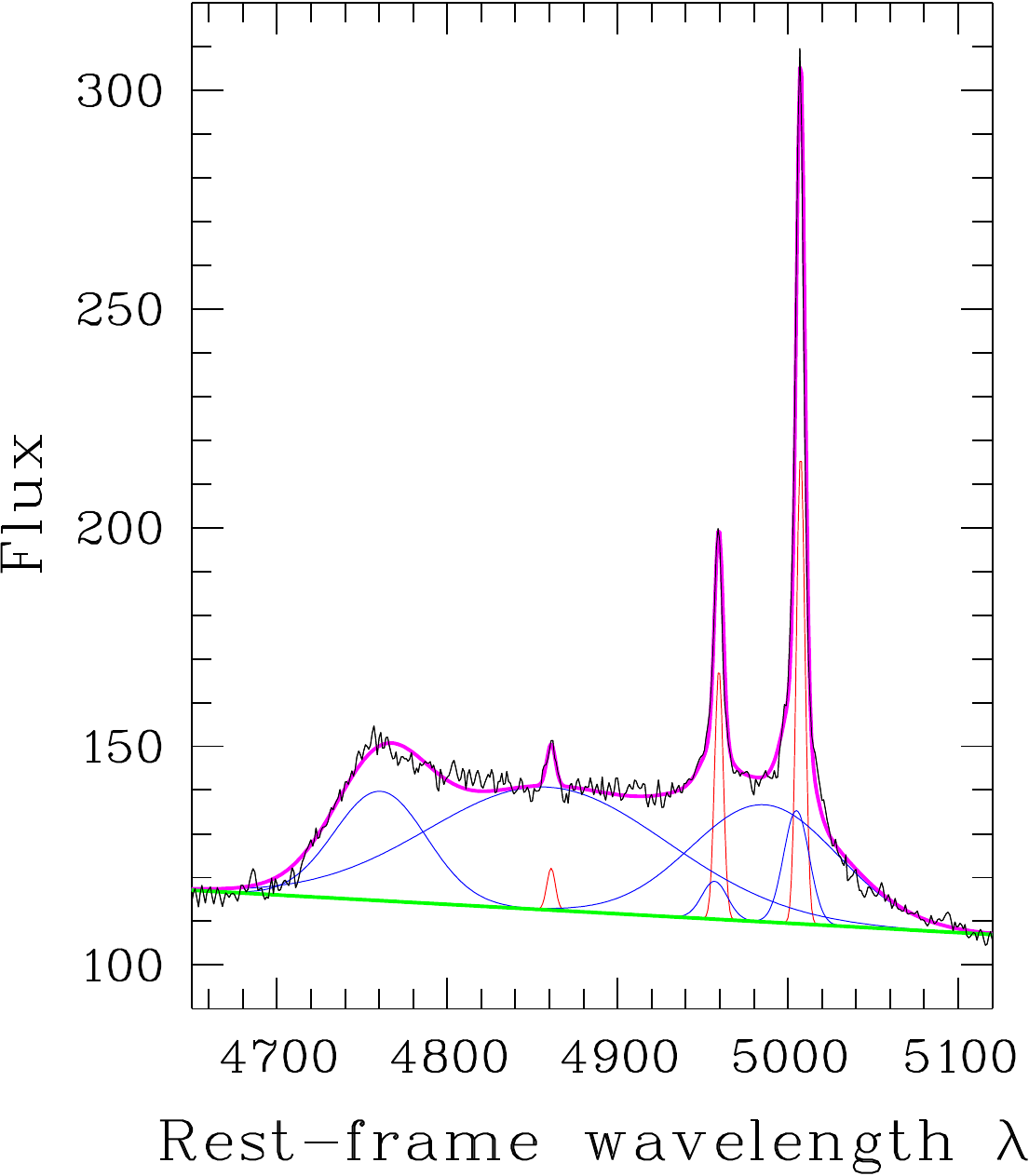}   
\includegraphics[viewport= 0 0 530 605,width=4.3cm,angle=0]{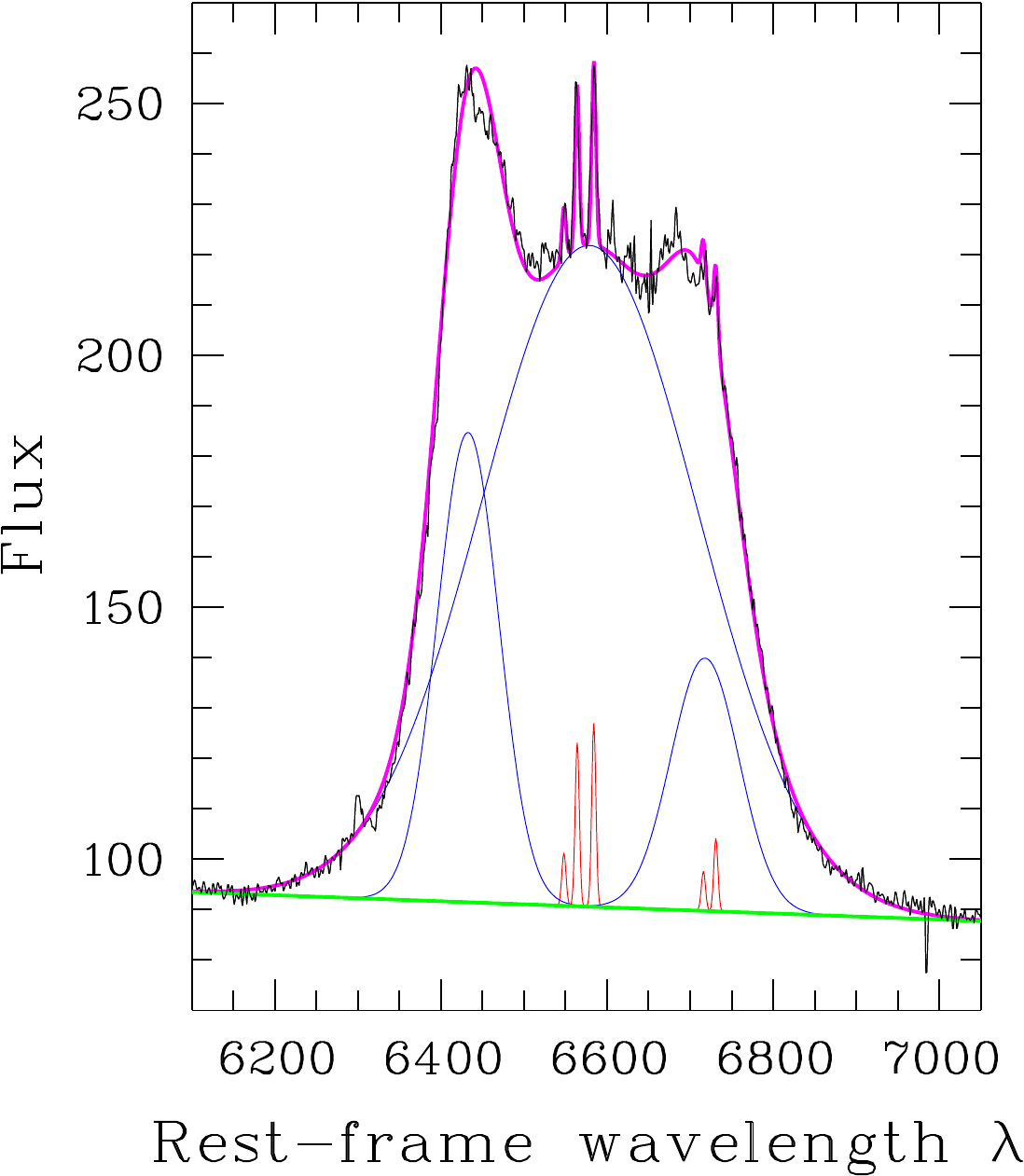}   \\
\vspace{0.3cm}
\includegraphics[viewport= 0 0 510 605,width=4.1cm,angle=0]{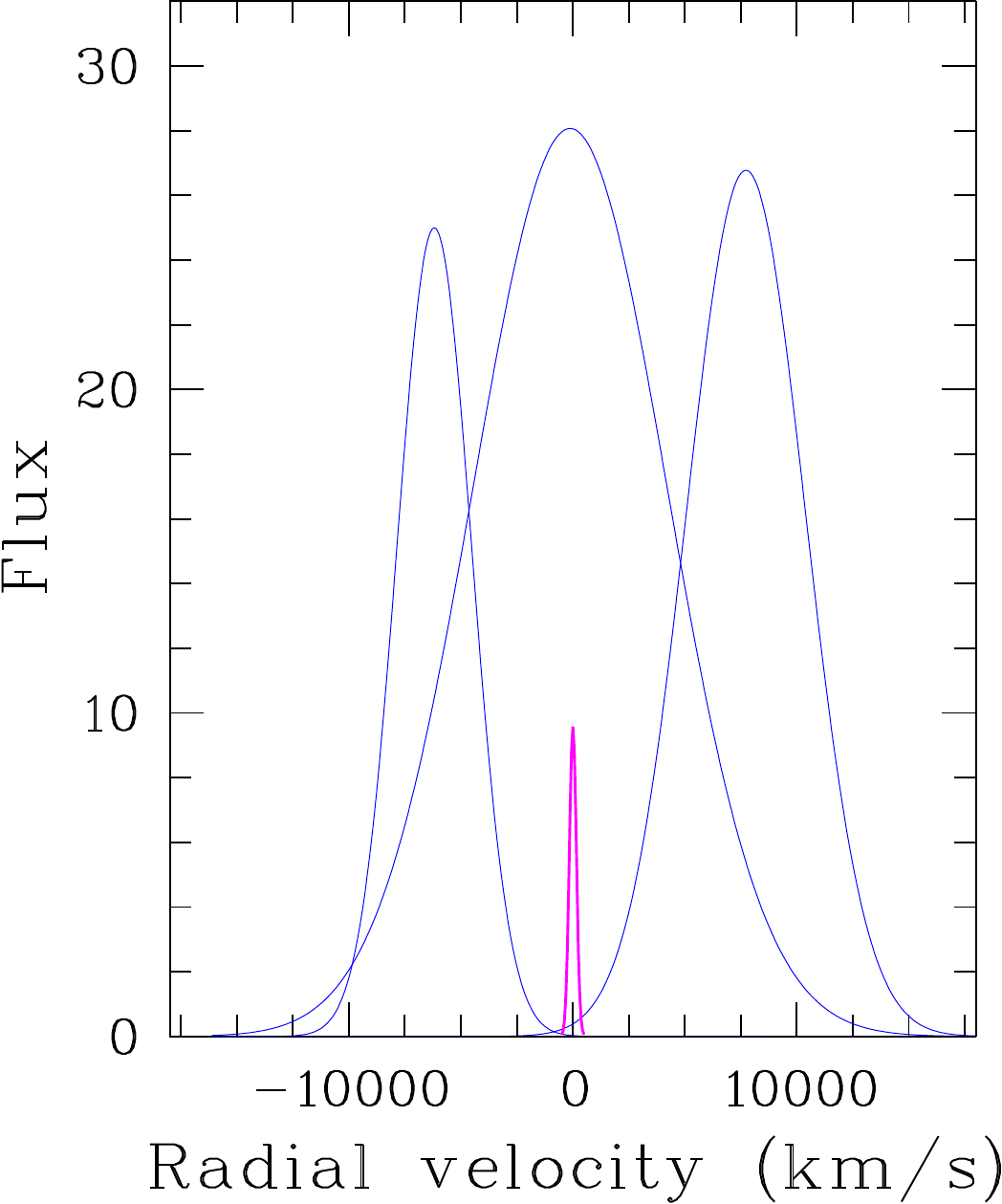}   
\includegraphics[viewport= 0 0 510 605,width=4.1cm,angle=0]{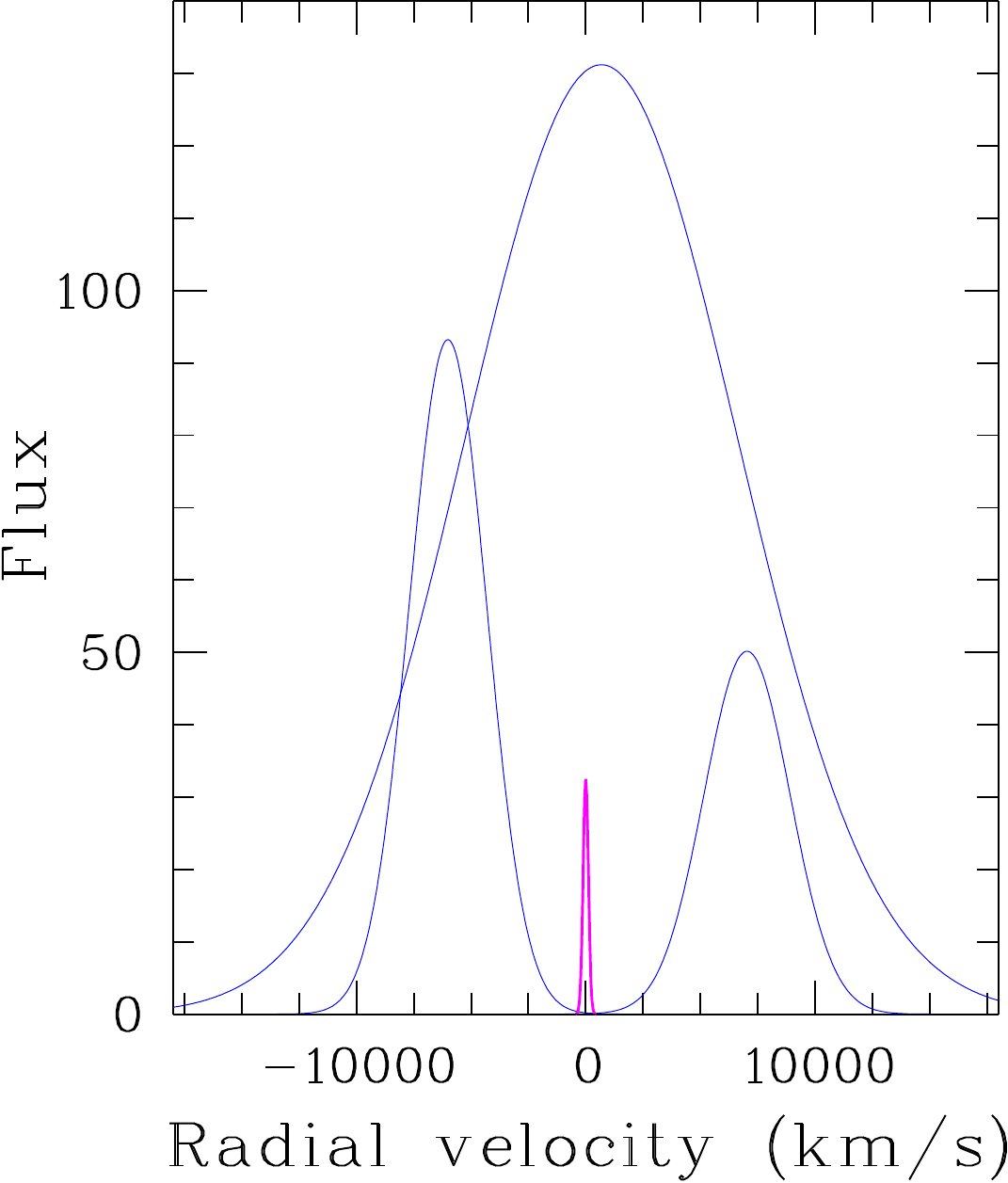}   \\
 \vspace{0.3cm}
\caption{Top: best model fit of the regions around  H$\beta$  (left) and H$\alpha$ (right) of \object{SDSS\,J161742.53+322234.4}. The fit of the  H$\beta$-[\ion{O}{iii}] complex includes eight Gaussian emission components:  three narrow lines (red), three broad H$\beta$ lines  (blue), and two broad [\ion{O}{iii}]  lines (blue).  Eight Gaussian components were also included for the H$\alpha$-[\ion{N}{ii}]-[\ion{S}{ii}] complex: five narrow lines (red) and three broad H$\alpha$ lines (blue).  A first-order polynomial for the continuum is shown in green,  the resulting composed model spectra in magenta, and the observed spectrum in black. Bottom: the line profiles for the H$\beta$  (left) and the H$\alpha$ (right) complexes  converted into relative velocities.}
\label{fig:Ha_J1617}
\end{center}
\end{figure}

Several explanations have been proposed and discussed for the double-peaked line profiles, including supermassive black hole binaries (SMBHBs),  bipolar outflows,  elliptical accretion disks caused by tidal disruption events, and unisotropically illuminated broad-line regions  \citep[e.g.,][]{Eracleous_2009, Liu_2016, Terwel_2022, Wevers_2022}.  The SMBHB model is no longer favoured  based on several reasons \citep[][and references therein]{Doan_2020, DeRosa_2019}.

Figure\,\ref{fig:Ha_J1617} shows that the line complexes around  H$\alpha$ and H$\beta$ of \object{SDSS\,J161742.53+322234.4} ($z = 0.151$)  are well described by a model with three broad Balmer components, one redshifted relative to the narrow lines, one blueshifted, and one at the redshift of the narrow lines. The velocity differences of the shifted lines relative to the non-shifted one is  $\sim 8000$\,km\,s$^{-1}$.  A similar result was found for other sources from this subsample. They are therefore actually `triple-peaked' emitters. 
Surprisingly, nine of the ten  QSOs in Fig.\,\ref{fig:dp_broad_Balmer} are associated with FIRST radio sources,  
none of which were selected for SDSS spectroscopy for this reason (Sect.\,\ref{sect:Discussion}).  The phenomenon of this high proportion of radio detections is the subject of a separate study based on a larger sample of  QSOs with double-peaked broad Balmer lines (Meusinger \& Andernach, in prep.).

\begin{figure}[htbp]
\begin{center}
\includegraphics[viewport= 00 00 608 450,width=9.0cm,angle=0]{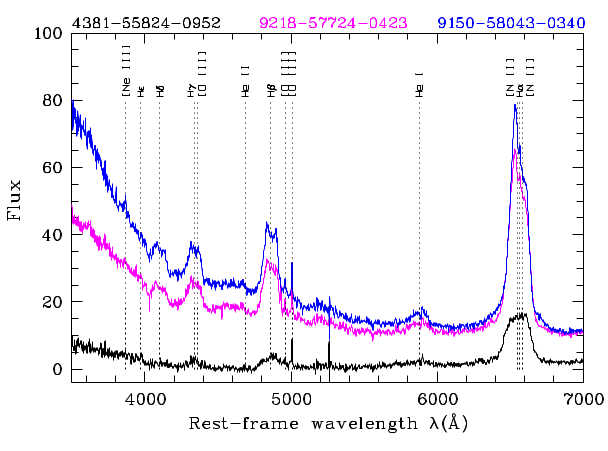} 
\caption{Spectra of the CLQ \object{SDSS\,J224113.54-012108.8}  at three different epochs after subtraction of an early-type galaxy template for the host galaxy. 
The plate-MJD-fibre designations are given at the top.}
\label{fig:J224113_3spectra}
\end{center}
\end{figure}

A particularly interesting case is \object{SDSS\,J224113.54-012108.8} (Fig.\,\ref{fig:dp_broad_Balmer}, top left), 
which has been classified as a CLQ  \citep{Green_2022, Panda_2024}.  In Fig.\,\ref{fig:J224113_3spectra}, the BOSS spectrum from the DR16 is compared with two eBOSS spectra of the same source from SDSS\,DR17 \citep{Abdurro_22}.  The profiles of the broad Balmer lines are variable on time scales of a few years. 
The UV continuum strongly increases with time, in agreement with the classification as `TurnOn' CLQ by \citet{Green_2022}.  In the WISE colour-colour diagram (Fig.\,\ref{fig:WISE_CCD}),
this source is surprisingly represented by a data point at ($W1-W2, W2-W3$) = (0.35, 1.37), i.e. in the region of passive galaxies (spheroids).  In the AllWISE catalogue \citep{Cutri_2021}, the source has the four-band variability flag (7,7,n,n),  which means highest probability of being variable in the first two WISE bands, whereas the data are insufficient to make a determination in the other two bands.  It is therefore very likely that the two WISE colours in 
Fig.\,\ref{fig:WISE_CCD} belong to the low state of the AGN.

\subsubsection{Narrow-line QSOs}\label{sect:type2}

33 spectra were categorised as type 2 QSOs because of their red continua in combination with a lack of evidence of broad-line components (subtype = type2).  A further seven noisy spectra with BALs are considered candidates for type 2 QSOs (remark =  type2?).  The redshifts of the whole sample of 40 possible type 2 QSOs are between $z = 0.77$ and 4.29 ($\overline{z} = 1.79 \pm 0.11$),  which is too high to utilise diagnostic line ratios to distinguish between AGNs and galaxies.  However, the classification as QSOs is justified due to their  bright absolute magnitudes\footnote{Corrected for Galactic foreground extinction,  but not for internal extinction and not $K$-corrected} $\overline{M_i} = -26.1 \pm 0.9$   (Fig.\,\ref{fig:Mi_z}) and their locations in the WISE colour plane (Fig.\,\ref{fig:WISE_CCD}).  The type 2 QSOs that belong to the two spectra  in Fig.\,\ref{fig:mix_QSOs_gals} have $M_i = -23.01$ and $M_i = -26.4$, respectively.   It cannot be ruled out that there are further type 2 QSOs among the sources classified as galaxies with emission lines (see below).

\subsection{Galaxies}\label{subsect:Rare_galaxies}

\subsubsection{Very red galaxies}

A conspicuous red continuum was noted for 51 galaxies, most of which (94\,\%) are emission line systems. 
Two examples are shown in Fig.\,\ref{fig:mix_QSOs_gals}. Remarkably, the spectra of all the red emission line galaxies originate from the plates 7260 and 7262  (Sect.\,\ref{sect:Sample}).  Adopting the extinction to reddening ratio $A_{\rm g}/E(B-V) = 3.3$ \citep{Yuan_2013} and $E(B-V)$ from \citet{Schlafly_2011},
the mean Galactic foreground extinction in the g band is  $A_{\rm g} = $ 4.44 and 3.64\,mag in these two fields, 
to be compared with  the mean value $A_{\rm g} = 0.10$\, mag for the spectra from the remaining plates. 
Obviously, the red spectra of these galaxies are not intrinsic properties.

\subsubsection{Galaxies at $z > 0.6$}

\begin{figure}[htbp]
\begin{center}
\includegraphics[viewport= 00 00 910 640,width=9.0cm,angle=0]{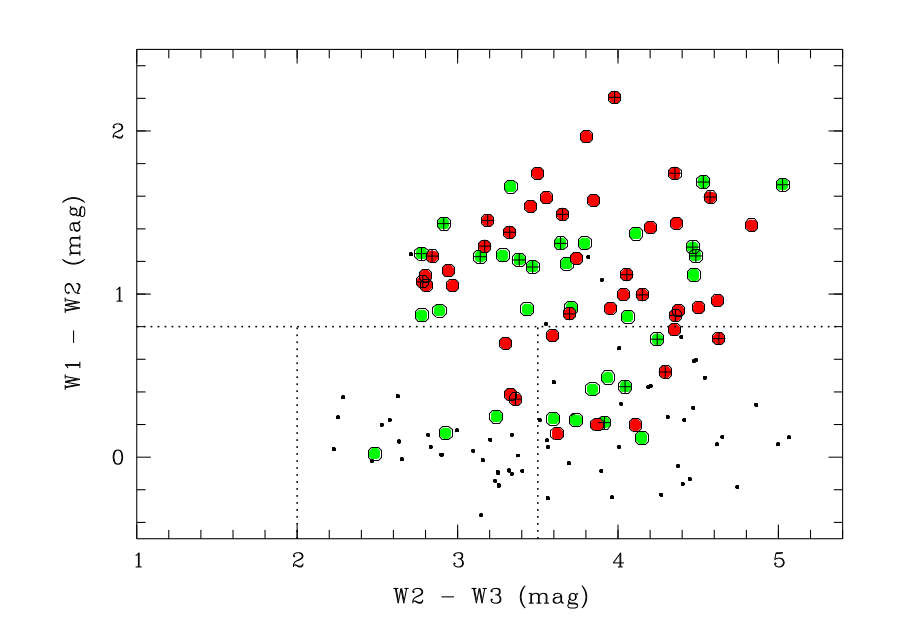} 
\caption{As Fig.\,\ref{fig:WISE_CCD}, but only for the 152 galaxies with $z \ge 0.6, r_z \ge 2$, and available magnitudes in the first three WISE bands.  Sources with uncertainties  $<0.5$ mag for $W3$ and $< 0.2$ mag for $W1$ and $W2$ are plotted as filled coloured circles ($z < 0.8$ green, $z  \ge 0.8$ red),  
all others as black dots. Radio-loud galaxies are marked by a black plus sign.}
\label{fig:WISE_CCD_galaxies}
\end{center}
\end{figure}

Redshifts greater than 0.6 were found for 205 galaxies. For 74\,\% of them,  photometric magnitudes are available in the first three WISE bands.   Figure\,\ref{fig:WISE_CCD_galaxies} shows the WISE colour-colour diagram. 
They mainly populate the area where one can find obscured AGNs, LINERs, and starbursts \citep[see][]{Jarrett_2017}:  39\,\% are located above the QSO threshold at $W1-W2=0.8$ and 40\,\% are in the region $W1-W2<0.8, W2-W3>3.5$.  For the 73 galaxies with uncertainties of $<0.5$\,mag in $W3$ and  $<0.2$\,mag in $W1$ and $W2$,   the percentage of galaxies in the QSO area is 66\,\% for  $z = 0.6 ... 0.8$ and 76\,\% for $z > 0.8$. 
However, the statistics are very limited and certainly biased towards galaxies that are bright in the MIR.

Eight galaxies with $r_z \ge 2$ have $z > 1$  (seven from eBOSS, one from BOSS),  all of them are emission line galaxies with reliable detection of the   [\ion{O}{ii}]\,3727,3729 doublet.  The highest redshift  is found for \object{SDSS\,J113833.30+324709.0} ($z = 1.488$),  based on the [\ion{O}{ii}] doublet and absorption lines from  \ion{Ca}{ii} H+K and \ion{Mg}{ii}.  The source is luminous ($M_i = -23.8$), has red MIR colours $(W1-W2, W2-W3) = (1.49, 3.75)$, and  is positionally coincident with a core-dominated FIRST radio source. These properties indicate either an AGN, a starburst, or both.  The higher values of $z_{\rm noqso}$ up to $\sim 2$ (Fig.\,\ref{fig:zHere_znoqso}) were not confirmed.

\subsubsection{Double-peaked narrow emission lines}

The phenomenon of double-peaked narrow emission lines in the spectra of galaxies is neither new  \citep{Heckman_1981}  nor particularly rare \citep[e.g.][]{Ge_2012, Maschmann_2020, Qiu_2024}. 
Recently, double-peaked galaxies (DPGs) have attracted particular attention in the context of the search for dual AGNs \citep{DeRosa_2019}.  However, the physical causes are still uncertain and may include other mechanisms, such as gas inflows and outflows, dynamic disturbances, and dust absorption.  There was no special focus on DPGs in the present work. However, for the sake of completeness, it should be mentioned that  double peaks were registered in eight galaxies at $z \approx 0.6$  to 0.9. Two examples are shown in Fig.\,\ref{fig:mix_QSOs_gals}.

\subsection{Stars}\label{subsect:Rare_stars}

\subsubsection{Supernovae}\label{subsect:SNe}

Four spectra appear unusual due to the significant contribution of a supernova (SN) to the host galaxies at $z \approx 0.02 - 0.22$.  Two of them were already known. \citet{Ofek_2011} discovered a type-II SN at $z=0.019$ in the spectrum spec-0339-51692-0524 of  \object{SDSS\,J130623.89-014033.6}, which is dominated by a strong H$\alpha$ emission line. Fig.\,\ref{fig:mix_QSOs_gals} (bottom left) shows the similar, albeit more noisy,  spectrum of \object{SDSS\,J013005.66+133617.2} with the newly discovered SN of type II.  The SN is clearly visible on the SDSS image in the outskirts of a starburst galaxy, but is no longer visible on the LS image.  The third SN was previously discovered by \citet{Baron_2017} in \object{SDSS\,J085436.69+180552.9} (spec-2281-53711-0156),  who classified it as type Ia. It  shows the characteristic broad \ion{Si}{ii} absorption at an outflow velocity of $\sim 1.7\, 10^4$ km\,s${^{-1}}$.   Fig.\,\ref{fig:mix_QSOs_gals} (bottom right) shows a similar spectrum of a hitherto unknown SN Ia that occurred in the centre of the face-on spiral galaxy  \object{SDSS\,J085449.24+180910.0} at $z = 0.189$.

\subsubsection{Polars}\label{subsect:Polars}

Polars are highly magnetic cataclysmic binary stars,  where the accretion of matter from the companion to the compact object is accompanied by the appearance of cyclotron radiation and variable brightness and polarisation 
\citep[e.g.][]{Downes_2001, Schmidt_2008, Parsons_2021, Inight_2023}.

Originally, five apparent high-$z$ QSOs from the present sample were identified here as stars with substantial contribution from cyclotron humps (Fig.\,\ref{fig:polars}).  Magnetic field strengths $B \approx 20$ to 50 MG were estimated from the wavelength $\lambda_m$ of the  $m$th harmonic adopting $B  = 107.1/(m\, {\it \lambda}_m)$ \citep{Ferrario_1996}, where $B$ is in MG. The subsequent analysis confirmed three of them as certain, one as unclear, and one as a misclassification.  The two sources  \object{SDSS\,J092122.83+203857.0} and \object{SDSS\,J154104.66+360252.9} were already known previously as  strongly magnetic  cataclysmic variables.   The cyclotron harmonics are clearly seen in both spectra. \citet{Schmidt_2008} identified the 5th harmonic in 
\object{SDSS\,J092122.83+203857.0} at 8020\,\AA. In the present study, it is assumed that $m = 6$ at 7980\,\AA\ provides a better fit. Polars are typically strongly variable in the optical with changes of the apparent magnitudes between one and several magnitudes.  In Fig.\,\ref{fig:polars}, the optical variability of these two sources is indicated by the differences between the SDSS spectra and the flux from the SDSS photometry.  The third source, \object{SDSS\,J010046.83+010859.1}, was not previously known. With its strong cyclotron features,  it
resembles the extreme polar \object{SDSS\,J132411.57+032050.5} \citep{Szkody_2003}. Interestingly, the spectrum shows no trace of the secondary star. The interpretation of the fourth source, \object{SDSS\,J$015707.95-005856.5}$, remains unclear.  In contrast to most polars, the Balmer lines are seen in absorption and are very narrow. 
This would only be possible with a  cool WD, if at all.  However, this does not match the very small Gaia parallax 
of $0.21\pm0.19$\,mas. There is one strong hump at $\sim 7470$\,\AA\ and perhaps a second, fainter one at $\sim 5600$\,\AA,  which would be best fitted by the 3rd and 4th harmonics.  However, the fluxes from the SDSS photometry do not confirm an elevated flux in the i band.  It is unlikely that this discrepancy is due to variability, as the source does not appear to be highly variable.  It is located in the SDSS stripe S82, where light curves are available, but it is not listed as variable in any of the corresponding catalogues  \citep{Bramich_2008, Becker_2011, Meusinger_2011, MacLeod_2012}.   Finally, \object{SDSS\,J144352.30+164503.3} proved to be a misclassification due to the erroneous SDSS spectrum of an M star in  DR16 (Fig.\,\ref{fig:polars}).

\section{Discussion}\label{sect:Discussion}

In relation to the enormous number of spectra provided by the SDSS (Sect. 1),  the number of misclassifications found here is small and probably irrelevant for most statistical analyses, except those that explicitly deal with high-z QSOs.  The most important finding from this study is that the SDSS DR16 spectroscopic pipeline tends to assign unusual spectra to high-redshift QSOs.  This provides an opportunity to expand small samples of known objects of rare types. 

\begin{figure}[htbp]
\begin{center}
\includegraphics[viewport= 20 0 530 520,width=4.3cm,angle=0]{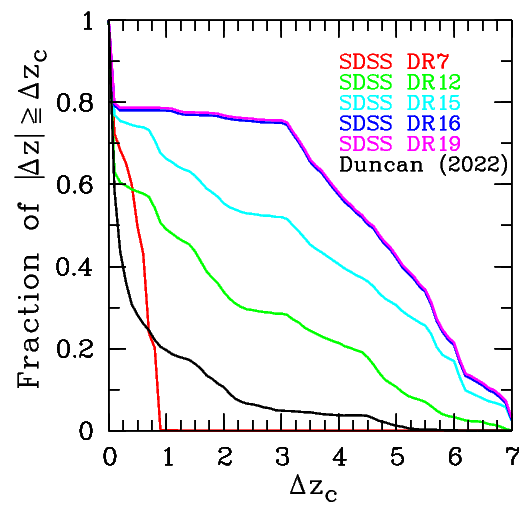} \
\includegraphics[viewport= 20 0 520 510,width=4.3cm,angle=0]{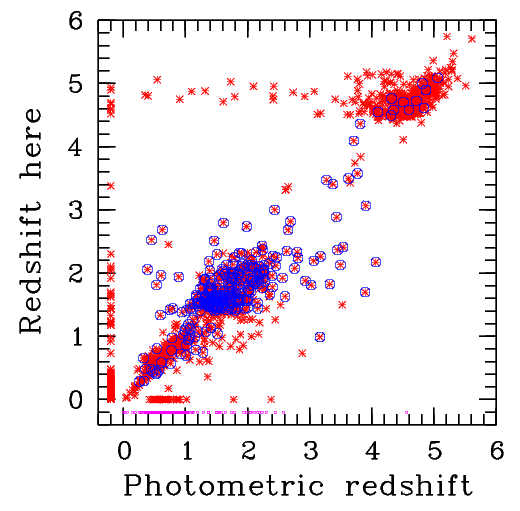} 
\caption{
Left: distributions of the differences $|\Delta z|$ between the revised redshift from the present study and the redshifts from different other sources.
Right:  revised redshift from the present work versus photometric redshift from \citet{Duncan_2022} 
for all objects with $r_z\ge 2$ (red), where peculiar types are marked by blue circles, as in Fig.\,\ref{fig:zHere_zSDSS}.  Small magenta symbols mark unclassified sources ($rz = 0$). 
}
\label{fig:dz_multiDR}
\end{center}
\end{figure}

Although it is not the aim of this paper to analyse the SDSS spectroscopic pipeline,  it is interesting to see how the proportion of misclassified high-$z$ QSOs varies with the version of the pipeline.  For this purpose, the redshifts from the SDSS DRs 7, 12, 15, 16 and 19 were compared with the revised $z$ for the sample of spectra examined here.  Figure\,\ref{fig:dz_multiDR} (left) shows how the proportion of the differences $|\Delta z|$ larger than a  given $ \Delta z_{\rm c}$ varies with $\Delta z_{\rm c}$.  The flatter the curve, the higher the proportion of large redshift errors.   Accordingly, DR7 shows the lowest proportion of strong redshift errors, while DR16 shows the highest.  The $z$-values in DR19 differ only slightly from  those in DR16.

\citet{Duncan_2022} provides photometric redshift, $z_{\rm ph}$, estimates for about 1.5 million sources from the LS DR8 based on a machine-learning approach.  For 63\% of the QSOs and galaxies of the present sample with reliable redshifts, $z_{\rm ph}$ of good quality (reliability flag fqual = 1) are available.  The comparison of the right panel of Fig.\,\ref{fig:dz_multiDR} with the left panel of Fig.\,\ref{fig:zHere_zSDSS}  shows that the revised redshift from the present paper correlate much better with $z_{\rm ph}$ than with the $z$ from SDSS DR16,  especially at $z < 4.5$  (see also the black curve in the left panel of Fig.\,\ref{fig:dz_multiDR}).  In particular, 98\% of the low-$z$ QSO sample has $z_{\rm ph} < 4.5$.  However, reliable photometric redshifts are only available for about half of the peculiar QSOs.  In Fig.\,\ref{fig:dz_multiDR}, the majority (63\%) of the low-$z$ QSOs with  $|z - z_{\rm ph}| > 1$ belong to peculiar types.

SDSS uses target flags which allow users to track why a particular object was selected for  spectroscopy.\footnote{The corresponding tables for the SDSS legacy, BOSS and eBOSS surveys and for the ancillary programs list altogether about 300 different target selection flags (see https://www.sdss4.org/dr16/algorithms/bitmasks/\#list).}   The target flags were downloaded via SQL query from the SDSS table specObjAll.  For clarity, the various flags have been divided into a manageable number of nine target classes,  four of which represent the object type (QSO, high $z$ QSO, galaxy, star) and five represent other properties (radio source, X-ray source, variable, BALs, others).

\begin{figure}[htbp]
\begin{center}
\includegraphics[viewport= 0 0 834 540,width=9.5cm,angle=0]{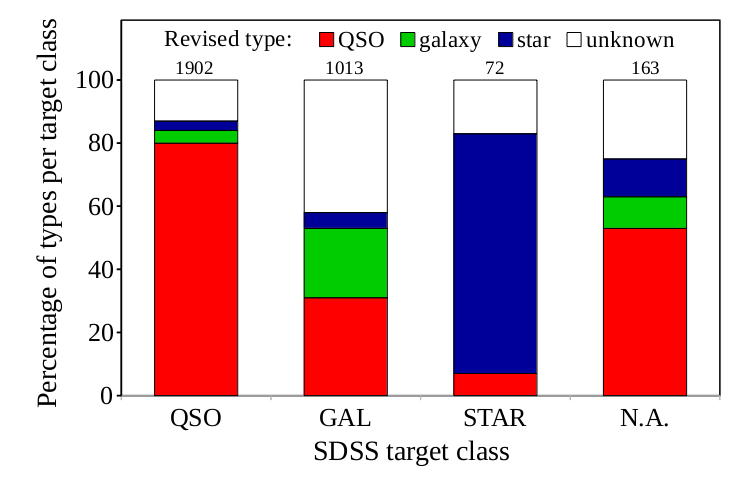} 
\caption{
Percentage of revised spectral types per SDSS DR16 target class. The total number of objects per target class is given at the top of each column.
}
\label{fig:target_flags}
\end{center}
\end{figure}

If the special plates are excluded, target information is available for 95\% of the spectra.  There is a clear correlation between the target class and the revised type (Fig.\,\ref{fig:target_flags}).  When only sources with reliable redshift are considered, ~90\% of the sources originally targeted as QSOs were confirmed as such here,  77\% of the sources in the target class  'high-$z$ QSO' have $z \ge 4.5$.  Conversely, 94\% of the sources classified here as QSOs were originally targeted as QSOs or galaxies.  The target class 'RADIO' contains 126 sources.  The percentage of sources selected for spectroscopy because they are radio sources among the LoBAL QSOs and the 3ABQs amounts 13\%.  It can therefore be ruled out that the significantly higher proportion of radio detections among these types (Sects.\,\ref{sect:FeLoBALs} and \ref{sect:3ABQs})  is due to such a selection effect.   The same applies to the QSOs with double-peaked broad Balmer lines (Sect.\,\ref{sect:DPBLs}).

Two objects with FIRST radio counterparts within 3 arcsec, \object{SDSS\,J157909.40+092641.2} and \object{SDSS\,J170743.97+342022.4}, were classified as stars.  The DR16 spectra clearly show stellar absorption lines at $z=0$ in the blue part. In both cases the radio emission can be attributed to a nearby galaxy.  The spectrum of J170743 is dominated by the star in the blue and by the galaxy ($z = 0.557$) in the red part. The red part of the spectrum of J157909 is unusual, but the same spectrum from DR15 clearly reveals a late-type star.

\section{Summary}\label{sect:Summary}

This paper  presents the results of a qualitative re-classification of spectra that were assigned to high-$z$ QSOs in SDSS DR16. The SQL query for QSOs with $z>4.5$ returns a list of 3955 spectra from 3351 sources, of which 22\,\% cannot be classified due to poor quality. If the non-classifiable spectra and those from special plates are excluded,  only 26\,\% of the sources turn out to be high-$z$ QSOs, whereas 50\,\% are QSOs at lower $z$, 16\,\% are galaxies, and 8\,\% are stars. The main scientific result of this paper is the discovery that the vast majority of the lower-$z$ QSOs belong to subtypes that are actually rare among the SDSS QSOs.  Compared to the total number of objects in the SDSS, the number of these misclassified sources is small and irrelevant for most statistical analyses.  However, their discovery contributes to expanding the available samples of rare types, which can contribute in particular to a better understanding of the evolution of AGNs and their interaction with their host galaxies. Among others, this re-classification reveals a considerable number of QSOs with a noticeable red continuum (247), unusual FeLoBALs (140), a 3000\,\AA\ break (12), weak emission lines (29),  double-peaked or strongly asymmetric broad Balmer lines (34),  and type 2 candidates (28). Four spectra could be assigned to galaxies with supernovae, of which only two were previously known. Particularly noteworthy is the discovery of distinct cyclotron humps in the spectra of three stars belonging to the rare class of polars. One of them, SDSS\,J010046.83+010859.1, is not yet mentioned in the literature.

\section{Data availability}

Table \ref{tab:cat} is only available in electronic form at the CDS via anonymous ftp to cdsarc.u-strasbg.fr or via http://cdsweb.u-strasbg.fr/cgi-bin/qcat?J/A+A/.

\begin{acknowledgements}

I thank the anonymous referee for their helpful report that has improved the quality of this work.
The author would also like to thank Boris G\"ansicke for his comments on the polar candidates, 
J\"org Br\"unecke for maintaining zshift, and Philipp Schalldach for his support with the Kohonen maps. 
Special thanks also go to Mathias Hoeft, Heinz Andernach, Aritra Basu, and the other members of the TLS Extragalactic Working Group for numerous discussions. \\

Funding for the Sloan Digital Sky Survey IV has been provided by the Alfred P. Sloan Foundation, 
the U.S. Department of Energy Office of Science, and the Participating Institutions. 
SDSS acknowledges support and resources from the Center for High-Performance Computing at the University of Utah. The SDSS web site is www.sdss4.org. 
SDSS is managed by the Astrophysical Research Consortium for the Participating Institutions of the SDSS Collaboration 
including the Brazilian Participation Group, the Carnegie Institution for Science, Carnegie Mellon University, 
Center for Astrophysics Harvard \& Smithsonian (CfA), the Chilean Participation Group, the French Participation Group, Instituto de Astrofísica de Canarias, 
The Johns Hopkins University, Kavli Institute for the Physics and Mathematics of the Universe (IPMU), the University of Tokyo, the Korean Participation Group, 
Lawrence Berkeley National Laboratory, Leibniz Institut f\"ur Astrophysik Potsdam (AIP), Max-Planck-Institut f\"ur Astronomie (MPIA Heidelberg), 
Max-Planck-Institut f\"ur Astrophysik (MPA Garching), Max-Planck-Institut f\"ur Extraterrestrische Physik (MPE), National Astronomical Observatories of China, 
New Mexico State University, New York University, University of Notre Dame, Observat\'orio Nacional / MCTI, The Ohio State University, Pennsylvania State University, 
Shanghai Astronomical Observatory, United Kingdom Participation Group, Universidad Nacional Autónoma de M\'exico, University of Arizona, 
University of Colorado Boulder, University of Oxford, University of Portsmouth, University of Utah, University of Virginia, University of Washington, 
University of Wisconsin, Vanderbilt University, and Yale University.\\

The Legacy Surveys consist of three individual and complementary projects: the Dark Energy Camera Legacy Survey 
(DECaLS; Proposal ID \#2014B-0404; PIs: David Schlegel and Arjun Dey), the Beijing-Arizona Sky Survey (BASS; NOAO Prop. ID \#2015A-0801; 
PIs: Zhou Xu and Xiaohui Fan), and the Mayall z-band Legacy Survey (MzLS; Prop. ID \#2016A-0453; PI: Arjun Dey). 
DECaLS, BASS and MzLS together include data obtained, respectively, at the Blanco telescope, 
Cerro Tololo Inter-American Observatory, NSF’s NOIRLab; the Bok telescope, Steward Observatory, University of Arizona; 
and the Mayall telescope, Kitt Peak National Observatory, NOIRLab. Pipeline processing 
and analyses of the data were supported by NOIRLab and the Lawrence Berkeley National Laboratory (LBNL). 
The Legacy Surveys project is honored to be permitted to conduct astronomical research on Iolkam Du’ag (Kitt Peak), 
a mountain with particular significance to the Tohono O’odham Nation. 
NOIRLab is operated by the Association of Universities for Research in Astronomy (AURA) under a cooperative agreement with the National Science Foundation. 
LBNL is managed by the Regents of the University of California under contract to the U.S. Department of Energy. 
This project used data obtained with the Dark Energy Camera (DECam), which was constructed by the Dark Energy Survey (DES) collaboration. 
Funding for the DES Projects has been provided by the U.S. Department of Energy, the U.S. National Science Foundation, 
the Ministry of Science and Education of Spain, the Science and Technology Facilities Council of the United Kingdom, the Higher Education Funding Council for England, 
the National Center for Supercomputing Applications at the University of Illinois at Urbana-Champaign, the Kavli Institute of Cosmological Physics at the University of Chicago, 
Center for Cosmology and Astro-Particle Physics at the Ohio State University, the Mitchell Institute for Fundamental Physics and Astronomy at Texas A\&M University, 
Financiadora de Estudos e Projetos, Fundacao Carlos Chagas Filho de Amparo, Financiadora de Estudos e Projetos, 
Fundacao Carlos Chagas Filho de Amparo a Pesquisa do Estado do Rio de Janeiro, Conselho Nacional de Desenvolvimento Cientifico e Tecnologico and the Ministerio da Ciencia, 
Tecnologia e Inovacao, the Deutsche Forschungsgemeinschaft and the Collaborating Institutions in the Dark Energy Survey. 
The Collaborating Institutions are Argonne National Laboratory, the University of California at Santa Cruz, the University of Cambridge, 
Centro de Investigaciones Energeticas, Medioambientales y Tecnologicas-Madrid, the University of Chicago, University College London, 
the DES-Brazil Consortium, the University of Edinburgh, the Eidgenossische Technische Hochschule (ETH) Zurich, 
Fermi National Accelerator Laboratory, the University of Illinois at Urbana-Champaign, the Institut de Ciencies de l’Espai (IEEC/CSIC), 
the Institut de Fisica d’Altes Energies, Lawrence Berkeley National Laboratory, the Ludwig Maximilians Universitat Munchen and the associated Excellence Cluster Universe, 
the University of Michigan, NSF’s NOIRLab, the University of Nottingham, the Ohio State University, the University of Pennsylvania, the University of Portsmouth, 
SLAC National Accelerator Laboratory, Stanford University, the University of Sussex, and Texas A\&M University. 
BASS is a key project of the Telescope Access Program (TAP), which has been funded by the National Astronomical Observatories of China, 
the Chinese Academy of Sciences (the Strategic Priority Research Program “The Emergence of Cosmological Structures” Grant \# XDB09000000), 
and the Special Fund for Astronomy from the Ministry of Finance. The BASS is also supported by the External Cooperation Program of Chinese Academy of Sciences 
(Grant \# 114A11KYSB20160057), and Chinese National Natural Science Foundation (Grant \# 12120101003, \# 11433005). 
The Legacy Survey team makes use of data products from the Near-Earth Object Wide-field Infrared Survey Explorer (NEOWISE), 
which is a project of the Jet Propulsion Laboratory/California 
Institute of Technology. NEOWISE is funded by the National Aeronautics and Space Administration. 
The Legacy Surveys imaging of the DESI footprint is supported by the Director, Office of Science, Office of High Energy Physics of the 
U.S. Department of Energy under Contract No. DE-AC02-05CH1123, 
by the National Energy Research Scientific Computing Center, a DOE Office of Science User Facility under the same contract; 
and by the U.S. National Science Foundation, Division of Astronomical Sciences under Contract No. AST-0950945 to NOAO.\\

This research has made use of images from the VLA FIRST survey of  the National Radio Astronomy Observatory, 
a facility of the National Science Foundation operated under cooperative agreement by Associated Universities, Inc. and of data products from WISE, 
which is a joint project of the University of California, Los Angeles, and the Jet Propulsion Laboratory/California Institute of Technology, 
funded by the National Aeronautics and Space Administration.\\

This work has also made use of observations made with the NASA/ESA Hubble Space Telescope,  obtained from the Hubble Legacy Archive, 
which is a joint project of the Space Telescope Science Institute (STScI),  the Space Telescope European Coordinating Facility (ST-ECF), 
and the Canadian Astronomy Data Centre (CADC).\\

This publication has made use of the VizieR catalogue access tool, CDS, Strasbourg, France, and of the NASA/IPAC Infrared Science Archive (IRSA), 
operated by the  Jet Propulsion Laboratories/California Institute of Technology, founded by the National Aeronautic and Space Administration.
In particular,   this publication makes use of data products from the Wide-field Infrared Survey Explorer, which is a joint project of the University of California, Los Angeles,  and the Jet Propulsion Laboratory/California Institute of Technology, funded by the National Aeronautics and Space Administration. 

\end{acknowledgements}

\bibliographystyle{aa} 
\bibliography{literature}

\begin{thebibliography}{104}
\expandafter\ifx\csname natexlab\endcsname\relax\def\natexlab#1{#1}\fi

\bibitem[{{Abdurro'uf} {et~al.}(2022){Abdurro'uf}, {Accetta}, {Aerts}, {Silva
  Aguirre}, {Ahumada}, {Ajgaonkar}, {Filiz Ak}, {Alam}, {Allende Prieto},
  {Almeida}, {Anders}, {Anderson}, {Andrews}, {Anguiano}, {Aquino-Ort{\'\i}z},
  {Arag{\'o}n-Salamanca}, {Argudo-Fern{\'a}ndez}, {Ata}, {Aubert},
  {Avila-Reese}, {Badenes}, {Barb{\'a}}, {Barger}, {Barrera-Ballesteros},
  {Beaton}, {Beers}, {Belfiore}, {Bender}, {Bernardi}, {Bershady}, {Beutler},
  {Bidin}, {Bird}, {Bizyaev}, {Blanc}, {Blanton}, {Boardman}, {Bolton},
  {Boquien}, {Borissova}, {Bovy}, {Brandt}, {Brown}, {Brownstein}, {Brusa},
  {Buchner}, {Bundy}, {Burchett}, {Bureau}, {Burgasser}, {Cabang}, {Campbell},
  {Cappellari}, {Carlberg}, {Wanderley}, {Carrera}, {Cash}, {Chen}, {Chen},
  {Cherinka}, {Chiappini}, {Choi}, {Chojnowski}, {Chung}, {Clerc}, {Cohen},
  {Comerford}, {Comparat}, {da Costa}, {Covey}, {Crane}, {Cruz-Gonzalez},
  {Culhane}, {Cunha}, {Dai}, {Damke}, {Darling}, {Davidson}, {Davies},
  {Dawson}, {De Lee}, {Diamond-Stanic}, {Cano-D{\'\i}az}, {S{\'a}nchez},
  {Donor}, {Duckworth}, {Dwelly}, {Eisenstein}, {Elsworth}, {Emsellem},
  {Eracleous}, {Escoffier}, {Fan}, {Farr}, {Feng}, {Fern{\'a}ndez-Trincado},
  {Feuillet}, {Filipp}, {Fillingham}, {Frinchaboy}, {Fromenteau}, {Galbany},
  {Garc{\'\i}a}, {Garc{\'\i}a-Hern{\'a}ndez}, {Ge}, {Geisler}, {Gelfand},
  {G{\'e}ron}, {Gibson}, {Goddy}, {Godoy-Rivera}, {Grabowski}, {Green},
  {Greener}, {Grier}, {Griffith}, {Guo}, {Guy}, {Hadjara}, {Harding},
  {Hasselquist}, {Hayes}, {Hearty}, {Hern{\'a}ndez}, {Hill}, {Hogg},
  {Holtzman}, {Horta}, {Hsieh}, {Hsu}, {Hsu}, {Huber}, {Huertas-Company},
  {Hutchinson}, {Hwang}, {Ibarra-Medel}, {Chitham}, {Ilha}, {Imig}, {Jaekle},
  {Jayasinghe}, {Ji}, {Johnson}, {Jones}, {J{\"o}nsson}, {Katkov}, {Khalatyan},
  {Kinemuchi}, {Kisku}, {Knapen}, {Kneib}, {Kollmeier}, {Kong}, {Kounkel},
  {Kreckel}, {Krishnarao}, {Lacerna}, {Lane}, {Langgin}, {Lavender}, {Law},
  {Lazarz}, {Leung}, {Leung}, {Lewis}, {Li}, {Li}, {Lian}, {Liang}, {Lin},
  {Lin}, {Lin}, {Lintott}, {Long}, {Longa-Pe{\~n}a}, {L{\'o}pez-Cob{\'a}},
  {Lu}, {Lundgren}, {Luo}, {Mackereth}, {de la Macorra}, {Mahadevan},
  {Majewski}, {Manchado}, {Mandeville}, {Maraston}, {Margalef-Bentabol},
  {Masseron}, {Masters}, {Mathur}, {McDermid}, {Mckay}, {Merloni},
  {Merrifield}, {Meszaros}, {Miglio}, {Di Mille}, {Minniti}, {Minsley}, \&
  {Monachesi}}]{Abdurro_22}
{Abdurro'uf}, {Accetta}, K., {Aerts}, C., {et~al.} 2022, \apjs, 259, 35

\bibitem[{{Ahumada} {et~al.}(2020){Ahumada}, {Prieto}, {Almeida}, {Anders},
  {Anderson}, {Andrews}, {Anguiano}, {Arcodia}, {Armengaud}, {Aubert}, {Avila},
  {Avila-Reese}, {Badenes}, {Balland}, {Barger}, {Barrera-Ballesteros}, {Basu},
  {Bautista}, {Beaton}, {Beers}, {Benavides}, {Bender}, {Bernardi}, {Bershady},
  {Beutler}, {Bidin}, {Bird}, {Bizyaev}, {Blanc}, {Blanton}, {Boquien},
  {Borissova}, {Bovy}, {Brandt}, {Brinkmann}, {Brownstein}, {Bundy}, {Bureau},
  {Burgasser}, {Burtin}, {Cano-D{\'\i}az}, {Capasso}, {Cappellari}, {Carrera},
  {Chabanier}, {Chaplin}, {Chapman}, {Cherinka}, {Chiappini}, {Doohyun Choi},
  {Chojnowski}, {Chung}, {Clerc}, {Coffey}, {Comerford}, {Comparat}, {da
  Costa}, {Cousinou}, {Covey}, {Crane}, {Cunha}, {Ilha}, {Dai}, {Damsted},
  {Darling}, {Davidson}, {Davies}, {Dawson}, {De}, {de la Macorra}, {De Lee},
  {Queiroz}, {Deconto Machado}, {de la Torre}, {Dell'Agli}, {du Mas des
  Bourboux}, {Diamond-Stanic}, {Dillon}, {Donor}, {Drory}, {Duckworth},
  {Dwelly}, {Ebelke}, {Eftekharzadeh}, {Davis Eigenbrot}, {Elsworth},
  {Eracleous}, {Erfanianfar}, {Escoffier}, {Fan}, {Farr},
  {Fern{\'a}ndez-Trincado}, {Feuillet}, {Finoguenov}, {Fofie},
  {Fraser-McKelvie}, {Frinchaboy}, {Fromenteau}, {Fu}, {Galbany}, {Garcia},
  {Garc{\'\i}a-Hern{\'a}ndez}, {Oehmichen}, {Ge}, {Maia}, {Geisler}, {Gelfand},
  {Goddy}, {Gonzalez-Perez}, {Grabowski}, {Green}, {Grier}, {Guo}, {Guy},
  {Harding}, {Hasselquist}, {Hawken}, {Hayes}, {Hearty}, {Hekker}, {Hogg},
  {Holtzman}, {Horta}, {Hou}, {Hsieh}, {Huber}, {Hunt}, {Chitham}, {Imig},
  {Jaber}, {Angel}, {Johnson}, {Jones}, {J{\"o}nsson}, {Jullo}, {Kim},
  {Kinemuchi}, {Kirkpatrick}, {Kite}, {Klaene}, {Kneib}, {Kollmeier}, {Kong},
  {Kounkel}, {Krishnarao}, {Lacerna}, {Lan}, {Lane}, {Law}, {Le Goff}, {Leung},
  {Lewis}, {Li}, {Lian}, {Lin}, {Long}, {Longa-Pe{\~n}a}, {Lundgren}, {Lyke},
  {Ted Mackereth}, {MacLeod}, {Majewski}, {Manchado}, {Maraston}, {Martini},
  {Masseron}, {Masters}, {Mathur}, {McDermid}, {Merloni}, {Merrifield},
  {M{\'e}sz{\'a}ros}, {Miglio}, {Minniti}, {Minsley}, {Miyaji}, {Mohammad},
  {Mosser}, {Mueller}, {Muna}, {Mu{\~n}oz-Guti{\'e}rrez}, {Myers}, {Nadathur},
  {Nair}, {Nandra}, {do Nascimento}, {Nevin}, {Newman}, {Nidever}, {Nitschelm},
  {Noterdaeme}, {O'Connell}, {Olmstead}, {Oravetz}, {Oravetz}, {Osorio},
  {Pace}, {Padilla}, {Palanque-Delabrouille}, {Palicio}, {Pan}, {Pan},
  {Parker}, {Paviot}, {Peirani}, {Ram{\'r}ez}, {Penny}, {Percival},
  {Perez-Fournon}, {P{\'e}rez-R{\`a}fols}, {Petitjean}, {Pieri},
  {Pinsonneault}, {Poovelil}, {Povick}, {Prakash}, {Price-Whelan}, {Raddick},
  {Raichoor}, {Ray}, {Rembold}, {Rezaie}, {Riffel}, {Riffel}, {Rix}, {Robin},
  {Roman-Lopes}, {Rom{\'a}n-Z{\'u}{\~n}iga}, {Rose}, {Ross}, {Rossi},
  {Rowlands}, {Rubin}, {Salvato}, {S{\'a}nchez}, {S{\'a}nchez-Menguiano},
  {S{\'a}nchez-Gallego}, {Sayres}, {Schaefer}, {Schiavon}, {Schimoia},
  {Schlafly}, {Schlegel}, {Schneider}, {Schultheis}, {Schwope}, {Seo},
  {Serenelli}, {Shafieloo}, {Shamsi}, {Shao}, {Shen}, {Shetrone}, {Shirley},
  {Aguirre}, {Simon}, {Skrutskie}, {Slosar}, {Smethurst}, {Sobeck}, {Sodi},
  {Souto}, {Stark}, {Stassun}, {Steinmetz}, {Stello}, {Stermer},
  {Storchi-Bergmann}, {Streblyanska}, {Stringfellow}, {Stutz}, {Su{\'a}rez},
  {Sun}, {Taghizadeh-Popp}, {Talbot}, {Tayar}, {Thakar}, {Theriault}, {Thomas},
  {Thomas}, {Tinker}, {Tojeiro}, {Toledo}, {Tremonti}, {Troup}, {Tuttle},
  {Unda-Sanzana}, {Valentini}, {Vargas-Gonz{\'a}lez}, {Vargas-Maga{\~n}a},
  {V{\'a}zquez-Mata}, {Vivek}, {Wake}, {Wang}, {Weaver}, {Weijmans}, {Wild},
  {Wilson}, {Wilson}, {Wolthuis}, {Wood-Vasey}, {Yan}, {Yang}, {Y{\`e}che},
  {Zamora}, {Zarrouk}, {Zasowski}, {Zhang}, {Zhao}, {Zhao}, {Zheng}, {Zheng},
  {Zhu}, \& {Zou}}]{Ahumada_2020}
{Ahumada}, R., {Prieto}, C.~A., {Almeida}, A., {et~al.} 2020, \apjs, 249, 3

\bibitem[{{Appenzeller} {et~al.}(2005){Appenzeller}, {Stahl}, {Tapken},
  {Mehlert}, \& {Noll}}]{Appenzeller_2005}
{Appenzeller}, I., {Stahl}, O., {Tapken}, C., {Mehlert}, D., \& {Noll}, S.
  2005, \aap, 435, 465

\bibitem[{{Baron} \& {Poznanski}(2017)}]{Baron_2017}
{Baron}, D. \& {Poznanski}, D. 2017, \mnras, 465, 4530

\bibitem[{{Becker} {et~al.}(2011){Becker}, {Bochanski}, {Hawley}, {Ivezi{\'c}},
  {Kowalski}, {Sesar}, \& {West}}]{Becker_2011}
{Becker}, A.~C., {Bochanski}, J.~J., {Hawley}, S.~L., {et~al.} 2011, \apj, 731,
  17

\bibitem[{{Bolton} {et~al.}(2012){Bolton}, {Schlegel}, {Aubourg}, {Bailey},
  {Bhardwaj}, {Brownstein}, {Burles}, {Chen}, {Dawson}, {Eisenstein}, {Gunn},
  {Knapp}, {Loomis}, {Lupton}, {Maraston}, {Muna}, {Myers}, {Olmstead},
  {Padmanabhan}, {P{\^a}ris}, {Percival}, {Petitjean}, {Rockosi}, {Ross},
  {Schneider}, {Shu}, {Strauss}, {Thomas}, {Tremonti}, {Wake}, {Weaver}, \&
  {Wood-Vasey}}]{Bolton_2012}
{Bolton}, A.~S., {Schlegel}, D.~J., {Aubourg}, {\'E}., {et~al.} 2012, \aj, 144,
  144

\bibitem[{{Bramich} {et~al.}(2008){Bramich}, {Vidrih}, {Wyrzykowski}, {Munn},
  {Lin}, {Evans}, {Smith}, {Belokurov}, {Gilmore}, {Zucker}, {Hewett},
  {Watkins}, {Faria}, {Fellhauer}, {Miknaitis}, {Bizyaev}, {Ivezi{\'c}},
  {Schneider}, {Snedden}, {Malanushenko}, {Malanushenko}, \&
  {Pan}}]{Bramich_2008}
{Bramich}, D.~M., {Vidrih}, S., {Wyrzykowski}, L., {et~al.} 2008, \mnras, 386,
  887

\bibitem[{{Calistro Rivera} {et~al.}(2024){Calistro Rivera}, {Alexander},
  {Harrison}, {Fawcett}, {Best}, {Williams}, {Hardcastle}, {Rosario}, {Smith},
  {Arnaudova}, {Escott}, {G{\"u}rkan}, {Kondapally}, {Miley}, {Morabito},
  {Petley}, {Prandoni}, {R{\"o}ttgering}, \& {Yue}}]{CalistroRivera_2024}
{Calistro Rivera}, G., {Alexander}, D.~M., {Harrison}, C.~M., {et~al.} 2024,
  \aap, 691, A191

\bibitem[{{Choi} {et~al.}(2022){Choi}, {Leighly}, {Dabbieri}, {Terndrup},
  {Gallagher}, \& {Richards}}]{Choi_2022}
{Choi}, H., {Leighly}, K.~M., {Dabbieri}, C., {et~al.} 2022, \apj, 936, 110

\bibitem[{{Choi} {et~al.}(2020){Choi}, {Leighly}, {Terndrup}, {Gallagher}, \&
  {Richards}}]{Choi_2020}
{Choi}, H., {Leighly}, K.~M., {Terndrup}, D.~M., {Gallagher}, S.~C., \&
  {Richards}, G.~T. 2020, \apj, 891, 53

\bibitem[{{Cutri} {et~al.}(2021){Cutri}, {Wright}, {Conrow}, {Fowler},
  {Eisenhardt}, {Grillmair}, {Kirkpatrick}, {Masci}, {McCallon}, {Wheelock},
  {Fajardo-Acosta}, {Yan}, {Benford}, {Harbut}, {Jarrett}, {Lake}, {Leisawitz},
  {Ressler}, {Stanford}, {Tsai}, {Liu}, {Helou}, {Mainzer}, {Gettngs},
  {Gonzalez}, {Hoffman}, {Marsh}, {Padgett}, {Skrutskie}, {Beck}, {Papin}, \&
  {Wittman}}]{Cutri_2021}
{Cutri}, R.~M., {Wright}, E.~L., {Conrow}, T., {et~al.} 2021, VizieR On-line
  Data Catalog: II/328.

\bibitem[{{Dai} {et~al.}(2012){Dai}, {Shankar}, \& {Sivakoff}}]{Dai_2012}
{Dai}, X., {Shankar}, F., \& {Sivakoff}, G.~R. 2012, \apj, 757, 180

\bibitem[{{Dawson} {et~al.}(2016){Dawson}, {Kneib}, {Percival}, {Alam},
  {Albareti}, {Anderson}, {Armengaud}, {Aubourg}, {Bailey}, {Bautista},
  {Berlind}, {Bershady}, {Beutler}, {Bizyaev}, {Blanton}, {Blomqvist},
  {Bolton}, {Bovy}, {Brandt}, {Brinkmann}, {Brownstein}, {Burtin}, {Busca},
  {Cai}, {Chuang}, {Clerc}, {Comparat}, {Cope}, {Croft}, {Cruz-Gonzalez}, {da
  Costa}, {Cousinou}, {Darling}, {de la Macorra}, {de la Torre}, {Delubac}, {du
  Mas des Bourboux}, {Dwelly}, {Ealet}, {Eisenstein}, {Eracleous}, {Escoffier},
  {Fan}, {Finoguenov}, {Font-Ribera}, {Frinchaboy}, {Gaulme}, {Georgakakis},
  {Green}, {Guo}, {Guy}, {Ho}, {Holder}, {Huehnerhoff}, {Hutchinson}, {Jing},
  {Jullo}, {Kamble}, {Kinemuchi}, {Kirkby}, {Kitaura}, {Klaene}, {Laher},
  {Lang}, {Laurent}, {Le Goff}, {Li}, {Liang}, {Lima}, {Lin}, {Lin}, {Lin},
  {Long}, {Lundgren}, {MacDonald}, {Geimba Maia}, {Malanushenko},
  {Malanushenko}, {Mariappan}, {McBride}, {McGreer}, {M{\'e}nard}, {Merloni},
  {Meza}, {Montero-Dorta}, {Muna}, {Myers}, {Nandra}, {Naugle}, {Newman},
  {Noterdaeme}, {Nugent}, {Ogando}, {Olmstead}, {Oravetz}, {Oravetz},
  {Padmanabhan}, {Palanque-Delabrouille}, {Pan}, {Parejko}, {P{\^a}ris},
  {Peacock}, {Petitjean}, {Pieri}, {Pisani}, {Prada}, {Prakash}, {Raichoor},
  {Reid}, {Rich}, {Ridl}, {Rodriguez-Torres}, {Carnero Rosell}, {Ross},
  {Rossi}, {Ruan}, {Salvato}, {Sayres}, {Schneider}, {Schlegel}, {Seljak},
  {Seo}, {Sesar}, {Shandera}, {Shu}, {Slosar}, {Sobreira}, {Streblyanska},
  {Suzuki}, {Taylor}, {Tao}, {Tinker}, {Tojeiro}, {Vargas-Maga{\~n}a}, {Wang},
  {Weaver}, {Weinberg}, {White}, {Wood-Vasey}, {Yeche}, {Zhai}, {Zhao}, {Zhao},
  {Zheng}, {Ben Zhu}, \& {Zou}}]{Dawson_2016}
{Dawson}, K.~S., {Kneib}, J.-P., {Percival}, W.~J., {et~al.} 2016, \aj, 151, 44

\bibitem[{{Dawson} {et~al.}(2013){Dawson}, {Schlegel}, {Ahn}, {Anderson},
  {Aubourg}, {Bailey}, {Barkhouser}, {Bautista}, {Beifiori}, {Berlind},
  {Bhardwaj}, {Bizyaev}, {Blake}, {Blanton}, {Blomqvist}, {Bolton}, {Borde},
  {Bovy}, {Brandt}, {Brewington}, {Brinkmann}, {Brown}, {Brownstein}, {Bundy},
  {Busca}, {Carithers}, {Carnero}, {Carr}, {Chen}, {Comparat}, {Connolly},
  {Cope}, {Croft}, {Cuesta}, {da Costa}, {Davenport}, {Delubac}, {de Putter},
  {Dhital}, {Ealet}, {Ebelke}, {Eisenstein}, {Escoffier}, {Fan}, {Filiz Ak},
  {Finley}, {Font-Ribera}, {G{\'e}nova-Santos}, {Gunn}, {Guo}, {Haggard},
  {Hall}, {Hamilton}, {Harris}, {Harris}, {Ho}, {Hogg}, {Holder}, {Honscheid},
  {Huehnerhoff}, {Jordan}, {Jordan}, {Kauffmann}, {Kazin}, {Kirkby}, {Klaene},
  {Kneib}, {Le Goff}, {Lee}, {Long}, {Loomis}, {Lundgren}, {Lupton}, {Maia},
  {Makler}, {Malanushenko}, {Malanushenko}, {Mandelbaum}, {Manera}, {Maraston},
  {Margala}, {Masters}, {McBride}, {McDonald}, {McGreer}, {McMahon}, {Mena},
  {Miralda-Escud{\'e}}, {Montero-Dorta}, {Montesano}, {Muna}, {Myers},
  {Naugle}, {Nichol}, {Noterdaeme}, {Nuza}, {Olmstead}, {Oravetz}, {Oravetz},
  {Owen}, {Padmanabhan}, {Palanque-Delabrouille}, {Pan}, {Parejko},
  {P{\^a}ris}, {Percival}, {P{\'e}rez-Fournon}, {P{\'e}rez-R{\`a}fols},
  {Petitjean}, {Pfaffenberger}, {Pforr}, {Pieri}, {Prada}, {Price-Whelan},
  {Raddick}, {Rebolo}, {Rich}, {Richards}, {Rockosi}, {Roe}, {Ross}, {Ross},
  {Rossi}, {Rubi{\~n}o-Martin}, {Samushia}, {S{\'a}nchez}, {Sayres}, {Schmidt},
  {Schneider}, {Sc{\'o}ccola}, {Seo}, {Shelden}, {Sheldon}, {Shen}, {Shu},
  {Slosar}, {Smee}, {Snedden}, {Stauffer}, {Steele}, {Strauss}, {Streblyanska},
  {Suzuki}, {Swanson}, {Tal}, {Tanaka}, {Thomas}, {Tinker}, {Tojeiro},
  {Tremonti}, {Vargas Maga{\~n}a}, {Verde}, {Viel}, {Wake}, {Watson}, {Weaver},
  {Weinberg}, {Weiner}, {West}, {White}, {Wood-Vasey}, {Yeche}, {Zehavi},
  {Zhao}, \& {Zheng}}]{Dawson_2013}
{Dawson}, K.~S., {Schlegel}, D.~J., {Ahn}, C.~P., {et~al.} 2013, \aj, 145, 10

\bibitem[{{De Rosa} {et~al.}(2019){De Rosa}, {Vignali}, {Bogdanovi{\'c}},
  {Capelo}, {Charisi}, {Dotti}, {Husemann}, {Lusso}, {Mayer}, {Paragi},
  {Runnoe}, {Sesana}, {Steinborn}, {Bianchi}, {Colpi}, {del Valle}, {Frey},
  {Gab{\'a}nyi}, {Giustini}, {Guainazzi}, {Haiman}, {Herrera Ruiz},
  {Herrero-Illana}, {Iwasawa}, {Komossa}, {Lena}, {Loiseau}, {Perez-Torres},
  {Piconcelli}, \& {Volonteri}}]{DeRosa_2019}
{De Rosa}, A., {Vignali}, C., {Bogdanovi{\'c}}, T., {et~al.} 2019, \nar, 86,
  101525

\bibitem[{{Dey} {et~al.}(2019){Dey}, {Schlegel}, {Lang}, {Blum}, {Burleigh},
  {Fan}, {Findlay}, {Finkbeiner}, {Herrera}, {Juneau}, {Landriau}, {Levi},
  {McGreer}, {Meisner}, {Myers}, {Moustakas}, {Nugent}, {Patej}, {Schlafly},
  {Walker}, {Valdes}, {Weaver}, {Y{\`e}che}, {Zou}, {Zhou}, {Abareshi},
  {Abbott}, {Abolfathi}, {Aguilera}, {Alam}, {Allen}, {Alvarez}, {Annis},
  {Ansarinejad}, {Aubert}, {Beechert}, {Bell}, {BenZvi}, {Beutler}, {Bielby},
  {Bolton}, {Brice{\~n}o}, {Buckley-Geer}, {Butler}, {Calamida}, {Carlberg},
  {Carter}, {Casas}, {Castander}, {Choi}, {Comparat}, {Cukanovaite}, {Delubac},
  {DeVries}, {Dey}, {Dhungana}, {Dickinson}, {Ding}, {Donaldson}, {Duan},
  {Duckworth}, {Eftekharzadeh}, {Eisenstein}, {Etourneau}, {Fagrelius},
  {Farihi}, {Fitzpatrick}, {Font-Ribera}, {Fulmer}, {G{\"a}nsicke},
  {Gaztanaga}, {George}, {Gerdes}, {Gontcho}, {Gorgoni}, {Green}, {Guy},
  {Harmer}, {Hernandez}, {Honscheid}, {Huang}, {James}, {Jannuzi}, {Jiang},
  {Joyce}, {Karcher}, {Karkar}, {Kehoe}, {Kneib}, {Kueter-Young}, {Lan},
  {Lauer}, {Le Guillou}, {Le Van Suu}, {Lee}, {Lesser}, {Perreault Levasseur},
  {Li}, {Mann}, {Marshall}, {Mart{\'\i}nez-V{\'a}zquez}, {Martini}, {du Mas des
  Bourboux}, {McManus}, {Meier}, {M{\'e}nard}, {Metcalfe},
  {Mu{\~n}oz-Guti{\'e}rrez}, {Najita}, {Napier}, {Narayan}, {Newman}, {Nie},
  {Nord}, {Norman}, {Olsen}, {Paat}, {Palanque-Delabrouille}, {Peng},
  {Poppett}, {Poremba}, {Prakash}, {Rabinowitz}, {Raichoor}, {Rezaie},
  {Robertson}, {Roe}, {Ross}, {Ross}, {Rudnick}, {Safonova}, {Saha},
  {S{\'a}nchez}, {Savary}, {Schweiker}, {Scott}, {Seo}, {Shan}, {Silva},
  {Slepian}, {Soto}, {Sprayberry}, {Staten}, {Stillman}, {Stupak}, {Summers},
  {Sien Tie}, {Tirado}, {Vargas-Maga{\~n}a}, {Vivas}, {Wechsler}, {Williams},
  {Yang}, {Yang}, {Yapici}, {Zaritsky}, {Zenteno}, {Zhang}, {Zhang}, {Zhou}, \&
  {Zhou}}]{Dey_2019}
{Dey}, A., {Schlegel}, D.~J., {Lang}, D., {et~al.} 2019, \aj, 157, 168

\bibitem[{{Diamond-Stanic} {et~al.}(2009){Diamond-Stanic}, {Fan}, {Brandt},
  {Shemmer}, {Strauss}, {Anderson}, {Carilli}, {Gibson}, {Jiang}, {Kim},
  {Richards}, {Schmidt}, {Schneider}, {Shen}, {Smith}, {Vestergaard}, \&
  {Young}}]{Diamond_2009}
{Diamond-Stanic}, A.~M., {Fan}, X., {Brandt}, W.~N., {et~al.} 2009, \apj, 699,
  782

\bibitem[{{Doan} {et~al.}(2020){Doan}, {Eracleous}, {Runnoe}, {Liu}, {Mathes},
  \& {Flohic}}]{Doan_2020}
{Doan}, A., {Eracleous}, M., {Runnoe}, J.~C., {et~al.} 2020, \mnras, 491, 1104

\bibitem[{{Doorenbos} {et~al.}(2021){Doorenbos}, {Cavuoti}, {Brescia},
  {D'Isanto}, \& {Longo}}]{Doorenbos_2021}
{Doorenbos}, L., {Cavuoti}, S., {Brescia}, M., {D'Isanto}, A., \& {Longo}, G.
  2021, in Intelligent Astrophysics, ed. I.~{Zelinka}, M.~{Brescia}, \&
  D.~{Baron}, Vol.~39, 197--223

\bibitem[{{Downes} {et~al.}(2001){Downes}, {Webbink}, {Shara}, {Ritter},
  {Kolb}, \& {Duerbeck}}]{Downes_2001}
{Downes}, R.~A., {Webbink}, R.~F., {Shara}, M.~M., {et~al.} 2001, \pasp, 113,
  764

\bibitem[{{Duncan}(2022)}]{Duncan_2022}
{Duncan}, K.~J. 2022, \mnras, 512, 3662

\bibitem[{{Eracleous} \& {Halpern}(1994)}]{Eracleous_1994}
{Eracleous}, M. \& {Halpern}, J.~P. 1994, \apjs, 90, 1

\bibitem[{{Eracleous} {et~al.}(2009){Eracleous}, {Lewis}, \&
  {Flohic}}]{Eracleous_2009}
{Eracleous}, M., {Lewis}, K.~T., \& {Flohic}, H. M.~L.~G. 2009, \nar, 53, 133

\bibitem[{{Fan} \& {SDSS Collaboration}(2002)}]{Fan_2002}
{Fan}, X. \& {SDSS Collaboration}. 2002, in Astronomical Society of the Pacific
  Conference Series, Vol. 280, Next Generation Wide-Field Multi-Object
  Spectroscopy, ed. M.~J.~I. {Brown} \& A.~{Dey}, 43

\bibitem[{{Fan} {et~al.}(1999){Fan}, {Strauss}, {Gunn}, {Lupton}, {Carilli},
  {Rupen}, {Schmidt}, {Moustakas}, {Davis}, {Annis}, {Bahcall}, {Brinkmann},
  {Brunner}, {Csabai}, {Doi}, {Fukugita}, {Heckman}, {Hennessy}, {Hindsley},
  {Ivezi{\'c} }, {Knapp}, {Lamb}, {Munn}, {Pauls}, {Pier}, {Rockosi},
  {Schneider}, {Szalay}, {Tucker}, \& {York}}]{Fan_1999}
{Fan}, X., {Strauss}, M.~A., {Gunn}, J.~E., {et~al.} 1999, \apjl, 526, L57

\bibitem[{{Fawcett} {et~al.}(2023){Fawcett}, {Alexander}, {Brodzeller}, {Edge},
  {Rosario}, {Myers}, {Aguilar}, {Ahlen}, {Alfarsy}, {Brooks}, {Canning},
  {Circosta}, {Dawson}, {de la Macorra}, {Doel}, {Fanning}, {Font-Ribera},
  {Forero-Romero}, {Gontcho A Gontcho}, {Guy}, {Harrison}, {Honscheid},
  {Juneau}, {Kehoe}, {Kisner}, {Kremin}, {Landriau}, {Manera}, {Meisner},
  {Miquel}, {Moustakas}, {Nie}, {Percival}, {Poppett}, {Pucha}, {Rossi},
  {Schlegel}, {Siudek}, {Tarl{\'e}}, {Weaver}, {Zhou}, \& {Zou}}]{Fawcett_2023}
{Fawcett}, V.~A., {Alexander}, D.~M., {Brodzeller}, A., {et~al.} 2023, \mnras,
  525, 5575

\bibitem[{{Fawcett} {et~al.}(2021){Fawcett}, {Alexander}, {Rosario}, \&
  {Klindt}}]{Fawcett_2021}
{Fawcett}, V.~A., {Alexander}, D.~M., {Rosario}, D.~J., \& {Klindt}, L. 2021,
  Galaxies, 9, 107

\bibitem[{{Fawcett} {et~al.}(2025){Fawcett}, {Harrison}, {Alexander},
  {Morabito}, {Kharb}, {Rosario}, {Baghel}, {Ghosh}, {Sasikumar}, {Petley},
  {Sargent}, \& {Calistro Rivera}}]{Fawcett_2025}
{Fawcett}, V.~A., {Harrison}, C.~M., {Alexander}, D.~M., {et~al.} 2025, \mnras,
  537, 2003

\bibitem[{{Ferrario} {et~al.}(1996){Ferrario}, {Bailey}, \&
  {Wickramasinghe}}]{Ferrario_1996}
{Ferrario}, L., {Bailey}, J., \& {Wickramasinghe}, D. 1996, \mnras, 282, 218

\bibitem[{{Flesch}(2021)}]{Flesch_2021}
{Flesch}, E.~W. 2021, \mnras, 504, 621

\bibitem[{{Fustes} {et~al.}(2013){Fustes}, {Manteiga}, {Dafonte}, {Arcay},
  {Ulla}, {Smith}, {Borrachero}, \& {Sordo}}]{Fustes_2013}
{Fustes}, D., {Manteiga}, M., {Dafonte}, C., {et~al.} 2013, \aap, 559, A7

\bibitem[{{Gaia Collaboration} {et~al.}(2023){Gaia Collaboration}, {Vallenari},
  {Brown}, {Prusti}, {de Bruijne}, {Arenou}, {Babusiaux}, {Biermann},
  {Creevey}, {Ducourant}, {Evans}, {Eyer}, {Guerra}, {Hutton}, {Jordi},
  {Klioner}, {Lammers}, {Lindegren}, {Luri}, {Mignard}, {Panem}, {Pourbaix},
  {Randich}, {Sartoretti}, {Soubiran}, {Tanga}, {Walton}, {Bailer-Jones},
  {Bastian}, {Drimmel}, {Jansen}, {Katz}, {Lattanzi}, {van Leeuwen}, {Bakker},
  {Cacciari}, {Casta{\~n}eda}, {De Angeli}, {Fabricius}, {Fouesneau},
  {Fr{\'e}mat}, {Galluccio}, {Guerrier}, {Heiter}, {Masana}, {Messineo},
  {Mowlavi}, {Nicolas}, {Nienartowicz}, {Pailler}, {Panuzzo}, {Riclet}, {Roux},
  {Seabroke}, {Sordo}, {Th{\'e}venin}, {Gracia-Abril}, {Portell}, {Teyssier},
  {Altmann}, {Andrae}, {Audard}, {Bellas-Velidis}, {Benson}, {Berthier},
  {Blomme}, {Burgess}, {Busonero}, {Busso}, {C{\'a}novas}, {Carry}, {Cellino},
  {Cheek}, {Clementini}, {Damerdji}, {Davidson}, {de Teodoro}, {Nu{\~n}ez
  Campos}, {Delchambre}, {Dell'Oro}, {Esquej}, {Fern{\'a}ndez-Hern{\'a}ndez},
  {Fraile}, {Garabato}, {Garc{\'\i}a-Lario}, {Gosset}, {Haigron}, {Halbwachs},
  {Hambly}, {Harrison}, {Hern{\'a}ndez}, {Hestroffer}, {Hodgkin}, {Holl},
  {Jan{\ss}en}, {Jevardat de Fombelle}, {Jordan}, {Krone-Martins}, {Lanzafame},
  {L{\"o}ffler}, {Marchal}, {Marrese}, {Moitinho}, {Muinonen}, {Osborne},
  {Pancino}, {Pauwels}, {Recio-Blanco}, {Reyl{\'e}}, {Riello}, {Rimoldini},
  {Roegiers}, {Rybizki}, {Sarro}, {Siopis}, {Smith}, {Sozzetti}, {Utrilla},
  {van Leeuwen}, {Abbas}, {{\'A}brah{\'a}m}, {Abreu Aramburu}, {Aerts},
  {Aguado}, {Ajaj}, {Aldea-Montero}, {Altavilla}, {{\'A}lvarez}, {Alves},
  {Anders}, {Anderson}, {Anglada Varela}, {Antoja}, {Baines}, {Baker},
  {Balaguer-N{\'u}{\~n}ez}, {Balbinot}, {Balog}, {Barache}, {Barbato},
  {Barros}, {Barstow}, {Bartolom{\'e}}, {Bassilana}, {Bauchet}, {Becciani},
  {Bellazzini}, {Berihuete}, {Bernet}, {Bertone}, {Bianchi}, {Binnenfeld},
  {Blanco-Cuaresma}, {Blazere}, {Boch}, {Bombrun}, {Bossini}, {Bouquillon},
  {Bragaglia}, {Bramante}, {Breedt}, {Bressan}, {Brouillet}, {Brugaletta},
  {Bucciarelli}, {Burlacu}, {Butkevich}, {Buzzi}, {Caffau}, {Cancelliere},
  {Cantat-Gaudin}, {Carballo}, {Carlucci}, {Carnerero}, {Carrasco},
  {Casamiquela}, {Castellani}, {Castro-Ginard}, {Chaoul}, {Charlot}, {Chemin},
  {Chiaramida}, {Chiavassa}, {Chornay}, {Comoretto}, {Contursi}, {Cooper},
  {Cornez}, {Cowell}, {Crifo}, {Cropper}, {Crosta}, {Crowley}, {Dafonte},
  {Dapergolas}, {David}, {David}, {de Laverny}, {De Luise}, \& {De
  March}}]{Gaia_2023}
{Gaia Collaboration}, {Vallenari}, A., {Brown}, A.~G.~A., {et~al.} 2023, \aap,
  674, A1

\bibitem[{{Gaskell} {et~al.}(2024){Gaskell}, {Gill}, \& {Singh}}]{Gaskell_2024}
{Gaskell}, C.~M., {Gill}, J. J.~M., \& {Singh}, J. 2024, \mnras, 533, 3676

\bibitem[{{Ge} {et~al.}(2012){Ge}, {Hu}, {Wang}, {Bai}, \& {Zhang}}]{Ge_2012}
{Ge}, J.-Q., {Hu}, C., {Wang}, J.-M., {Bai}, J.-M., \& {Zhang}, S. 2012, \apjs,
  201, 31

\bibitem[{{Green}(2006)}]{Green_2006}
{Green}, P.~J. 2006, \apj, 644, 733

\bibitem[{{Green} {et~al.}(2022){Green}, {Pulgarin-Duque}, {Anderson},
  {MacLeod}, {Eracleous}, {Ruan}, {Runnoe}, {Graham}, {Roulston}, {Schneider},
  {Ahlf}, {Bizyaev}, {Brownstein}, {del Casal}, {Dodd}, {Hoover}, {Matt},
  {Merloni}, {Pan}, {Ramirez}, {Ridder}, \& {Moseley}}]{Green_2022}
{Green}, P.~J., {Pulgarin-Duque}, L., {Anderson}, S.~F., {et~al.} 2022, \apj,
  933, 180

\bibitem[{{Hall} {et~al.}(2002){Hall}, {Anderson}, {Strauss}, {York},
  {Richards}, {Fan}, {Knapp}, {Schneider}, {Vanden Berk}, {Geballe}, {Bauer},
  {Becker}, {Davis}, {Rix}, {Nichol}, {Bahcall}, {Brinkmann}, {Brunner},
  {Connolly}, {Csabai}, {Doi}, {Fukugita}, {Gunn}, {Haiman}, {Harvanek},
  {Heckman}, {Hennessy}, {Inada}, {Ivezi{\'c}}, {Johnston}, {Kleinman},
  {Krolik}, {Krzesinski}, {Kunszt}, {Lamb}, {Long}, {Lupton}, {Miknaitis},
  {Munn}, {Narayanan}, {Neilsen}, {Newman}, {Nitta}, {Okamura}, {Pentericci},
  {Pier}, {Schlegel}, {Snedden}, {Szalay}, {Thakar}, {Tsvetanov}, {White}, \&
  {Zheng}}]{Hall_2002}
{Hall}, P.~B., {Anderson}, S.~F., {Strauss}, M.~A., {et~al.} 2002, \apjs, 141,
  267

\bibitem[{{Harrison} \& {Ramos Almeida}(2024)}]{Harrison_2024}
{Harrison}, C.~M. \& {Ramos Almeida}, C. 2024, Galaxies, 12, 17

\bibitem[{{Heckman} {et~al.}(1981){Heckman}, {Miley}, {van Breugel}, \&
  {Butcher}}]{Heckman_1981}
{Heckman}, T.~M., {Miley}, G.~K., {van Breugel}, W.~J.~M., \& {Butcher}, H.~R.
  1981, \apj, 247, 403

\bibitem[{{Helfand} {et~al.}(2015){Helfand}, {White}, \&
  {Becker}}]{Helfand_2015}
{Helfand}, D.~J., {White}, R.~L., \& {Becker}, R.~H. 2015, \apj, 801, 26

\bibitem[{{Hopkins} {et~al.}(2005){Hopkins}, {Hernquist}, {Martini}, {Cox},
  {Robertson}, {Di Matteo}, \& {Springel}}]{Hopkins_2005}
{Hopkins}, P.~F., {Hernquist}, L., {Martini}, P., {et~al.} 2005, \apjl, 625,
  L71

\bibitem[{{in der Au} {et~al.}(2012){in der Au}, {Meusinger}, {Schalldach}, \&
  {Newholm}}]{inderAu_2012}
{in der Au}, A., {Meusinger}, H., {Schalldach}, P.~F., \& {Newholm}, M. 2012,
  \aap, 547, A115

\bibitem[{{Inight} {et~al.}(2023){Inight}, {G{\"a}nsicke}, {Breedt}, {Israel},
  {Littlefair}, {Manser}, {Marsh}, {Mulvany}, {Pala}, \&
  {Thorstensen}}]{Inight_2023}
{Inight}, K., {G{\"a}nsicke}, B.~T., {Breedt}, E., {et~al.} 2023, \mnras, 524,
  4867

\bibitem[{{Ivezi{\'c}} {et~al.}(2002){Ivezi{\'c}}, {Menou}, {Knapp}, {Strauss},
  {Lupton}, {Vanden Berk}, {Richards}, {Tremonti}, {Weinstein}, {Anderson},
  {Bahcall}, {Becker}, {Bernardi}, {Blanton}, {Eisenstein}, {Fan},
  {Finkbeiner}, {Finlator}, {Frieman}, {Gunn}, {Hall}, {Kim}, {Kinkhabwala},
  {Narayanan}, {Rockosi}, {Schlegel}, {Schneider}, {Strateva}, {SubbaRao},
  {Thakar}, {Voges}, {White}, {Yanny}, {Brinkmann}, {Doi}, {Fukugita},
  {Hennessy}, {Munn}, {Nichol}, \& {York}}]{Ivezic_2002}
{Ivezi{\'c}}, {\v{Z}}., {Menou}, K., {Knapp}, G.~R., {et~al.} 2002, \aj, 124,
  2364

\bibitem[{{Jarrett} {et~al.}(2017){Jarrett}, {Cluver}, {Magoulas}, {Bilicki},
  {Alpaslan}, {Bland-Hawthorn}, {Brough}, {Brown}, {Croom}, {Driver},
  {Holwerda}, {Hopkins}, {Loveday}, {Norberg}, {Peacock}, {Popescu}, {Sadler},
  {Taylor}, {Tuffs}, \& {Wang}}]{Jarrett_2017}
{Jarrett}, T.~H., {Cluver}, M.~E., {Magoulas}, C., {et~al.} 2017, \apj, 836,
  182

\bibitem[{{Jiang} {et~al.}(2013){Jiang}, {Zhou}, {Ji}, {Shu}, {Liu}, {Wang},
  {Dong}, {Bai}, {Wang}, \& {Wang}}]{Jiang_2013}
{Jiang}, P., {Zhou}, H., {Ji}, T., {et~al.} 2013, \aj, 145, 157

\bibitem[{{Kennefick} \& {Bursick}(2008)}]{Kennefick_2008}
{Kennefick}, J. \& {Bursick}, S. 2008, \aj, 136, 1799

\bibitem[{{Kim} \& {Im}(2018)}]{Kim_2018}
{Kim}, D. \& {Im}, M. 2018, \aap, 610, A31

\bibitem[{{Kohonen}(2001)}]{Kohonen_2001}
{Kohonen}, T. 2001, {Self-Organizing Maps}

\bibitem[{{Komossa} {et~al.}(2024){Komossa}, {Grupe}, {Marziani}, {Popovic},
  {Marceta-Mandic}, {Bon}, {Ilic}, {Kovacevic}, {Kraus}, {Haiman}, {Petrecca},
  {De Cicco}, {Dimitrijevic}, {Sreckovic}, {Kovacevic Dojcinovic},
  {Pannikkote}, {Bon}, {Gupta}, \& {Iacob}}]{Komossa_2024}
{Komossa}, S., {Grupe}, D., {Marziani}, P., {et~al.} 2024, arXiv e-prints,
  arXiv:2408.00089

\bibitem[{{Kumar} {et~al.}(2025){Kumar}, {Krishna}, {Chand}, \&
  {Negi}}]{Kumar_2025}
{Kumar}, R., {Krishna}, G., {Chand}, H., \& {Negi}, V. 2025, \mnras, 538, L83

\bibitem[{{LaMassa} {et~al.}(2015){LaMassa}, {Cales}, {Moran}, {Myers},
  {Richards}, {Eracleous}, {Heckman}, {Gallo}, \& {Urry}}]{LaMassa_2015}
{LaMassa}, S.~M., {Cales}, S., {Moran}, E.~C., {et~al.} 2015, \apj, 800, 144

\bibitem[{{Leighly} {et~al.}(2024){Leighly}, {Choi}, {Eracleous}, {Terndrup},
  {Gallagher}, \& {Richards}}]{Leighly_2024}
{Leighly}, K.~M., {Choi}, H., {Eracleous}, M., {et~al.} 2024, \apj, 966, 87

\bibitem[{{Leighly} {et~al.}(2014){Leighly}, {Terndrup}, {Baron}, {Lucy},
  {Dietrich}, \& {Gallagher}}]{Leighly_2014}
{Leighly}, K.~M., {Terndrup}, D.~M., {Baron}, E., {et~al.} 2014, \apj, 788, 123

\bibitem[{{Leighly} {et~al.}(2018){Leighly}, {Terndrup}, {Gallagher},
  {Richards}, \& {Dietrich}}]{Leighly_2018}
{Leighly}, K.~M., {Terndrup}, D.~M., {Gallagher}, S.~C., {Richards}, G.~T., \&
  {Dietrich}, M. 2018, \apj, 866, 7

\bibitem[{{Liu} {et~al.}(2016){Liu}, {Eracleous}, \& {Halpern}}]{Liu_2016}
{Liu}, J., {Eracleous}, M., \& {Halpern}, J.~P. 2016, \apj, 817, 42

\bibitem[{{Lyke} {et~al.}(2020){Lyke}, {Higley}, {McLane}, {Schurhammer},
  {Myers}, {Ross}, {Dawson}, {Chabanier}, {Martini}, {Busca}, {Mas des
  Bourboux}, {Salvato}, {Streblyanska}, {Zarrouk}, {Burtin}, {Anderson},
  {Bautista}, {Bizyaev}, {Brandt}, {Brinkmann}, {Brownstein}, {Comparat},
  {Green}, {de la Macorra}, {Mu{\~n}oz Guti{\'e}rrez}, {Hou}, {Newman},
  {Palanque-Delabrouille}, {P{\^a}ris}, {Percival}, {Petitjean}, {Rich},
  {Rossi}, {Schneider}, {Smith}, {Vivek}, \& {Weaver}}]{Lyke_2020}
{Lyke}, B.~W., {Higley}, A.~N., {McLane}, J.~N., {et~al.} 2020, \apjs, 250, 8

\bibitem[{{Lynds}(1967)}]{Lynds_1967}
{Lynds}, C.~R. 1967, \apj, 147, 396

\bibitem[{{MacLeod} {et~al.}(2012){MacLeod}, {Ivezi{\'c}}, {Sesar}, {de Vries},
  {Kochanek}, {Kelly}, {Becker}, {Lupton}, {Hall}, {Richards}, {Anderson}, \&
  {Schneider}}]{MacLeod_2012}
{MacLeod}, C.~L., {Ivezi{\'c}}, {\v{Z}}., {Sesar}, B., {et~al.} 2012, \apj,
  753, 106

\bibitem[{{Maschmann} {et~al.}(2020){Maschmann}, {Melchior}, {Mamon},
  {Chilingarian}, \& {Katkov}}]{Maschmann_2020}
{Maschmann}, D., {Melchior}, A.-L., {Mamon}, G.~A., {Chilingarian}, I.~V., \&
  {Katkov}, I.~Y. 2020, \aap, 641, A171

\bibitem[{{Massaro} {et~al.}(2015){Massaro}, {Maselli}, {Leto}, {Marchegiani},
  {Perri}, {Giommi}, \& {Piranomonte}}]{Massaro_2015}
{Massaro}, E., {Maselli}, A., {Leto}, C., {et~al.} 2015, \apss, 357, 75

\bibitem[{{McDowell} {et~al.}(1995){McDowell}, {Canizares}, {Elvis},
  {Lawrence}, {Markoff}, {Mathur}, \& {Wilkes}}]{McDowell_1995}
{McDowell}, J.~C., {Canizares}, C., {Elvis}, M., {et~al.} 1995, \apj, 450, 585

\bibitem[{{Meusinger} \& {Balafkan}(2014)}]{Meusinger_2014}
{Meusinger}, H. \& {Balafkan}, N. 2014, \aap, 568, A114

\bibitem[{{Meusinger} {et~al.}(2017){Meusinger}, {Br{\"u}necke}, {Schalldach},
  \& {in der Au}}]{Meusinger_2017}
{Meusinger}, H., {Br{\"u}necke}, J., {Schalldach}, P., \& {in der Au}, A. 2017,
  \aap, 597, A134

\bibitem[{{Meusinger} {et~al.}(2011){Meusinger}, {Hinze}, \& {de
  Hoon}}]{Meusinger_2011}
{Meusinger}, H., {Hinze}, A., \& {de Hoon}, A. 2011, \aap, 525, A37

\bibitem[{{Meusinger} {et~al.}(2016){Meusinger}, {Schalldach}, {Mirhosseini},
  \& {Pertermann}}]{Meusinger_2016}
{Meusinger}, H., {Schalldach}, P., {Mirhosseini}, A., \& {Pertermann}, F. 2016,
  \aap, 587, A83

\bibitem[{{Meusinger} {et~al.}(2012){Meusinger}, {Schalldach}, {Scholz}, {in
  der Au}, {Newholm}, {de Hoon}, \& {Kaminsky}}]{Meusinger_2012}
{Meusinger}, H., {Schalldach}, P., {Scholz}, R.~D., {et~al.} 2012, \aap, 541,
  A77

\bibitem[{{Nair} \& {Vivek}(2022)}]{Nair_2022}
{Nair}, A. \& {Vivek}, M. 2022, \mnras, 511, 4946

\bibitem[{{Ofek} {et~al.}(2011){Ofek}, {Gal-Yam}, \& {Groot}}]{Ofek_2011}
{Ofek}, E.~O., {Gal-Yam}, A., \& {Groot}, P. 2011, The Astronomer's Telegram,
  3474, 1

\bibitem[{{Panda} \& {{\'S}niegowska}(2024)}]{Panda_2024}
{Panda}, S. \& {{\'S}niegowska}, M. 2024, \apjs, 272, 13

\bibitem[{{Parsons} {et~al.}(2021){Parsons}, {G{\"a}nsicke}, {Schreiber},
  {Marsh}, {Ashley}, {Breedt}, {Littlefair}, \& {Meusinger}}]{Parsons_2021}
{Parsons}, S.~G., {G{\"a}nsicke}, B.~T., {Schreiber}, M.~R., {et~al.} 2021,
  \mnras, 502, 4305

\bibitem[{{Paul} {et~al.}(2022){Paul}, {Plotkin}, {Shemmer}, {Anderson},
  {Brandt}, {Fan}, {Gallo}, {Luo}, {Ni}, {Richards}, {Schneider}, {Wu}, \&
  {Yi}}]{Paul_2022}
{Paul}, J.~D., {Plotkin}, R.~M., {Shemmer}, O., {et~al.} 2022, \apj, 929, 78

\bibitem[{{Peng} {et~al.}(2024){Peng}, {Chen}, {He}, {Pang}, \&
  {Wang}}]{Peng_2024}
{Peng}, X.-L., {Chen}, Z.-F., {He}, Z.-C., {Pang}, T.-T., \& {Wang}, Z.-W.
  2024, \apj, 963, 3

\bibitem[{{Plotkin} {et~al.}(2010){Plotkin}, {Anderson}, {Brandt},
  {Diamond-Stanic}, {Fan}, {MacLeod}, {Schneider}, \& {Shemmer}}]{Plotkin_2010}
{Plotkin}, R.~M., {Anderson}, S.~F., {Brandt}, W.~N., {et~al.} 2010, \apj, 721,
  562

\bibitem[{{Plotkin} {et~al.}(2008){Plotkin}, {Anderson}, {Hall}, {Margon},
  {Voges}, {Schneider}, {Stinson}, \& {York}}]{Plotkin_2008}
{Plotkin}, R.~M., {Anderson}, S.~F., {Hall}, P.~B., {et~al.} 2008, \aj, 135,
  2453

\bibitem[{{Polletta} {et~al.}(2007){Polletta}, {Tajer}, {Maraschi},
  {Trinchieri}, {Lonsdale}, {Chiappetti}, {Andreon}, {Pierre}, {Le F{\`e}vre},
  {Zamorani}, {Maccagni}, {Garcet}, {Surdej}, {Franceschini}, {Alloin},
  {Shupe}, {Surace}, {Fang}, {Rowan-Robinson}, {Smith}, \&
  {Tresse}}]{Polletta_2007}
{Polletta}, M., {Tajer}, M., {Maraschi}, L., {et~al.} 2007, \apj, 663, 81

\bibitem[{{Potts} \& {Villforth}(2021)}]{Potts_2021}
{Potts}, B. \& {Villforth}, C. 2021, \aap, 650, A33

\bibitem[{{Qiu} {et~al.}(2024){Qiu}, {Shen}, {Feng}, {Chen}, {Chang}, {Zhao},
  \& {Zeng}}]{Qiu_2024}
{Qiu}, J., {Shen}, S., {Feng}, S., {et~al.} 2024, \apj, 976, 15

\bibitem[{{Reichard} {et~al.}(2003){Reichard}, {Richards}, {Hall}, {Schneider},
  {Vanden Berk}, {Fan}, {York}, {Knapp}, \& {Brinkmann}}]{Reichard_2003}
{Reichard}, T.~A., {Richards}, G.~T., {Hall}, P.~B., {et~al.} 2003, \aj, 126,
  2594

\bibitem[{{Reis} {et~al.}(2021){Reis}, {Rotman}, {Poznanski}, {Prochaska}, \&
  {Wolf}}]{Reis_2021}
{Reis}, I., {Rotman}, M., {Poznanski}, D., {Prochaska}, J.~X., \& {Wolf}, L.
  2021, Astronomy and Computing, 34, 100437

\bibitem[{{Schlafly} \& {Finkbeiner}(2011)}]{Schlafly_2011}
{Schlafly}, E.~F. \& {Finkbeiner}, D.~P. 2011, \apj, 737, 103

\bibitem[{{Schlafly} {et~al.}(2019){Schlafly}, {Meisner}, \&
  {Green}}]{Schlafly_2019}
{Schlafly}, E.~F., {Meisner}, A.~M., \& {Green}, G.~M. 2019, \apjs, 240, 30

\bibitem[{{Schmidt} {et~al.}(2008){Schmidt}, {Smith}, {Szkody}, \&
  {Anderson}}]{Schmidt_2008}
{Schmidt}, G.~D., {Smith}, P.~S., {Szkody}, P., \& {Anderson}, S.~F. 2008,
  \pasp, 120, 160

\bibitem[{{Schneider} {et~al.}(2010){Schneider}, {Richards}, {Hall}, {Strauss},
  {Anderson}, {Boroson}, {Ross}, {Shen}, {Brandt}, {Fan}, {Inada}, {Jester},
  {Knapp}, {Krawczyk}, {Thakar}, {Vanden Berk}, {Voges}, {Yanny}, {York},
  {Bahcall}, {Bizyaev}, {Blanton}, {Brewington}, {Brinkmann}, {Eisenstein},
  {Frieman}, {Fukugita}, {Gray}, {Gunn}, {Hibon}, {Ivezi{\'c}}, {Kent}, {Kron},
  {Lee}, {Lupton}, {Malanushenko}, {Malanushenko}, {Oravetz}, {Pan}, {Pier},
  {Price}, {Saxe}, {Schlegel}, {Simmons}, {Snedden}, {SubbaRao}, {Szalay}, \&
  {Weinberg}}]{Schneider_2010}
{Schneider}, D.~P., {Richards}, G.~T., {Hall}, P.~B., {et~al.} 2010, \aj, 139,
  2360

\bibitem[{{Shen} {et~al.}(2011){Shen}, {Richards}, {Strauss}, {Hall},
  {Schneider}, {Snedden}, {Bizyaev}, {Brewington}, {Malanushenko},
  {Malanushenko}, {Oravetz}, {Pan}, \& {Simmons}}]{Shen_2011}
{Shen}, Y., {Richards}, G.~T., {Strauss}, M.~A., {et~al.} 2011, \apjs, 194, 45

\bibitem[{{Shimwell} {et~al.}(2019){Shimwell}, {Tasse}, {Hardcastle}, {Mechev},
  {Williams}, {Best}, {R{\"o}ttgering}, {Callingham}, {Dijkema}, {de Gasperin},
  {Hoang}, {Hugo}, {Mirmont}, {Oonk}, {Prandoni}, {Rafferty}, {Sabater},
  {Smirnov}, {van Weeren}, {White}, {Atemkeng}, {Bester}, {Bonnassieux},
  {Br{\"u}ggen}, {Brunetti}, {Chy{\.z}y}, {Cochrane}, {Conway}, {Croston},
  {Danezi}, {Duncan}, {Haverkorn}, {Heald}, {Iacobelli}, {Intema}, {Jackson},
  {Jamrozy}, {Jarvis}, {Lakhoo}, {Mevius}, {Miley}, {Morabito}, {Morganti},
  {Nisbet}, {Orr{\'u}}, {Perkins}, {Pizzo}, {Schrijvers}, {Smith}, {Vermeulen},
  {Wise}, {Alegre}, {Bacon}, {van Bemmel}, {Beswick}, {Bonafede}, {Botteon},
  {Bourke}, {Brienza}, {Calistro Rivera}, {Cassano}, {Clarke}, {Conselice},
  {Dettmar}, {Drabent}, {Dumba}, {Emig}, {En{\ss}lin}, {Ferrari}, {Garrett},
  {G{\'e}nova-Santos}, {Goyal}, {G{\"u}rkan}, {Hale}, {Harwood}, {Heesen},
  {Hoeft}, {Horellou}, {Jackson}, {Kokotanekov}, {Kondapally},
  {Kunert-Bajraszewska}, {Mahatma}, {Mahony}, {Mandal}, {McKean}, {Merloni},
  {Mingo}, {Miskolczi}, {Mooney}, {Nikiel-Wroczy{\'n}ski}, {O'Sullivan},
  {Quinn}, {Reich}, {Roskowi{\'n}ski}, {Rowlinson}, {Savini}, {Saxena},
  {Schwarz}, {Shulevski}, {Sridhar}, {Stacey}, {Urquhart}, {van der Wiel},
  {Varenius}, {Webster}, \& {Wilber}}]{Shimwell_2019}
{Shimwell}, T.~W., {Tasse}, C., {Hardcastle}, M.~J., {et~al.} 2019, \aap, 622,
  A1

\bibitem[{{Sprayberry} \& {Foltz}(1992)}]{Sprayberry_1992}
{Sprayberry}, D. \& {Foltz}, C.~B. 1992, \apj, 390, 39

\bibitem[{{Stern} {et~al.}(2012){Stern}, {Assef}, {Benford}, {Blain}, {Cutri},
  {Dey}, {Eisenhardt}, {Griffith}, {Jarrett}, {Lake}, {Masci}, {Petty},
  {Stanford}, {Tsai}, {Wright}, {Yan}, {Harrison}, \& {Madsen}}]{Stern_2012}
{Stern}, D., {Assef}, R.~J., {Benford}, D.~J., {et~al.} 2012, \apj, 753, 30

\bibitem[{{Strateva} {et~al.}(2003){Strateva}, {Strauss}, {Hao}, {Schlegel},
  {Hall}, {Gunn}, {Li}, {Ivezi{\'c}}, {Richards}, {Zakamska}, {Voges},
  {Anderson}, {Lupton}, {Schneider}, {Brinkmann}, \& {Nichol}}]{Strateva_2003}
{Strateva}, I.~V., {Strauss}, M.~A., {Hao}, L., {et~al.} 2003, \aj, 126, 1720

\bibitem[{{Szkody} {et~al.}(2003){Szkody}, {Anderson}, {Schmidt}, {Hall},
  {Margon}, {Miceli}, {SubbaRao}, {Frith}, {Harris}, {Hawley}, {Lawton},
  {Covarrubias}, {Covey}, {Fan}, {Murphy}, {Narayanan}, {Raymond}, {Rest},
  {Strauss}, {Stubbs}, {Turner}, {Voges}, {Bauer}, {Brinkmann}, {Knapp}, \&
  {Schneider}}]{Szkody_2003}
{Szkody}, P., {Anderson}, S.~F., {Schmidt}, G., {et~al.} 2003, \apj, 583, 902

\bibitem[{{Terwel} \& {Jonker}(2022)}]{Terwel_2022}
{Terwel}, J.~H. \& {Jonker}, P.~G. 2022, \mnras, 512, L80

\bibitem[{{Tiwari} \& {Vivek}(2025)}]{Tiwari_2025}
{Tiwari}, A. \& {Vivek}, M. 2025, \aap, 699, A132

\bibitem[{{Trump} {et~al.}(2006){Trump}, {Hall}, {Reichard}, {Richards},
  {Schneider}, {Vanden Berk}, {Knapp}, {Anderson}, {Fan}, {Brinkman},
  {Kleinman}, \& {Nitta}}]{Trump_2006}
{Trump}, J.~R., {Hall}, P.~B., {Reichard}, T.~A., {et~al.} 2006, \apjs, 165, 1

\bibitem[{{Urrutia} {et~al.}(2009){Urrutia}, {Becker}, {White}, {Glikman},
  {Lacy}, {Hodge}, \& {Gregg}}]{Urrutia_2009}
{Urrutia}, T., {Becker}, R.~H., {White}, R.~L., {et~al.} 2009, \apj, 698, 1095

\bibitem[{{Vanden Berk} {et~al.}(2001){Vanden Berk}, {Richards}, {Bauer},
  {Strauss}, {Schneider}, {Heckman}, {York}, {Hall}, {Fan}, {Knapp},
  {Anderson}, {Annis}, {Bahcall}, {Bernardi}, {Briggs}, {Brinkmann}, {Brunner},
  {Burles}, {Carey}, {Castander}, {Connolly}, {Crocker}, {Csabai}, {Doi},
  {Finkbeiner}, {Friedman}, {Frieman}, {Fukugita}, {Gunn}, {Hennessy},
  {Ivezi{\'c}}, {Kent}, {Kunszt}, {Lamb}, {Leger}, {Long}, {Loveday}, {Lupton},
  {Meiksin}, {Merelli}, {Munn}, {Newberg}, {Newcomb}, {Nichol}, {Owen}, {Pier},
  {Pope}, {Rockosi}, {Schlegel}, {Siegmund}, {Smee}, {Snir}, {Stoughton},
  {Stubbs}, {SubbaRao}, {Szalay}, {Szokoly}, {Tremonti}, {Uomoto}, {Waddell},
  {Yanny}, \& {Zheng}}]{VandenBerk_2001}
{Vanden Berk}, D.~E., {Richards}, G.~T., {Bauer}, A., {et~al.} 2001, \aj, 122,
  549

\bibitem[{{Voit} {et~al.}(1993){Voit}, {Weymann}, \& {Korista}}]{Voit_1993}
{Voit}, G.~M., {Weymann}, R.~J., \& {Korista}, K.~T. 1993, \apj, 413, 95

\bibitem[{{Wang} {et~al.}(2017){Wang}, {Xu}, \& {Wei}}]{Wang_2017}
{Wang}, J., {Xu}, D., \& {Wei}, J. 2017, Frontiers in Astronomy and Space
  Sciences, 4, 40

\bibitem[{{Wei} {et~al.}(2013){Wei}, {Luo}, {Li}, {Pan}, {Tu}, {Jiang}, {Kong},
  {Shi}, {Yi}, {Wang}, {Liu}, \& {Zhao}}]{Wei_2013}
{Wei}, P., {Luo}, A., {Li}, Y., {et~al.} 2013, \mnras, 431, 1800

\bibitem[{{Wevers} {et~al.}(2022){Wevers}, {Nicholl}, {Guolo},
  {Charalampopoulos}, {Gromadzki}, {Reynolds}, {Kankare}, {Leloudas},
  {Anderson}, {Arcavi}, {Cannizzaro}, {Chen}, {Ihanec}, {Inserra},
  {Guti{\'e}rrez}, {Jonker}, {Lawrence}, {Magee}, {M{\"u}ller-Bravo}, {Onori},
  {Ridley}, {Schulze}, {Short}, {Hiramatsu}, {Newsome}, {Terwel}, {Yang}, \&
  {Young}}]{Wevers_2022}
{Wevers}, T., {Nicholl}, M., {Guolo}, M., {et~al.} 2022, \aap, 666, A6

\bibitem[{{Weymann} {et~al.}(1991){Weymann}, {Morris}, {Foltz}, \&
  {Hewett}}]{Weymann_1991}
{Weymann}, R.~J., {Morris}, S.~L., {Foltz}, C.~B., \& {Hewett}, P.~C. 1991,
  \apj, 373, 23

\bibitem[{{Wright} {et~al.}(2010){Wright}, {Eisenhardt}, {Mainzer}, {Ressler},
  {Cutri}, {Jarrett}, {Kirkpatrick}, {Padgett}, {McMillan}, {Skrutskie},
  {Stanford}, {Cohen}, {Walker}, {Mather}, {Leisawitz}, {Gautier}, {McLean},
  {Benford}, {Lonsdale}, {Blain}, {Mendez}, {Irace}, {Duval}, {Liu}, {Royer},
  {Heinrichsen}, {Howard}, {Shannon}, {Kendall}, {Walsh}, {Larsen}, {Cardon},
  {Schick}, {Schwalm}, {Abid}, {Fabinsky}, {Naes}, \& {Tsai}}]{Wright_2010}
{Wright}, E.~L., {Eisenhardt}, P. R.~M., {Mainzer}, A.~K., {et~al.} 2010, \aj,
  140, 1868

\bibitem[{{Yi} {et~al.}(2022){Yi}, {Brandt}, {Ni}, {Ho}, {Luo}, {Yan},
  {Schneider}, {Paul}, {Plotkin}, {Yang}, {Wang}, {He}, {Chen}, {Wu}, \&
  {Bai}}]{Yi_2022}
{Yi}, W., {Brandt}, W.~N., {Ni}, Q., {et~al.} 2022, \apj, 930, 5

\bibitem[{{York} {et~al.}(2000){York}, {Adelman}, {Anderson}, {Anderson},
  {Annis}, {Bahcall}, {Bakken}, {Barkhouser}, {Bastian}, {Berman}, {Boroski},
  {Bracker}, {Briegel}, {Briggs}, {Brinkmann}, {Brunner}, {Burles}, {Carey},
  {Carr}, {Castander}, {Chen}, {Colestock}, {Connolly}, {Crocker}, {Csabai},
  {Czarapata}, {Davis}, {Doi}, {Dombeck}, {Eisenstein}, {Ellman}, {Elms},
  {Evans}, {Fan}, {Federwitz}, {Fiscelli}, {Friedman}, {Frieman}, {Fukugita},
  {Gillespie}, {Gunn}, {Gurbani}, {de Haas}, {Haldeman}, {Harris}, {Hayes},
  {Heckman}, {Hennessy}, {Hindsley}, {Holm}, {Holmgren}, {Huang}, {Hull},
  {Husby}, {Ichikawa}, {Ichikawa}, {Ivezi{\'c}}, {Kent}, {Kim}, {Kinney},
  {Klaene}, {Kleinman}, {Kleinman}, {Knapp}, {Korienek}, {Kron}, {Kunszt},
  {Lamb}, {Lee}, {Leger}, {Limmongkol}, {Lindenmeyer}, {Long}, {Loomis},
  {Loveday}, {Lucinio}, {Lupton}, {MacKinnon}, {Mannery}, {Mantsch}, {Margon},
  {McGehee}, {McKay}, {Meiksin}, {Merelli}, {Monet}, {Munn}, {Narayanan},
  {Nash}, {Neilsen}, {Neswold}, {Newberg}, {Nichol}, {Nicinski}, {Nonino},
  {Okada}, {Okamura}, {Ostriker}, {Owen}, {Pauls}, {Peoples}, {Peterson},
  {Petravick}, {Pier}, {Pope}, {Pordes}, {Prosapio}, {Rechenmacher}, {Quinn},
  {Richards}, {Richmond}, {Rivetta}, {Rockosi}, {Ruthmansdorfer}, {Sandford},
  {Schlegel}, {Schneider}, {Sekiguchi}, {Sergey}, {Shimasaku}, {Siegmund},
  {Smee}, {Smith}, {Snedden}, {Stone}, {Stoughton}, {Strauss}, {Stubbs},
  {SubbaRao}, {Szalay}, {Szapudi}, {Szokoly}, {Thakar}, {Tremonti}, {Tucker},
  {Uomoto}, {Vanden Berk}, {Vogeley}, {Waddell}, {Wang}, {Watanabe},
  {Weinberg}, {Yanny}, {Yasuda}, \& {SDSS Collaboration}}]{York_2000}
{York}, D.~G., {Adelman}, J., {Anderson}, John~E., J., {et~al.} 2000, \aj, 120,
  1579

\bibitem[{{Yuan} {et~al.}(2013){Yuan}, {Liu}, \& {Xiang}}]{Yuan_2013}
{Yuan}, H.~B., {Liu}, X.~W., \& {Xiang}, M.~S. 2013, \mnras, 430, 2188

\end{thebibliography}

\begin{appendix}

\section{The catalogue}\label{sect:Catalogue}

\begin{table}[htb!]
\begin{minipage}[t]{1.0\textwidth}
\caption{Re-classified high-$z$ spectra from SDSS DR16. }
\begin{tabular}{rrrcccccccc} 
\hline\hline 
\noalign{\smallskip}
& RA
& DEC
& Spectrum
& Target
& $n_{\rm sp}$
& $z_{\rm DR16}$
& Error $z_{\rm DR16}$
& Survey 
& Class
& Subclass\\
& (1)
& (2)
& (3)
& (4) 
& (5)
& (6) 
& (7) 
& (8)
& (9)
& (10) \\
\hline
\noalign{\smallskip}
  1 &   0.0348 &  -3.8380 & \ 7895-57659-0889 &   QSO &   1 &  5.3719 & 0.0014 & eBOSS & QSO &      -        \\
  2 &   0.0638 &   6.1766 & \ 8740-57367-0686 &   QSO &   1 &  6.5350 & 0.0003 & eBOSS & QSO &       -       \\
  3 &   0.2621 &  35.4040 & \ 7750-58402-0841 &   QSO &   1 &  5.7173 & 0.0010 & eBOSS & QSO &      -          \\
  4 &   0.3064 &   6.4368 & 11279-58449-0028&   galaxy &   1 &  7.0305 & 0.0237 & eBOSS & QSO &      -          \\
\noalign{\smallskip}
\end{tabular}
\begin{tabular}{ccccccccl} 
\hline\hline 
\noalign{\smallskip}
& $z_{\rm QC}$
& $z $
& $r_z$
& Sel
& Spectype
& Subtype
& Quality
& Remark \\
& (11)
& (12)
& (13)
& (14) 
& (15)
& (16)  
& (17)  
& (18) \\
\hline
\noalign{\smallskip}
    1   & 1.760  &   1.757 &  3 &  1 & QSO     &    -    &    -    & red, IALs    \\        
    2   & 0.833  &   0.832 &  3 &  1 & galaxy  & ELs     &  -      &     -        \\        
    3   &     -    &      *  &  0 &  1 & unknown &   -     & noisy   &     -        \\        
    4   &     -   &      *  &  0 &  1 & unknown &   -     & noisy   &     -        \\   
\hline                 
\noalign{\smallskip}
\end{tabular}     
\tablefoot{
(1) - (2) right ascension and declination J2000  (degrees);
(3)  name of spectrum (plate-mjd-fiber);
(4)  SDSS target type (see Sect.\,\ref{sect:Discussion});
(5)  number of spectra at this position;
(6) - (7) spectroscopic redshift and error from SDSS DR16;
(8)  survey name;
(9) - (10) spectroscopic class and subclass from SDSS DR16;
(11) redshift from QCDR16 \citep{Lyke_2020}; 
(12) revised redshift from present study; 
(13) reliability of revised redshift  (3 = good, 2 = reasonable, 1 = tentative, 0 = unknown);
(14) selected as best version of SDSS spectrum at this location among the spectra in the present study;
(15) revised spectroscopic type; 
(16) revised spectroscopic subtype  (see Table\,\ref{tab:subtypes});
(17) remark on bad quality;                                               
(18) remarks on special features in the spectrum (see Table\,\ref{tab:remarks}).
Only the first four rows of the table are shown here for illustration.  The full list  is available in electronic form  from the VizieR Service of the CDS Strasbourg.                                                
}                                                                                    
\label{tab:cat}
\end{minipage}
\end{table}

\begin{table}[!h]
\vspace{0.0cm}
\begin{minipage}[b]{0.5\textwidth}
\caption{Acronyms used in the subtype column of Table\ref{tab:cat}.}
\begin{tabular}{ll} 
\hline\hline 
\noalign{\smallskip}
Acronym  &  Explanation \\
\noalign{\smallskip}
\hline                 
3ABQ    & 3000\,\AA\ break QSO                                          \\
ALs     & absorption lines                                          \\
C       & Carbon star                                               \\
ELs     & emission lines                                            \\
FeLoBAL & FeLoBAL QSO                                            \\
HiBAL   & HighBAL QSO                                               \\
LoBAL   & LowBAL QSO                                                \\
LT      & late-type star                                            \\
sdB     & subdwarf B star                                           \\
SN      & supernova                                                 \\
type2   & narrow-emission line QSO (type 2)                         \\
WD      & white dwarf      \\
        & (spectral type, if known, in parenthesis) \\
WDLT    & white dwarf plus late-type star                           \\
WLQ     & QSO with weak emission lines                              \\
\noalign{\smallskip}
\end{tabular}                                               
\label{tab:subtypes}
\end{minipage}
\end{table}

\begin{table}[h]
\vspace{10.9cm}
\begin{minipage}[b]{0.5\textwidth}
\caption{Acronyms used in the remark column of Table\ref{tab:cat}.}
\begin{tabular}{ll} 
\hline\hline 
\noalign{\smallskip}
Acronym  &  Explanation \\
\noalign{\smallskip}
\hline                 
AALs     & associated absorption lines                                 \\
AsBLs    & asymmetric broad Balmer lines                               \\
CyHs     & cyclotron humps in stellar spectrum                         \\
DBR      & displacement between blue and red part                      \\
DLAS     & damped Ly$\alpha$ system                                   \\
DLs      & double narrow emission lines                                \\
DPBLs    & double-peaked broad Balmer lines                            \\
Fe-em    & strong Fe II emission                                       \\
FeIII-em & strong FeIII emission                                       \\
GP       & possibly Gunn-Petterson trough                              \\
HBD      & high Balmer decrement                                       \\
IALs     & intermediate absorption lines                \\
Lya abs  & Ly$\alpha$ line significantly absorbed            \\
MATs     & many absorption troughs                                     \\
NBO      & bright object nearby                                        \\
NBS      & blue part dominated by nearby blue source   \\
NERP     & non-genuine emission peak near red edge      \\
PM       & significant proper motion (Gaia DR3)                        \\
PSB      & post-starburst                                              \\
red      & unusually red continuum                                     \\
RWA      & red wing of emission line absorbed                          \\
SELs     & star with remarkably strong emission lines                  \\
sLya     & very strong Ly$\alpha$ emission line                       \\
UC       & unusual continuum                                           \\
unusual  & spectral type identified, but unusual                       \\
\noalign{\smallskip}
\end{tabular}                                               
\label{tab:remarks}
\end{minipage}
\end{table}

\FloatBarrier

\section{Selected SDSS spectra}\label{sect:spectra}

\begin{figure*}[bhtp]
\includegraphics[viewport= 0 -30 1185 535,width=9.1cm,angle=0]{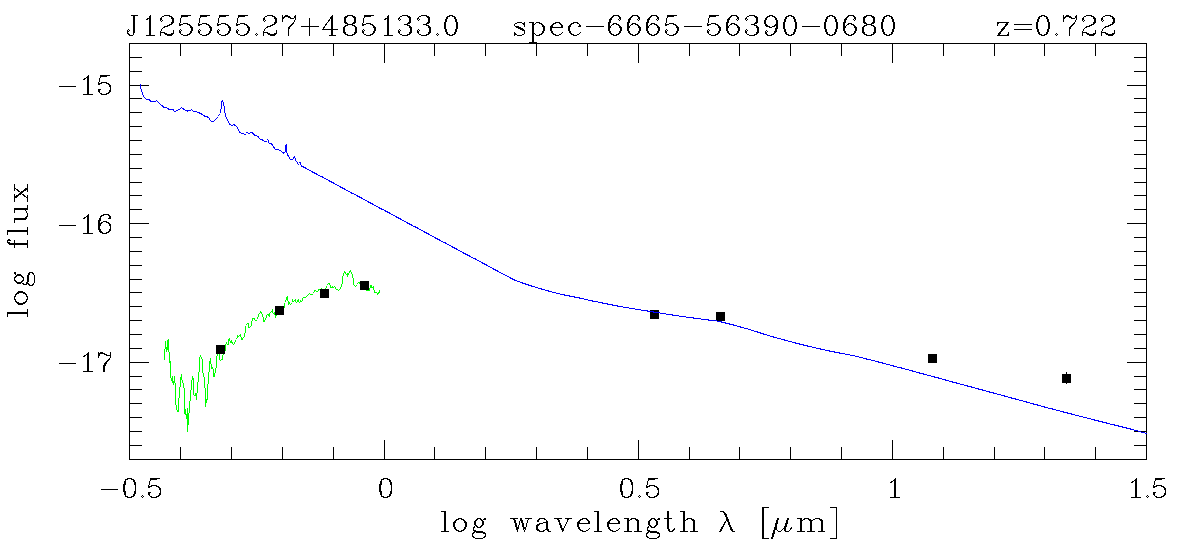} \
\includegraphics[viewport= 0 -30 1185 535,width=9.1cm,angle=0]{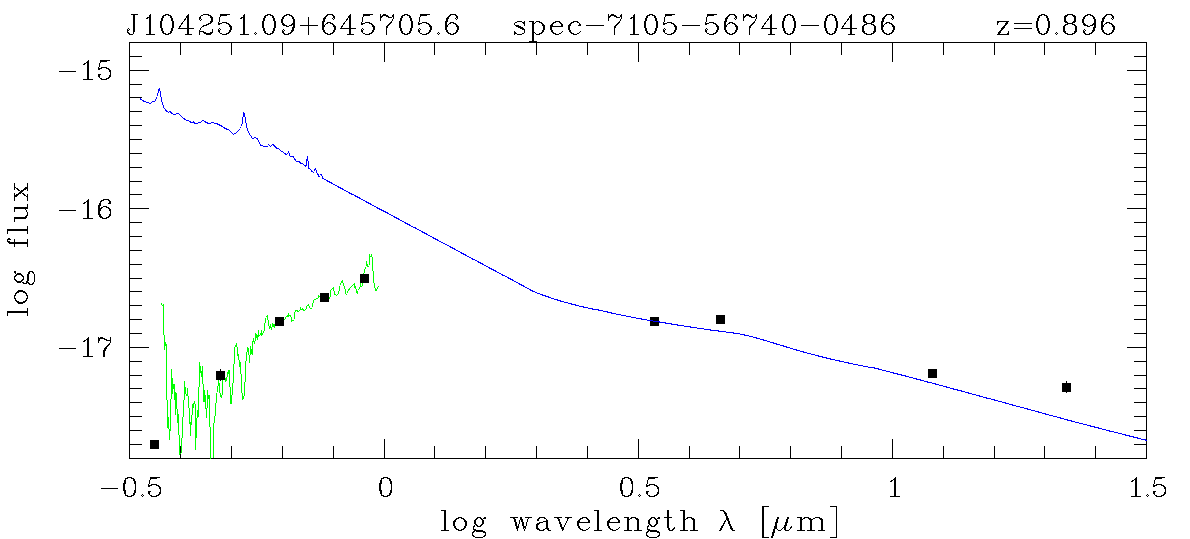} \\
\includegraphics[viewport= 0 -30 1185 535,width=9.1cm,angle=0]{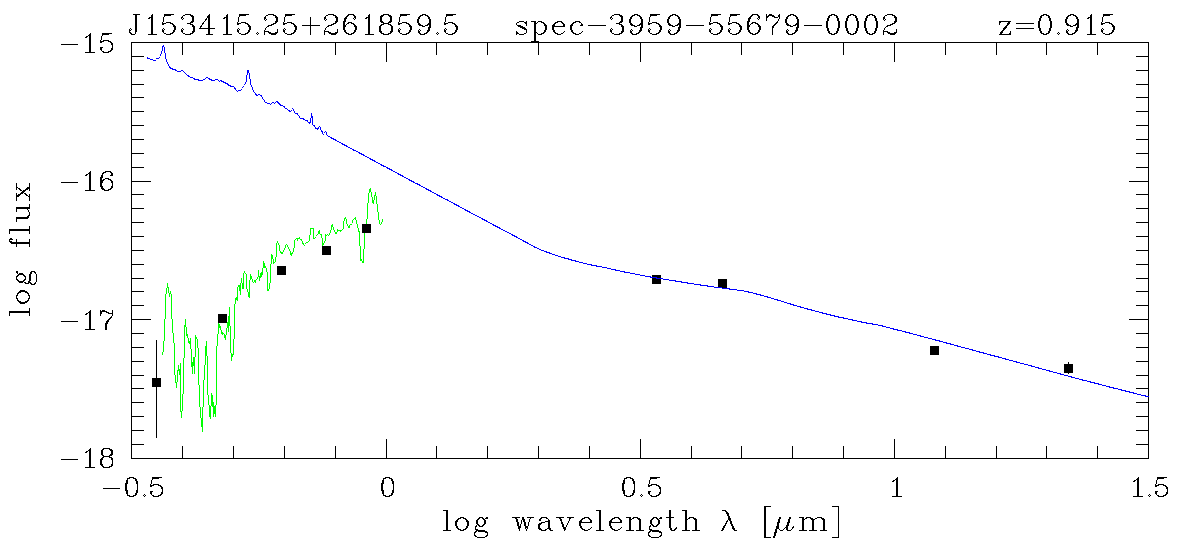} \
\includegraphics[viewport= 0 -30 1185 535,width=9.1cm,angle=0]{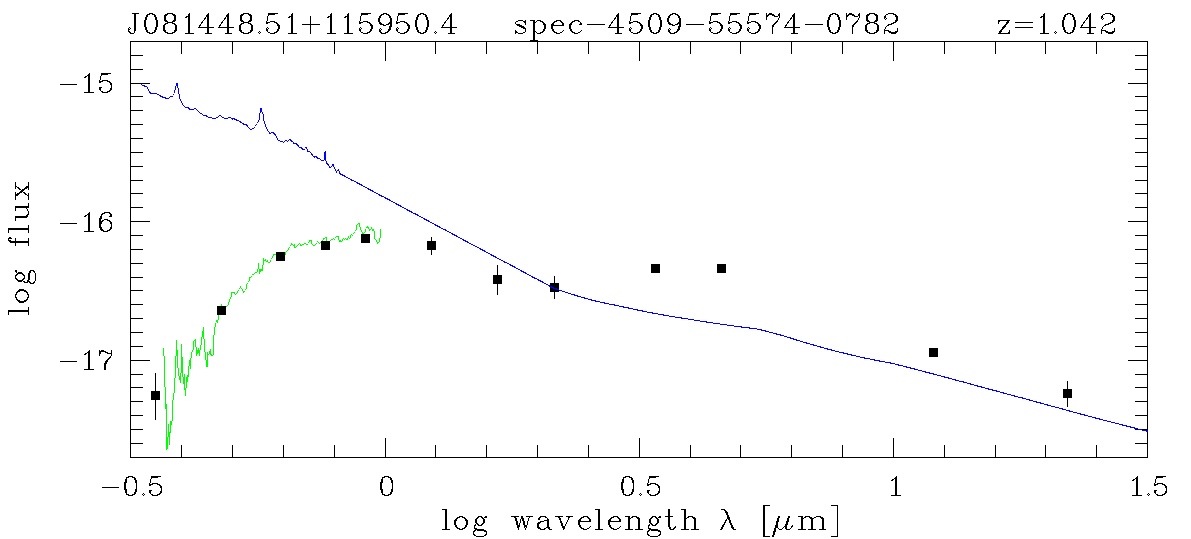} \\
\includegraphics[viewport= 0 -30 1185 535,width=9.1cm,angle=0]{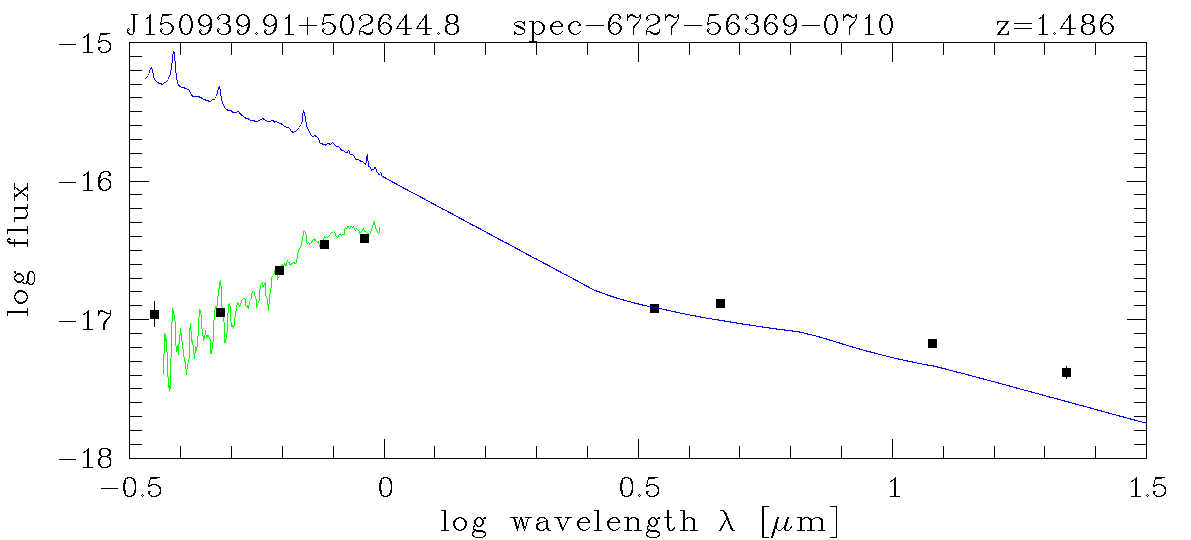} \
\includegraphics[viewport= 0 -30 1185 535,width=9.1cm,angle=0]{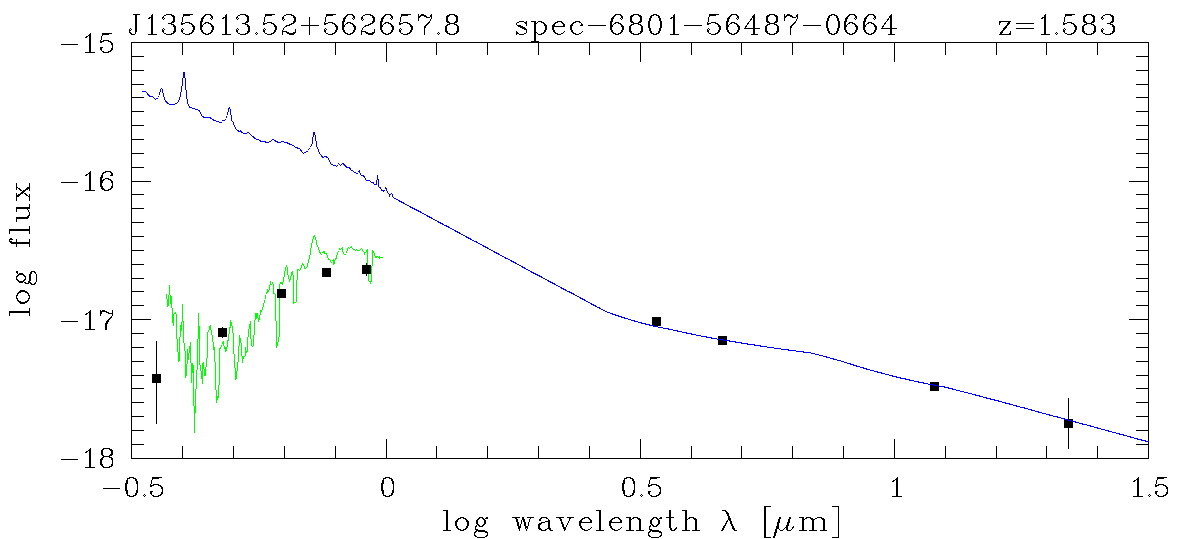} \\
\includegraphics[viewport= 0 -30 1185 535,width=9.1cm,angle=0]{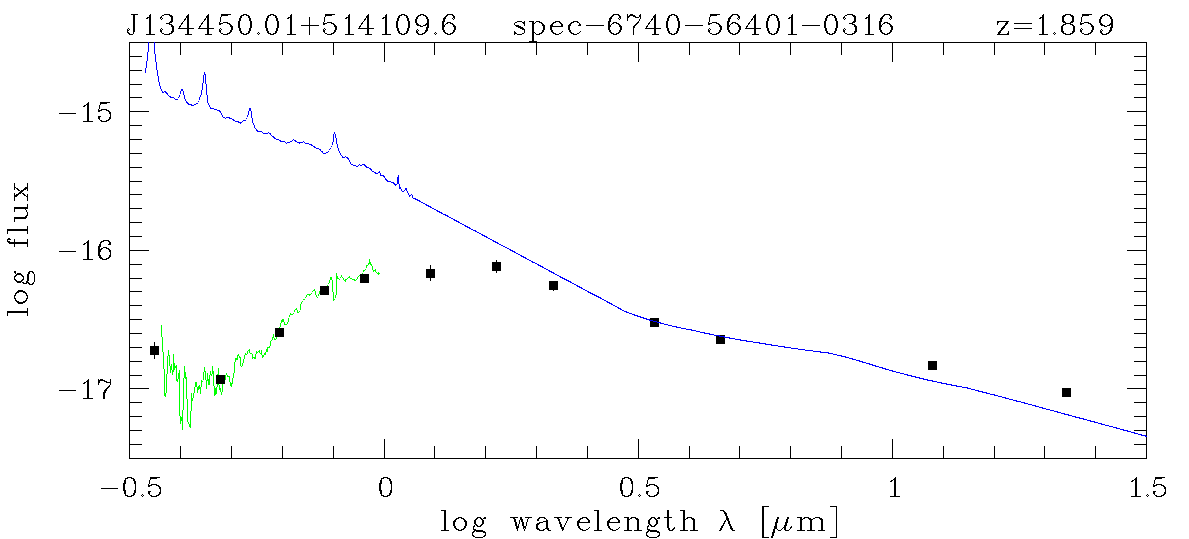} \
\includegraphics[viewport= 0 -30 1185 535,width=9.1cm,angle=0]{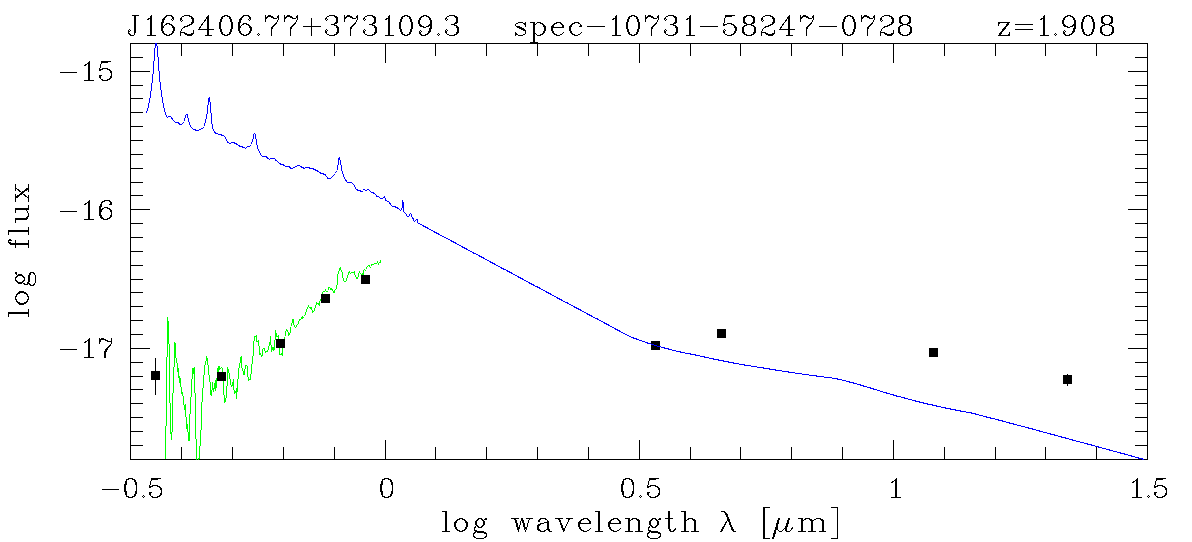} \\
\includegraphics[viewport= 0 -30 1185 535,width=9.1cm,angle=0]{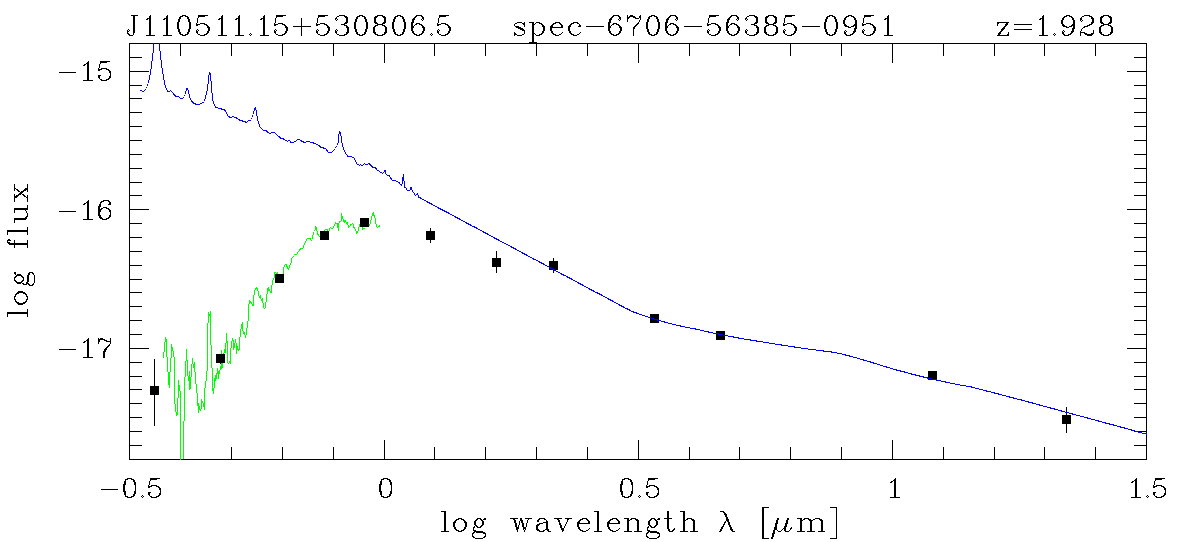} \
\includegraphics[viewport= 0 -30 1185 535,width=9.1cm,angle=0]{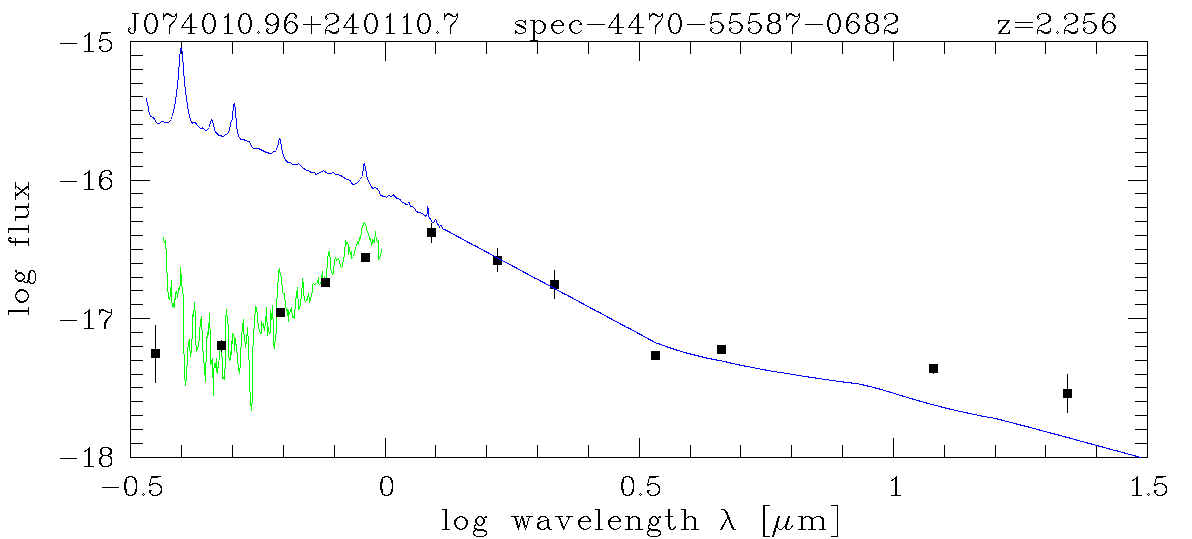} \\
\caption{MBSEDs (in the observer's frame) for the ten QSOs with the highest colour excess $E(g-z)$. 
The flux is in  units of erg\,s$^{-1}$\,cm$^{-2}$\,\AA$^{-1}$. Black squares with error bars are the fluxes from photometric data, the SDSS spectra are plotted in green (both corrected for Galactic foreground extinction). 
The template spectrum for bright QSOs of type 1 from \citet{Polletta_2007} is shown in blue.}
\label{fig:reddest_NonBALQSOs}
\end{figure*}

\begin{figure*}[bhtp]
\includegraphics[viewport= 0 -30 1185 535,width=9.1cm,angle=0]{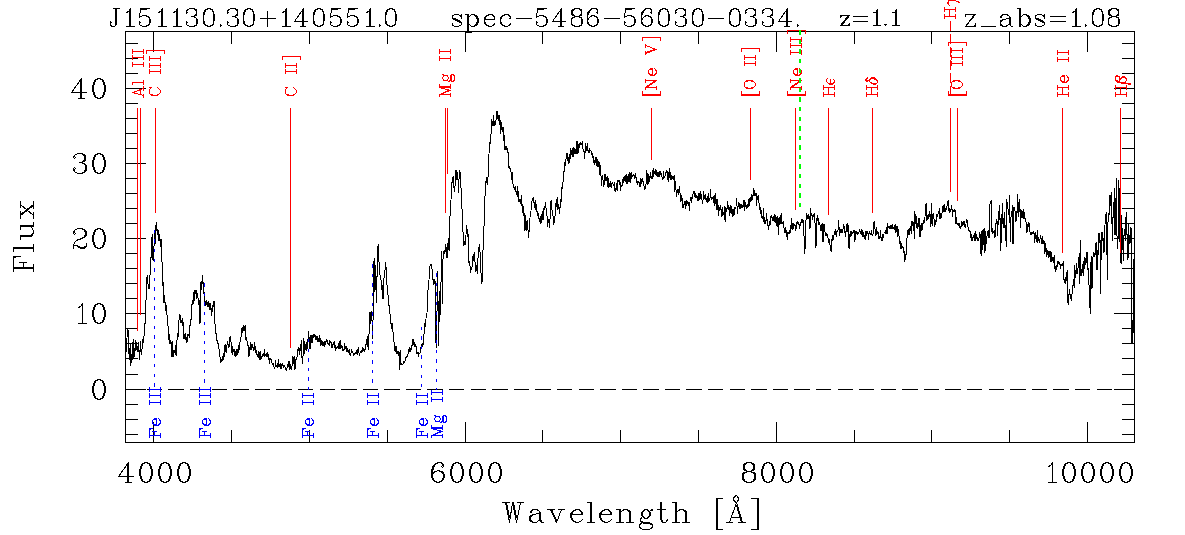} \
\includegraphics[viewport= 0 -30 1185 535,width=9.1cm,angle=0]{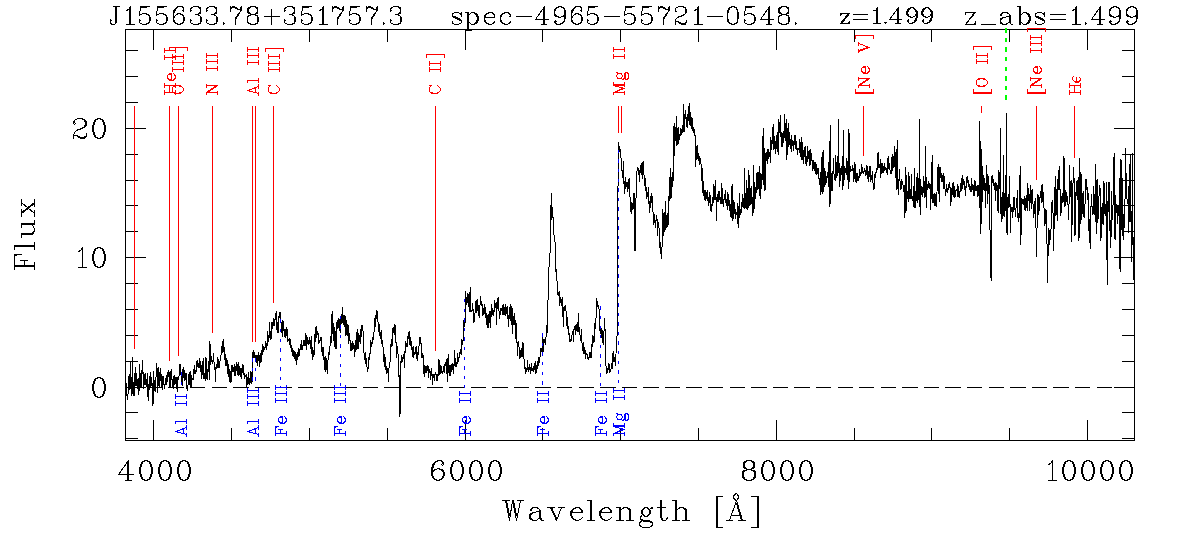} \\
\includegraphics[viewport= 0 -30 1185 535,width=9.1cm,angle=0]{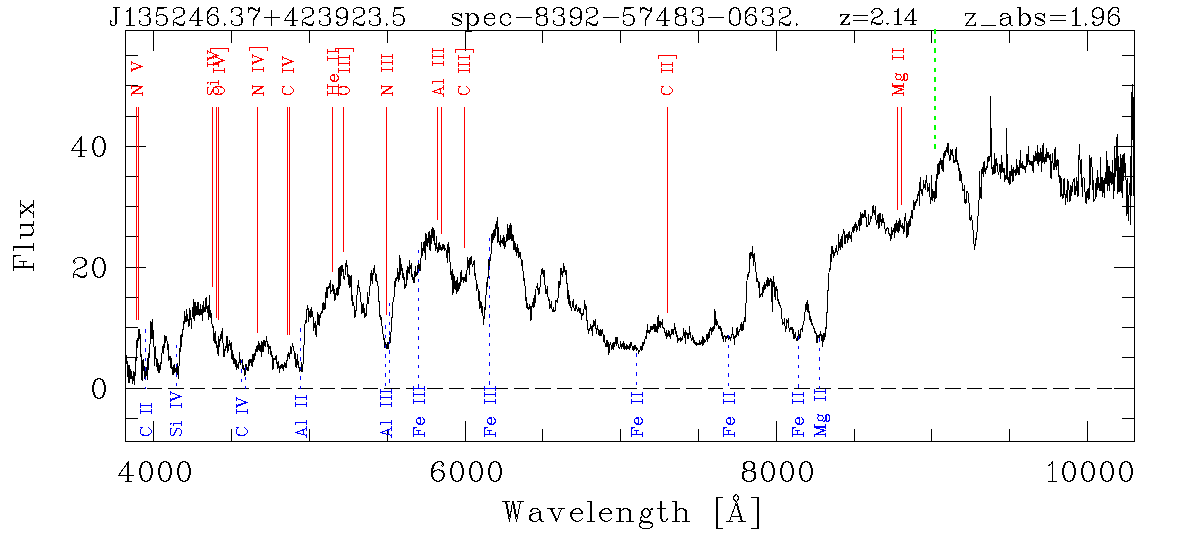} \
\includegraphics[viewport= 0 -30 1185 535,width=9.1cm,angle=0]{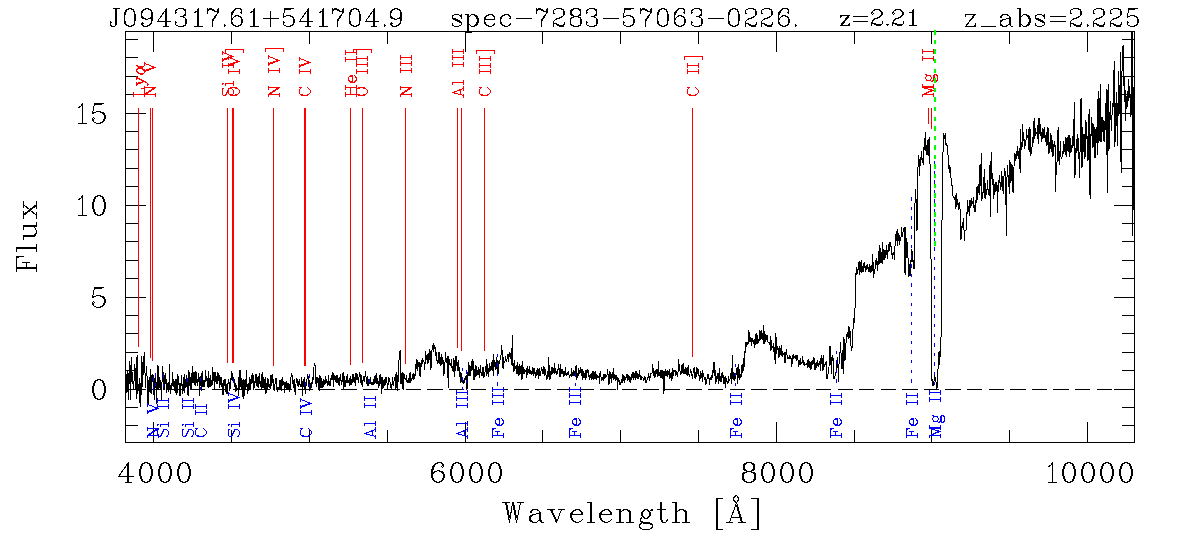} \\
\includegraphics[viewport= 0 -30 1185 535,width=9.1cm,angle=0]{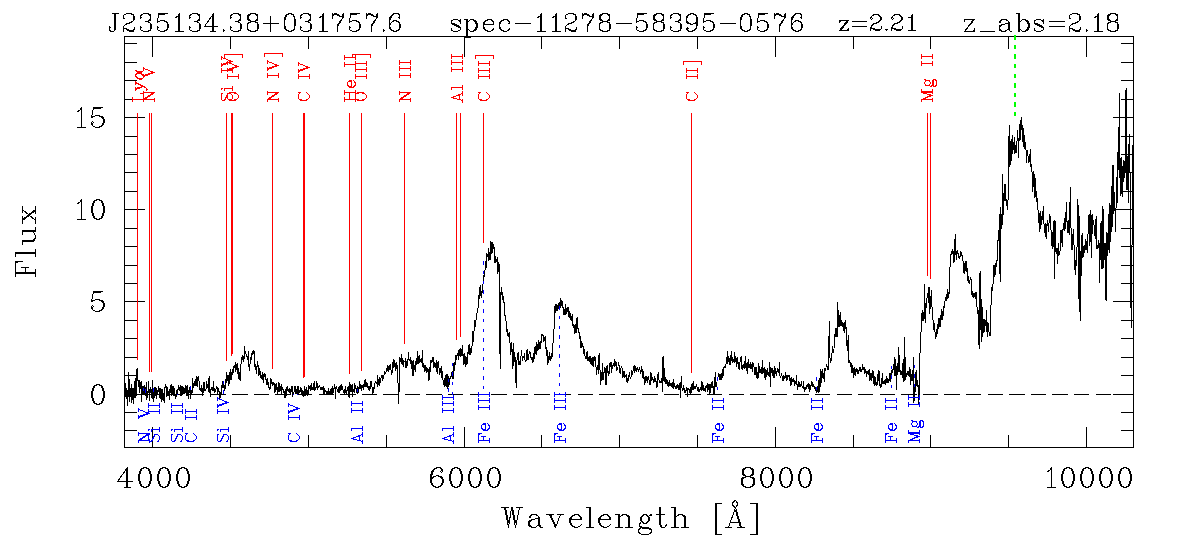} \
\includegraphics[viewport= 0 -30 1185 535,width=9.1cm,angle=0]{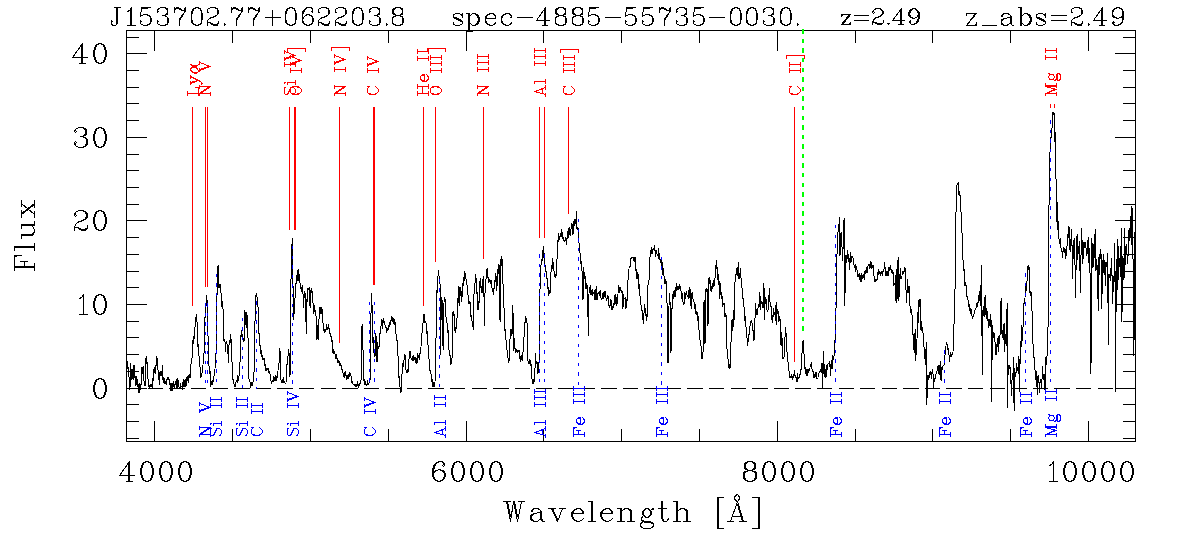} \\
\includegraphics[viewport= 0 -30 1185 535,width=9.1cm,angle=0]{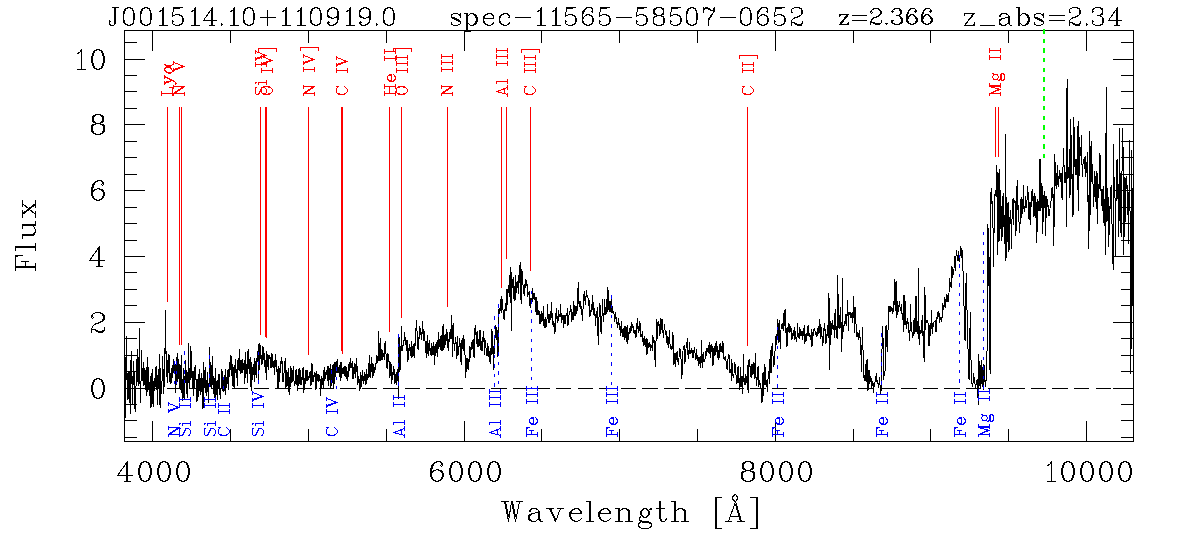} \
\includegraphics[viewport= 0 -30 1185 535,width=9.1cm,angle=0]{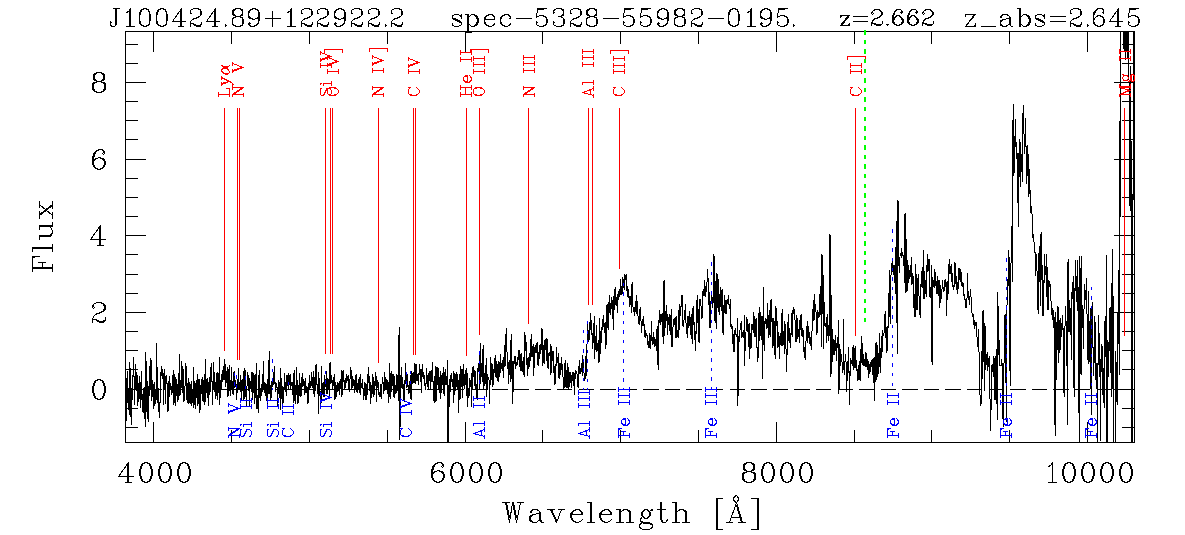} \\
\includegraphics[viewport= 0 -30 1185 535,width=9.1cm,angle=0]{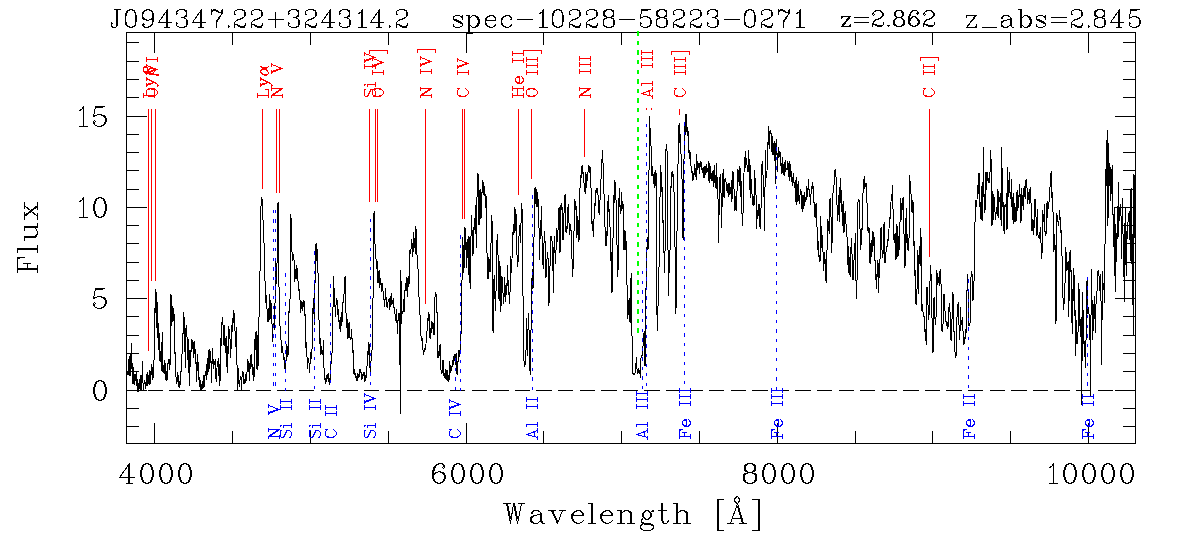} \
\includegraphics[viewport= 0 -30 1185 535,width=9.1cm,angle=0]{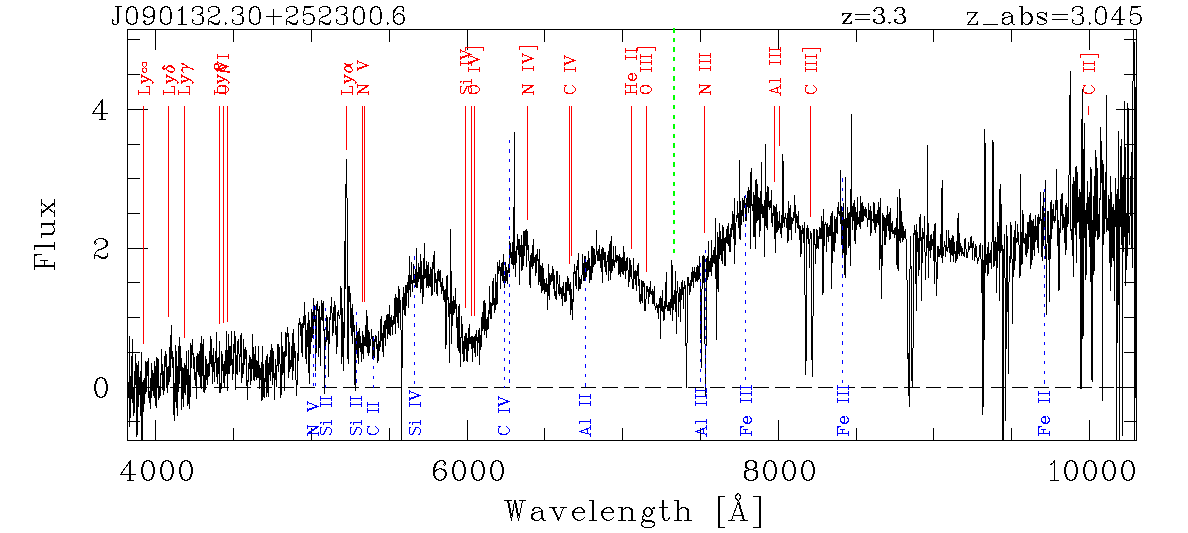} \\
\caption{Examples of FeLoBAL QSOs. The red solid vertical lines indicate the positions of the emission lines that are typical for QSO spectra,  the blue dotted lines mark the positions of typical absorption lines. The best estimates for the emission line redshift $z$ and the absorption line redshift $z_{\rm abs}$ are given at the top of each panel. 
The vertical green dotted line marks the position of the Ly\,$\alpha$ line for the redshift given by the SDSS DR16. 
The flux is in units of $10^{-17}$\,erg\,s$^{-1}$\,cm$^{-2}$\,\AA$^{-1}$.
The spectrum of SDSS\,J090132.30+252400.6 (bottom right) is the result of the coaddition of three SDSS spectra.
}
\label{fig:LoBALs}
\end{figure*}

\begin{figure*}[bhtp]
\includegraphics[viewport= 0 -30 1185 535,width=9.1cm,angle=0]{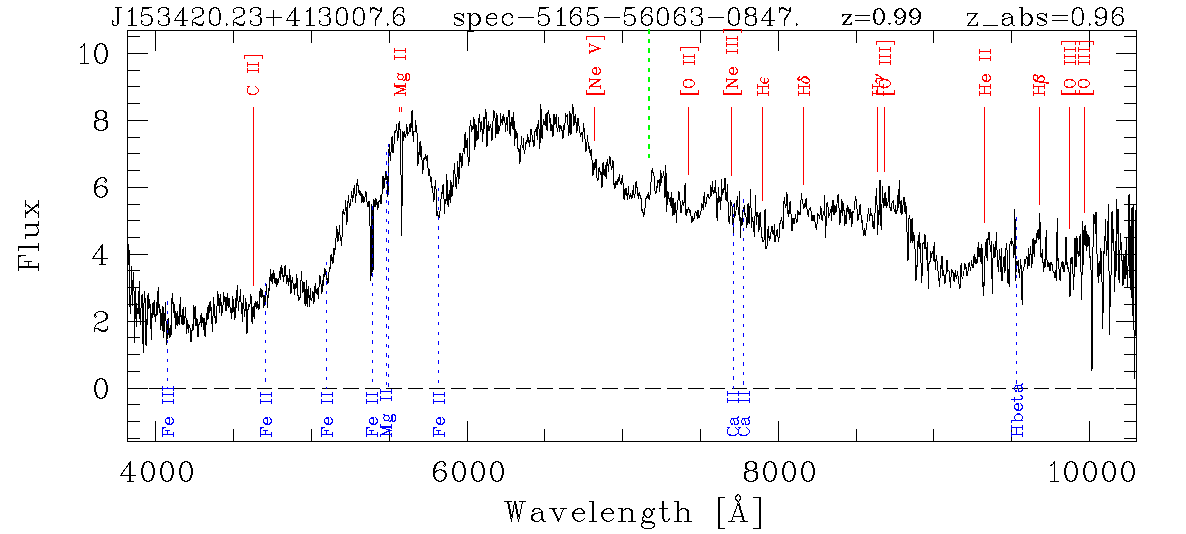} \
\includegraphics[viewport= 0 -30 1185 535,width=9.1cm,angle=0]{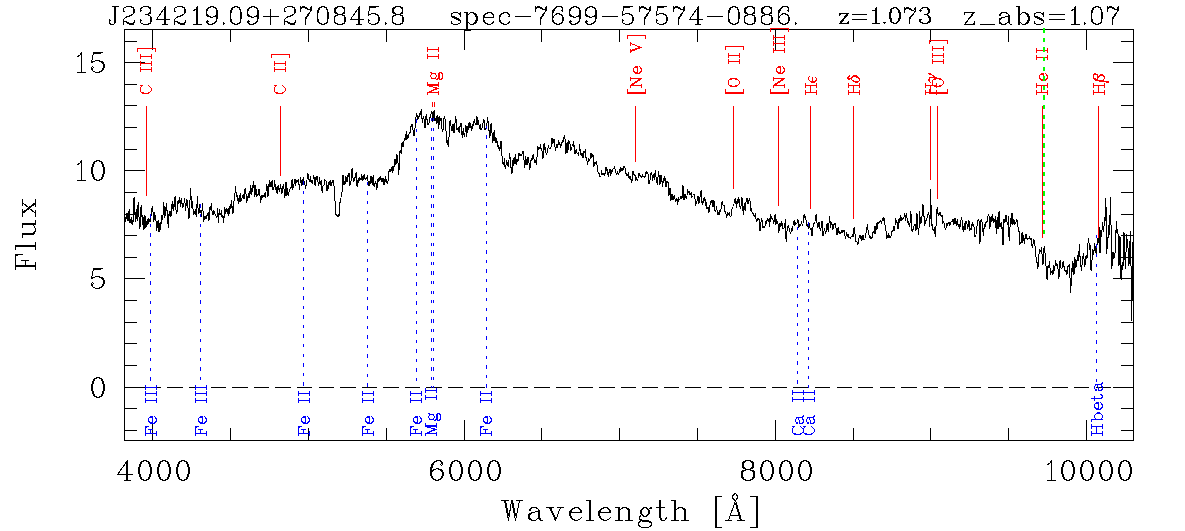} \\
\includegraphics[viewport= 0 -30 1185 535,width=9.1cm,angle=0]{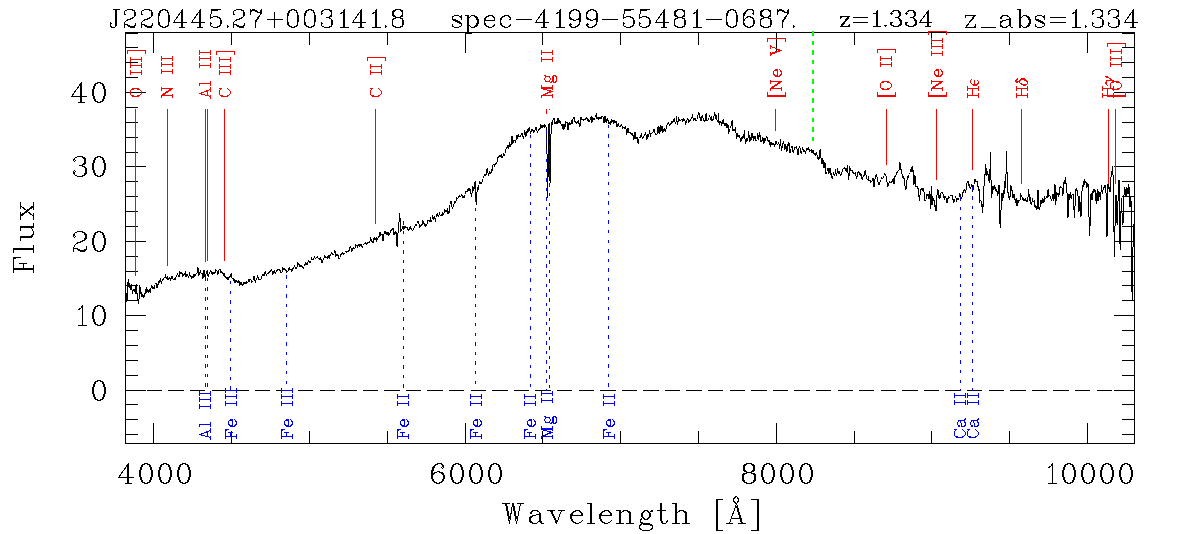} \
\includegraphics[viewport= 0 -30 1185 535,width=9.1cm,angle=0]{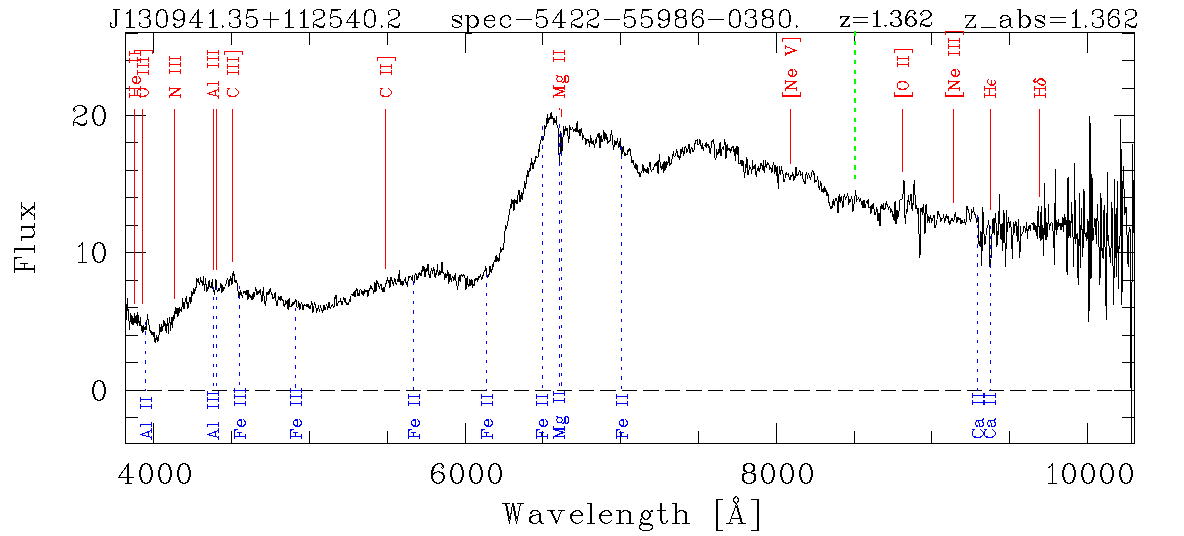} \\
\includegraphics[viewport= 0 -30 1185 535,width=9.1cm,angle=0]{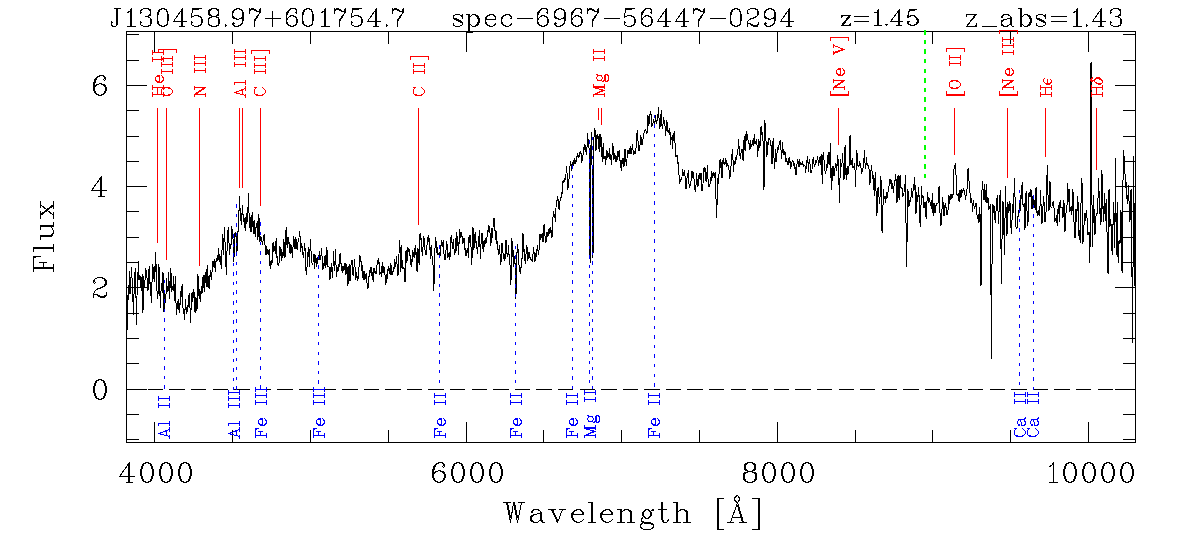} \
\includegraphics[viewport= 0 -30 1185 535,width=9.1cm,angle=0]{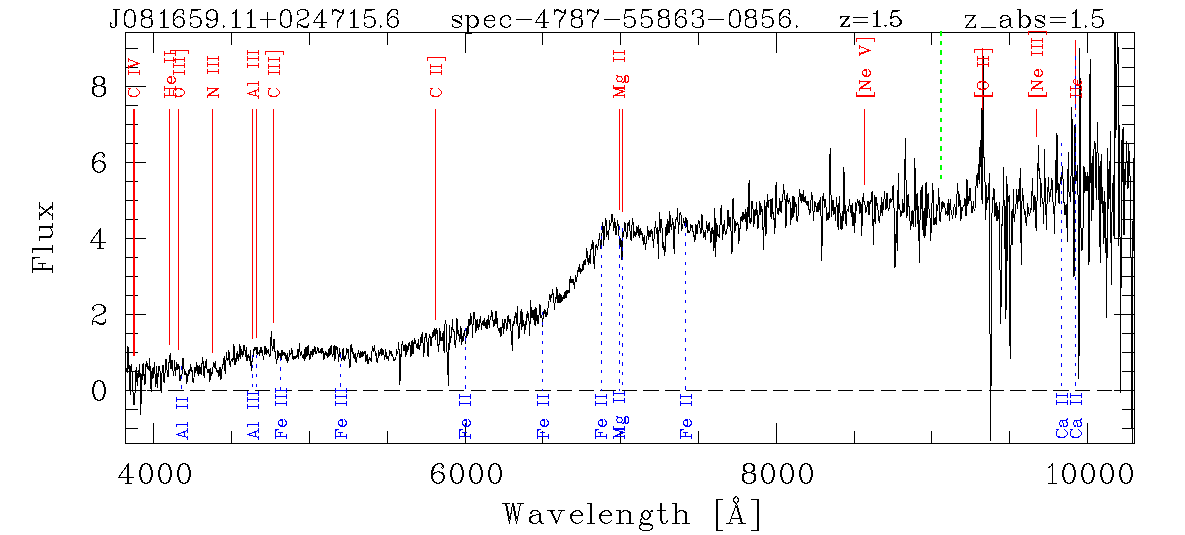}  \\
\includegraphics[viewport= 0 -30 1185 535,width=9.1cm,angle=0]{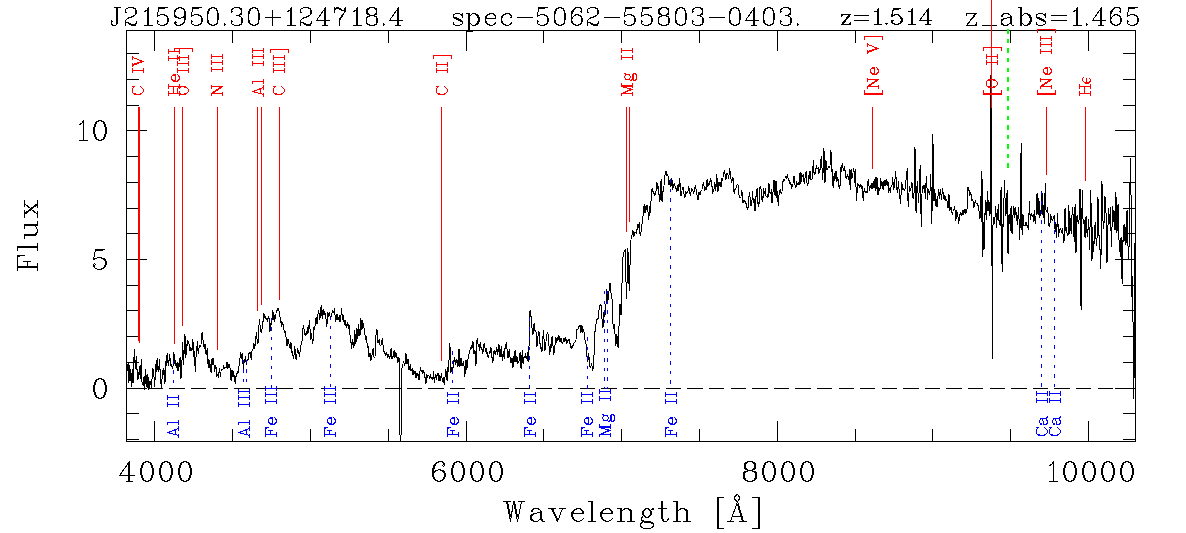} \
\includegraphics[viewport= 0 -30 1185 535,width=9.1cm,angle=0]{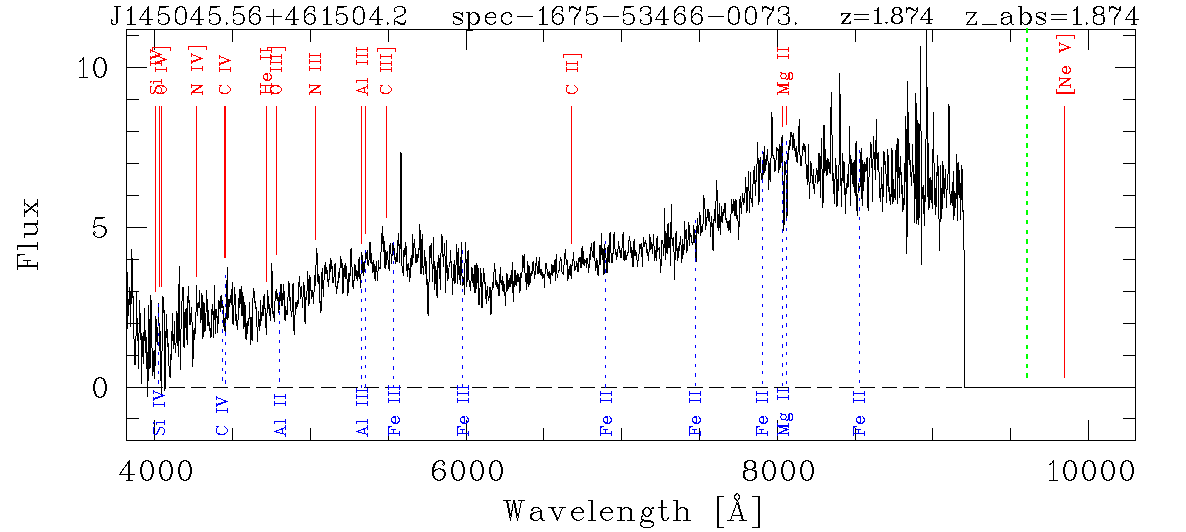} \\
\includegraphics[viewport= 0 -30 1185 535,width=9.1cm,angle=0]{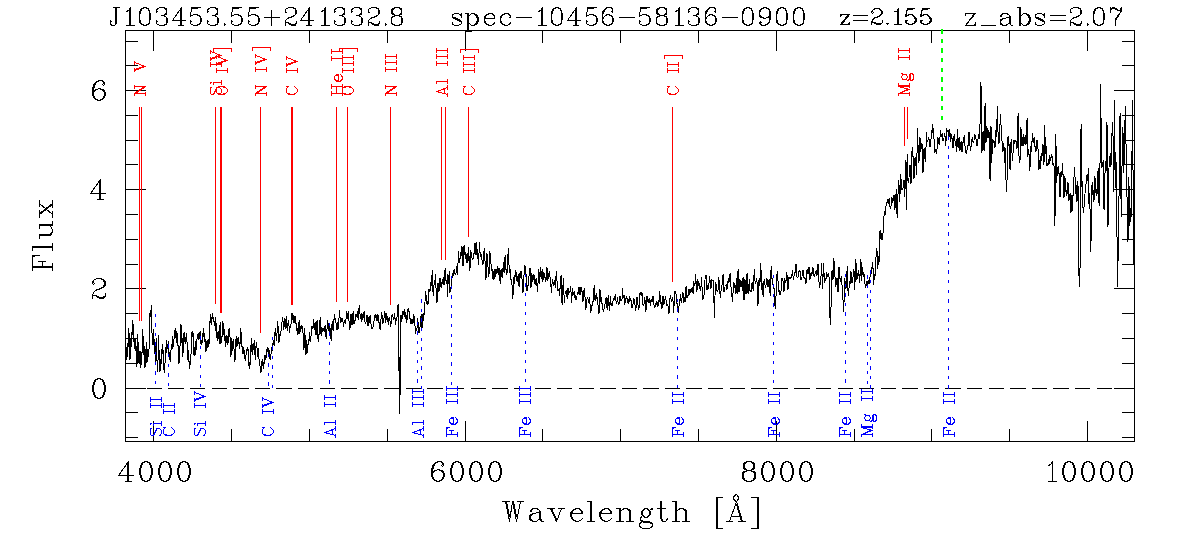} \
\includegraphics[viewport= 0 -30 1185 535,width=9.1cm,angle=0]{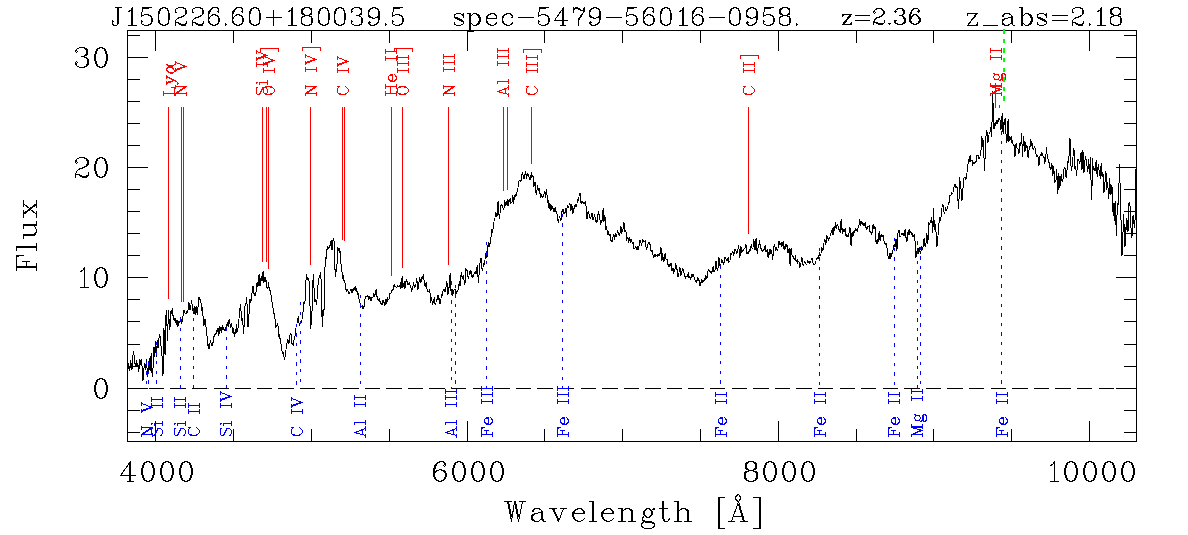}  \\
\caption{Examples of 3000\,\AA\ break QSOs or similar objects.   Some show structures reminiscent of FeLoBALs. 
In fact, SDSS J153420.23+413007.6 (top left) was previously classified as FeLoBAL QSO.
The unit of the flux and the meaning of the vertical lines are the same as in Fig.\,\ref{fig:LoBALs}.
}
\label{fig:3ABQs}
\end{figure*}

\begin{figure*}[bhtp]
\includegraphics[viewport= 0 -30 1185 535,width=9.1cm,angle=0]{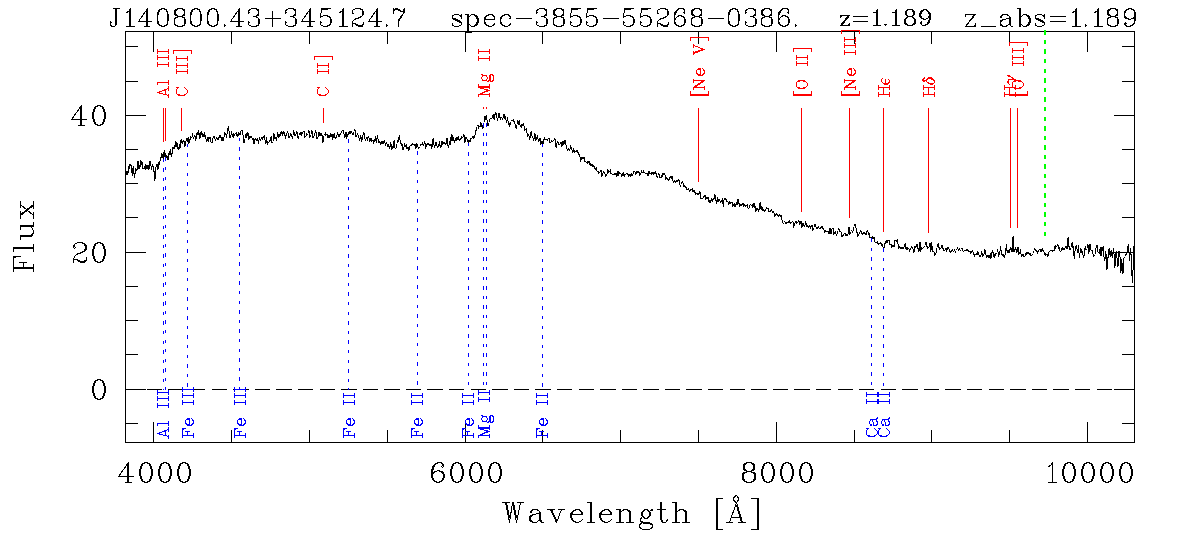} \
\includegraphics[viewport= 0 -30 1185 535,width=9.1cm,angle=0]{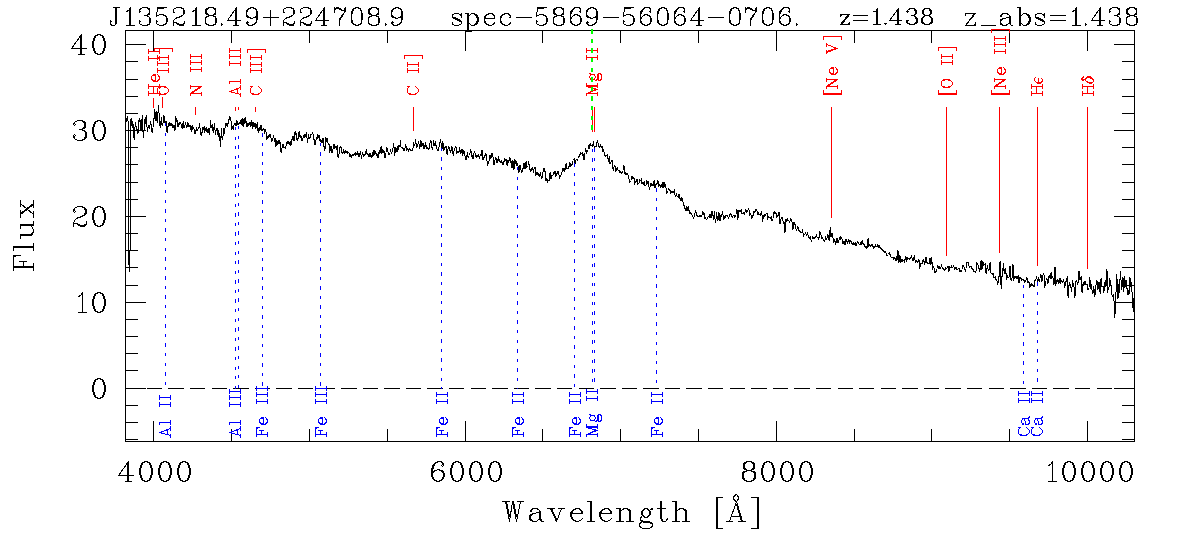} \\
\includegraphics[viewport= 0 -30 1185 535,width=9.1cm,angle=0]{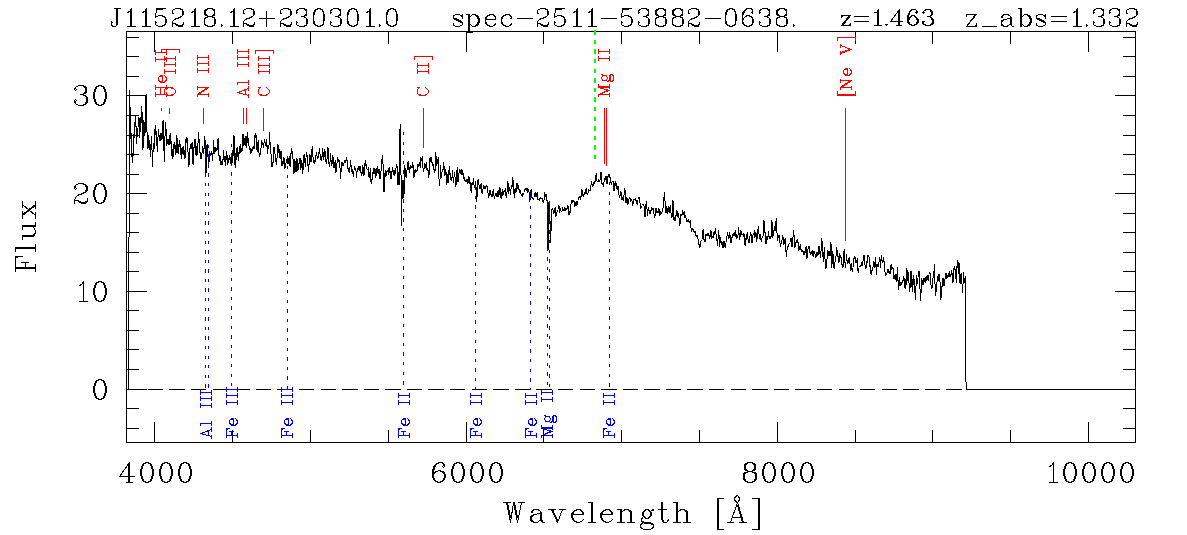} \
\includegraphics[viewport= 0 -30 1185 535,width=9.1cm,angle=0]{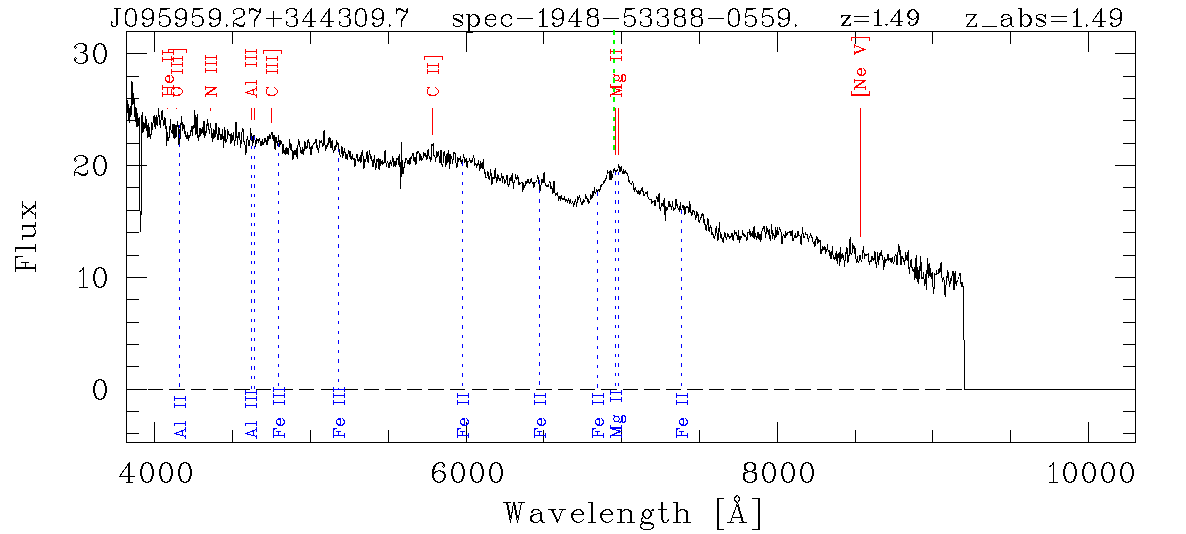} \\
\includegraphics[viewport= 0 -30 1185 535,width=9.1cm,angle=0]{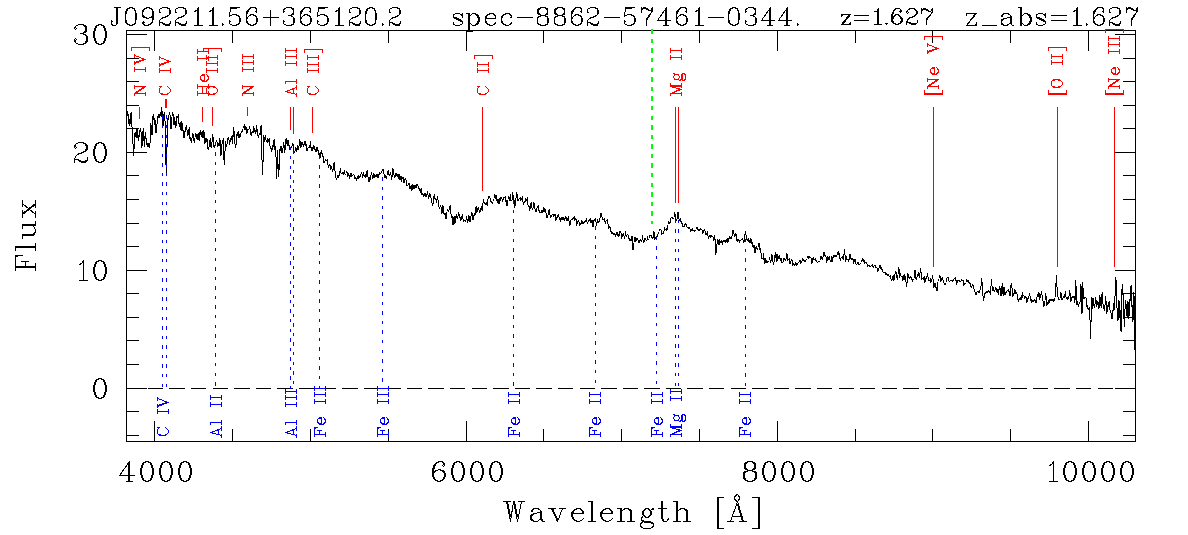} \
\includegraphics[viewport= 0 -30 1185 535,width=9.1cm,angle=0]{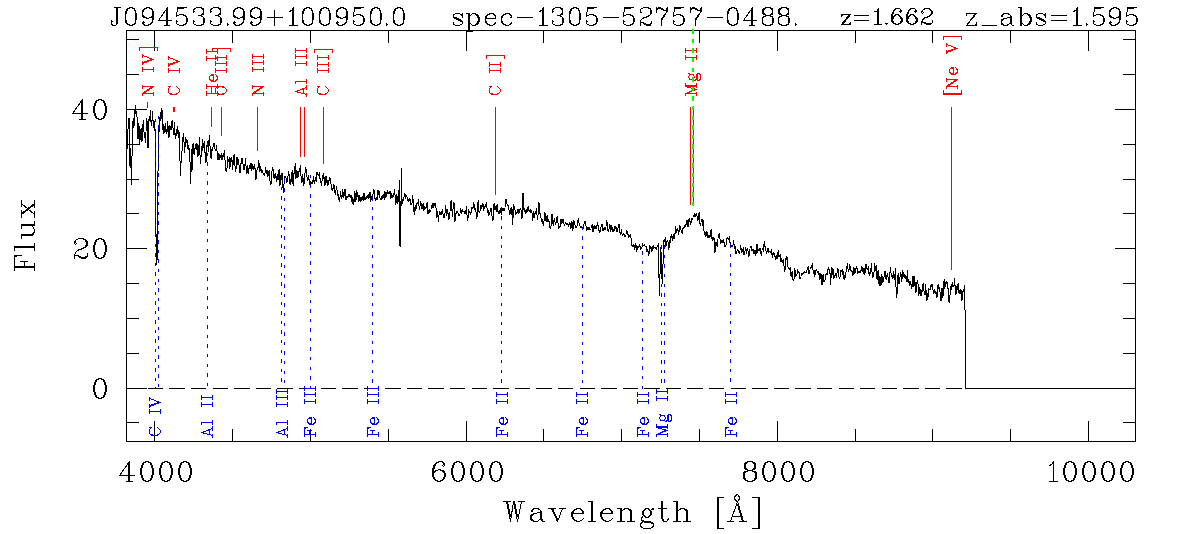} \\
\includegraphics[viewport= 0 -30 1185 535,width=9.1cm,angle=0]{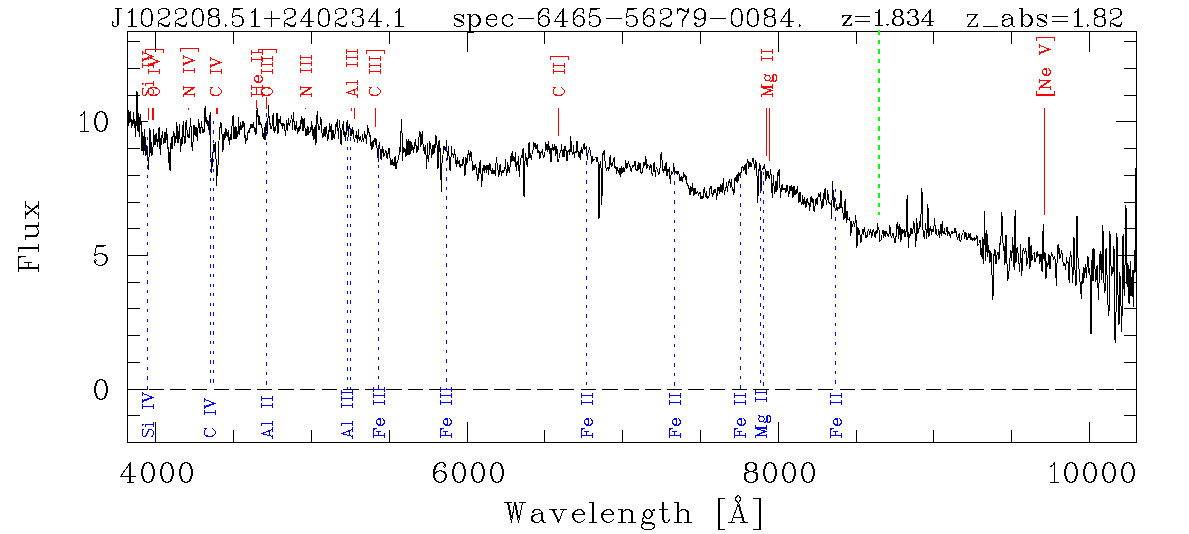} \
\includegraphics[viewport= 0 -30 1185 535,width=9.1cm,angle=0]{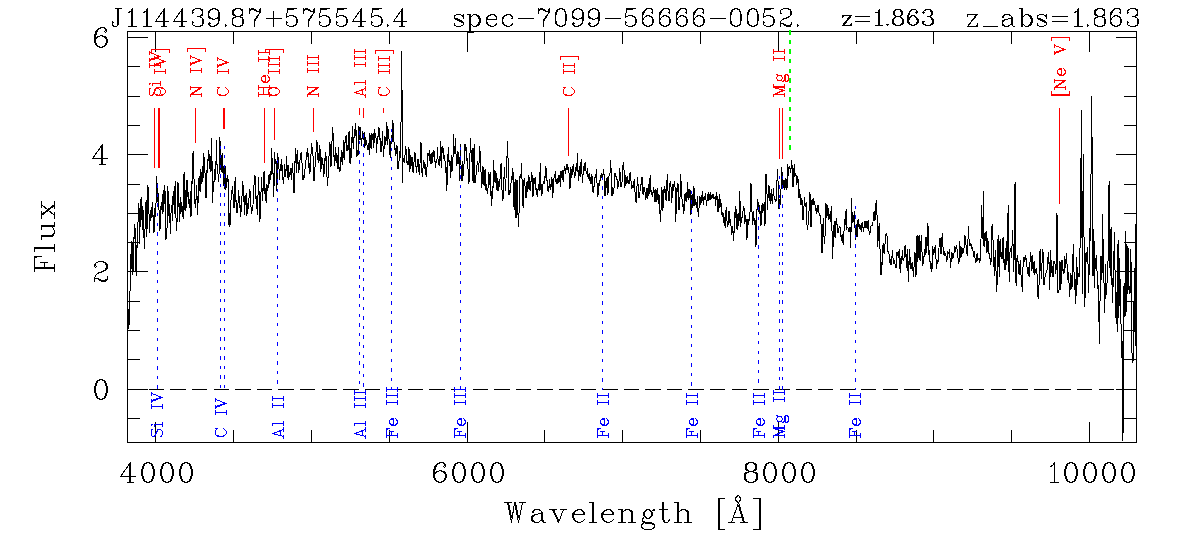} \\
\includegraphics[viewport= 0 -30 1185 535,width=9.1cm,angle=0]{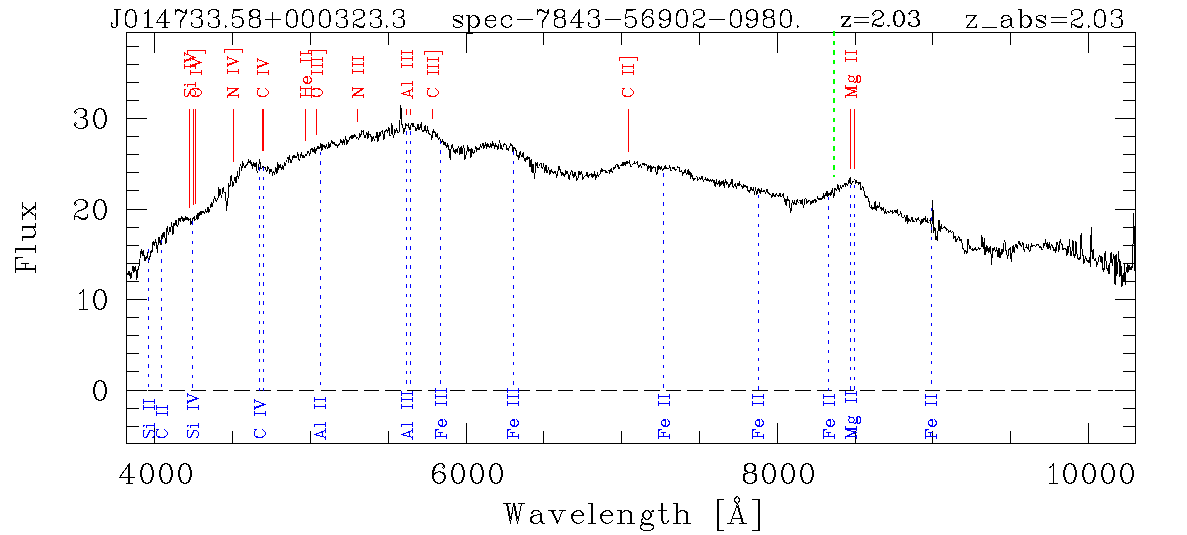} \
\includegraphics[viewport= 0 -30 1185 535,width=9.1cm,angle=0]{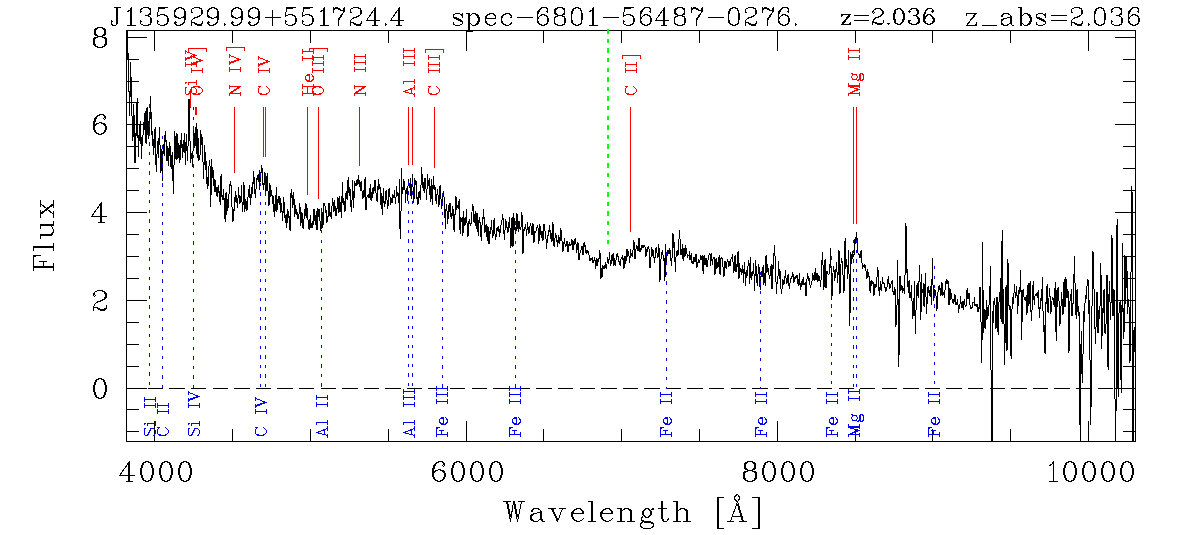} \\
\caption{Examples of WLQs. 
The unit of the flux and the meaning of the vertical lines are the same as in Fig.\,\ref{fig:LoBALs}.
}
\label{fig:WLQs}
\end{figure*}

\begin{figure*}[bhtp]
\begin{center}
\includegraphics[viewport= 0 -30 1185 535,width=9.1cm,angle=0]{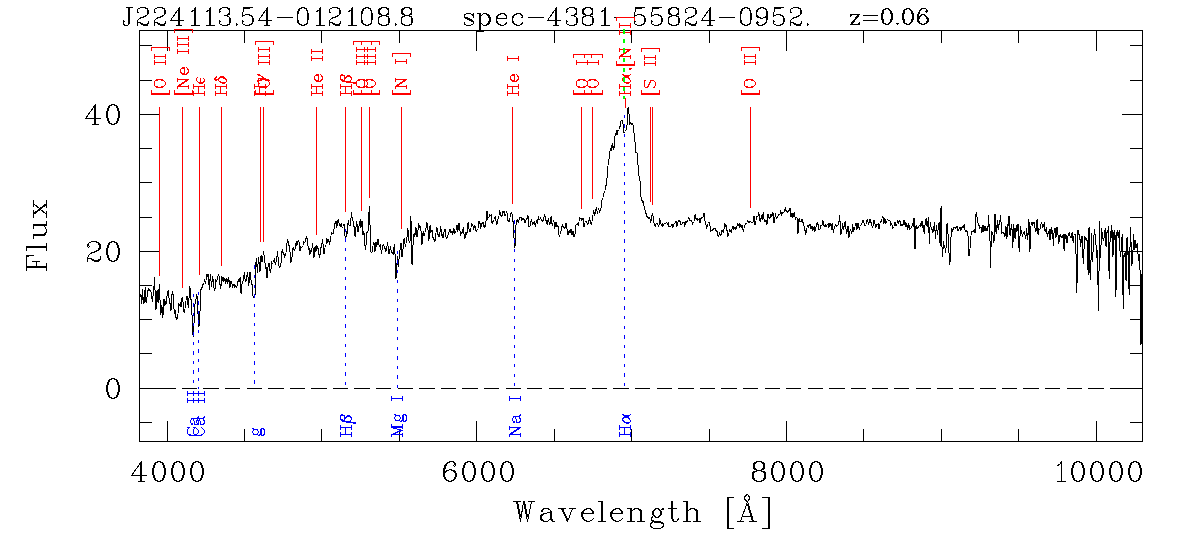} \
\includegraphics[viewport= 0 -30 1185 535,width=9.1cm,angle=0]{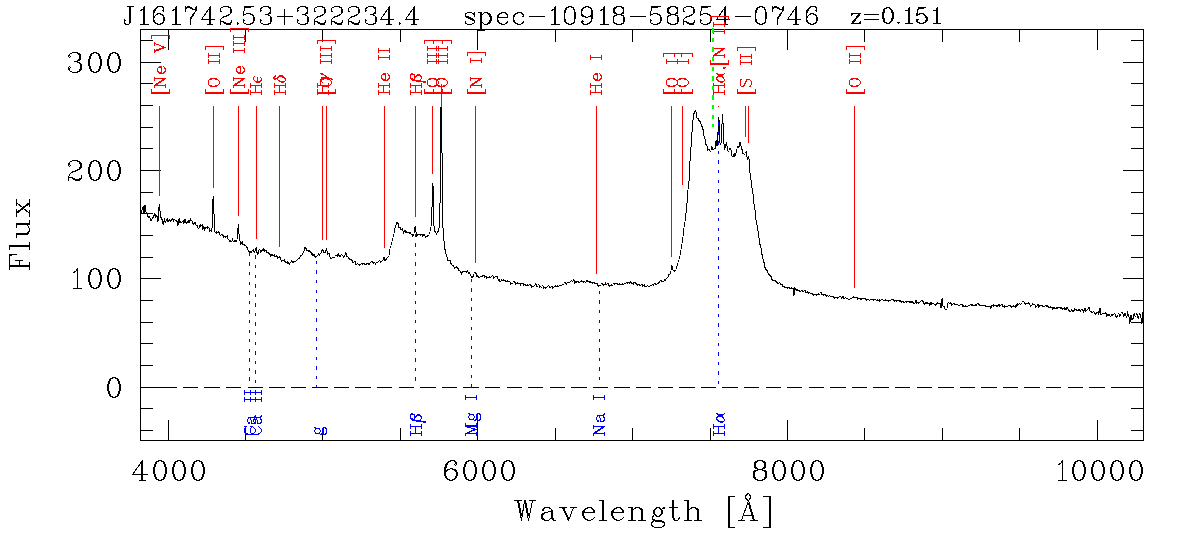} \\
\includegraphics[viewport= 0 -30 1185 535,width=9.1cm,angle=0]{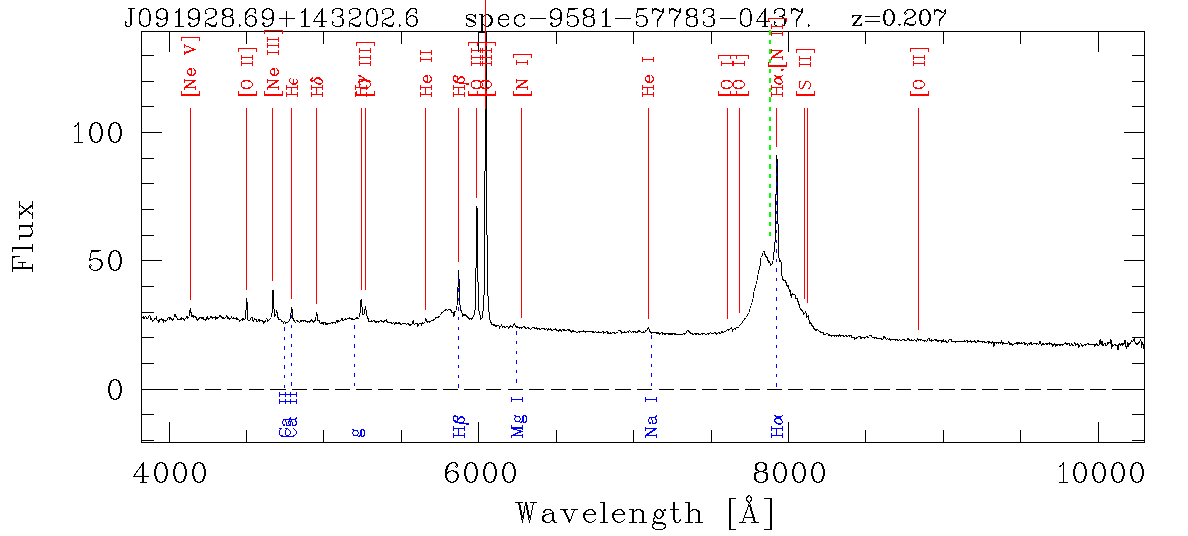} \
\includegraphics[viewport= 0 -30 1185 535,width=9.1cm,angle=0]{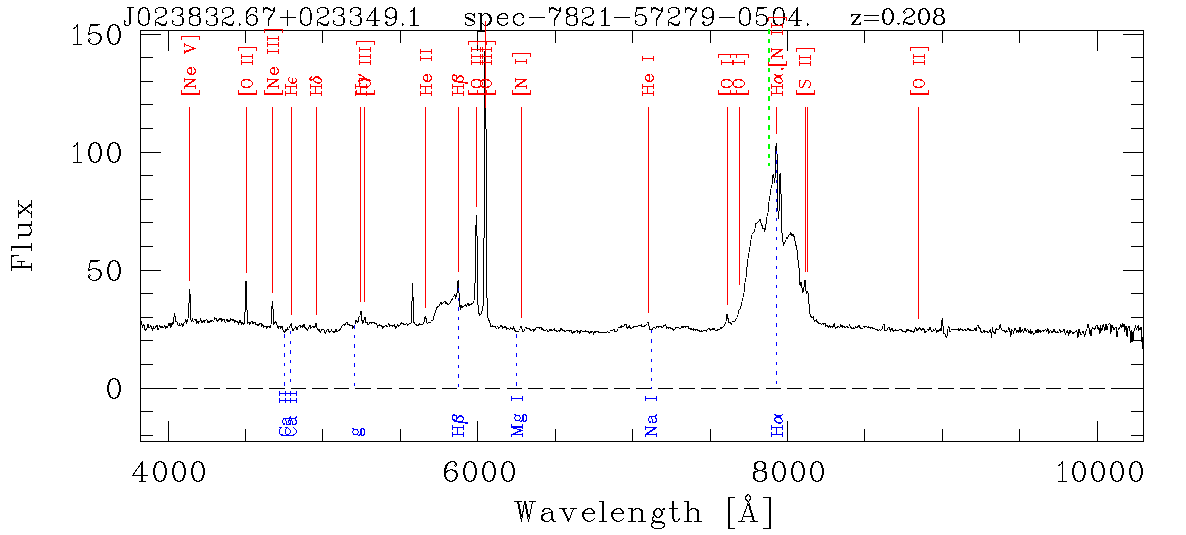} \\
\includegraphics[viewport= 0 -30 1185 535,width=9.1cm,angle=0]{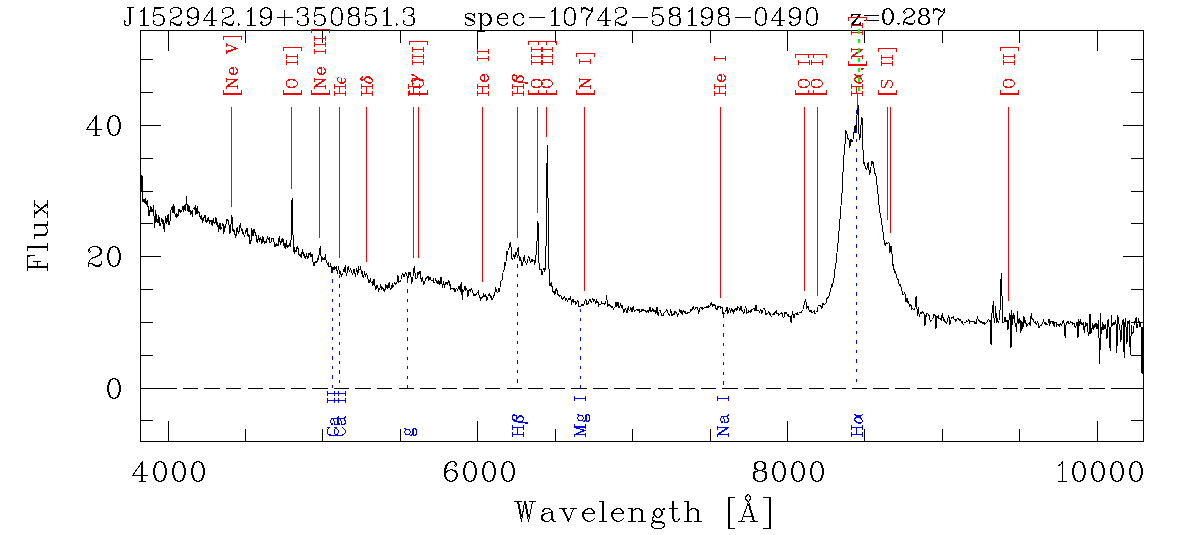} \
\includegraphics[viewport= 0 -30 1185 535,width=9.1cm,angle=0]{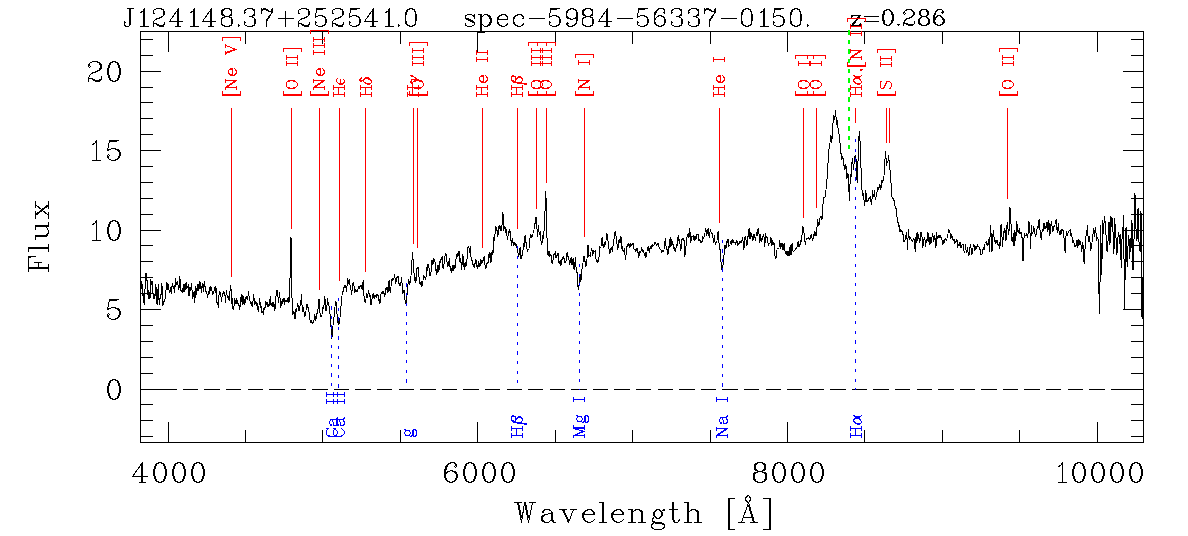} \\
\includegraphics[viewport= 0 -30 1185 535,width=9.1cm,angle=0]{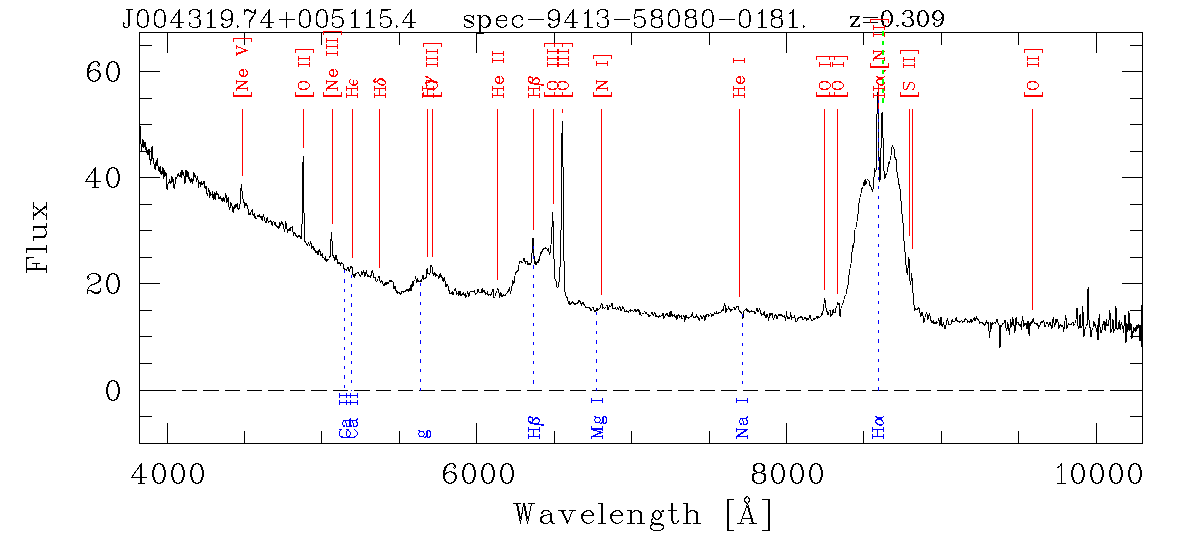} \
\includegraphics[viewport= 0 -30 1185 535,width=9.1cm,angle=0]{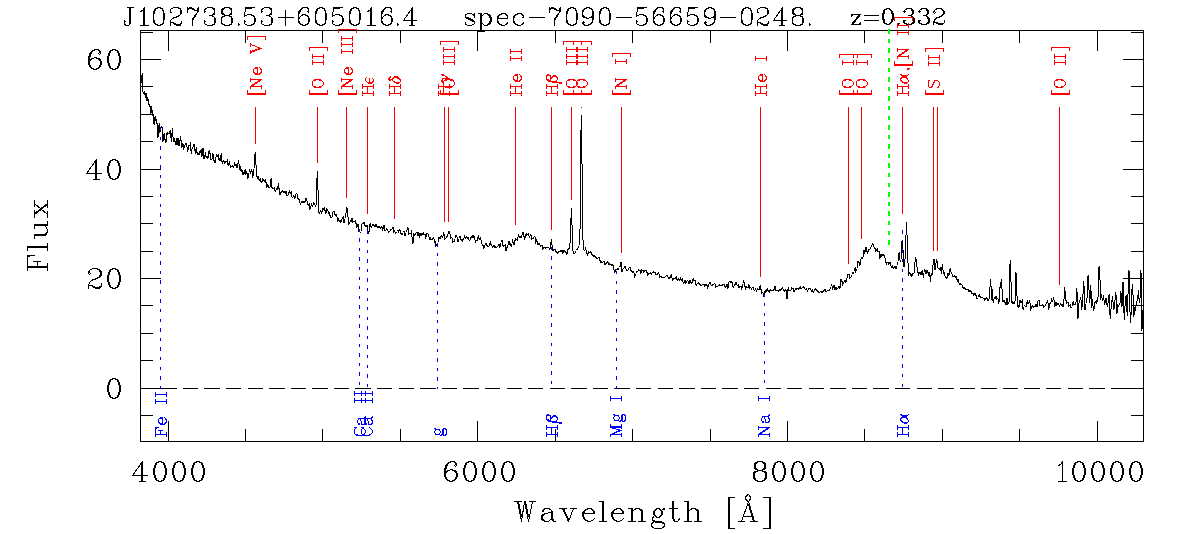} \\
\includegraphics[viewport= 0 -30 1185 535,width=9.1cm,angle=0]{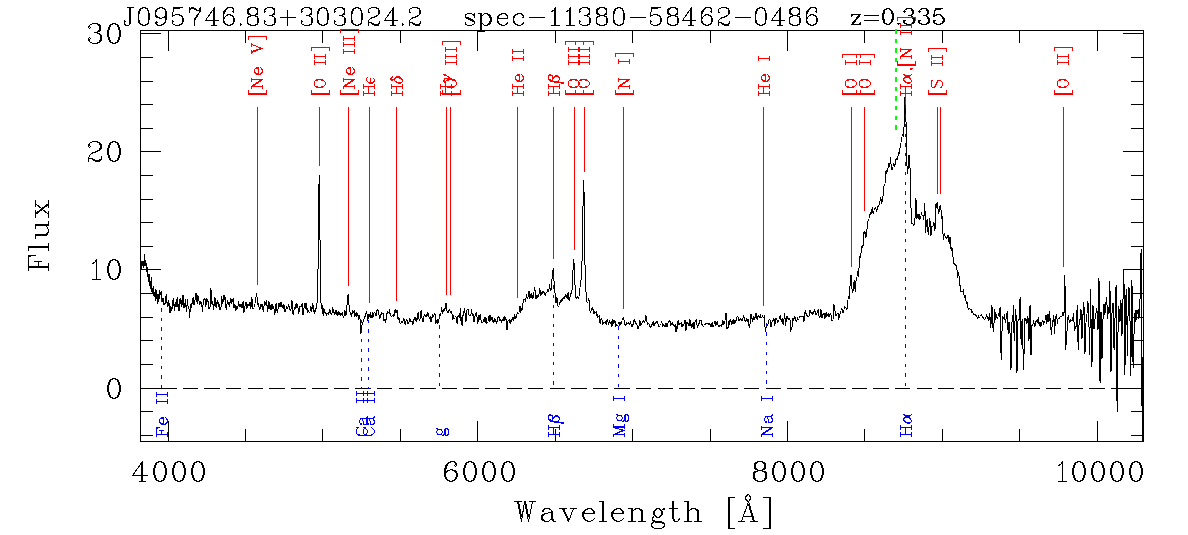} \
\includegraphics[viewport= 0 -30 1185 535,width=9.1cm,angle=0]{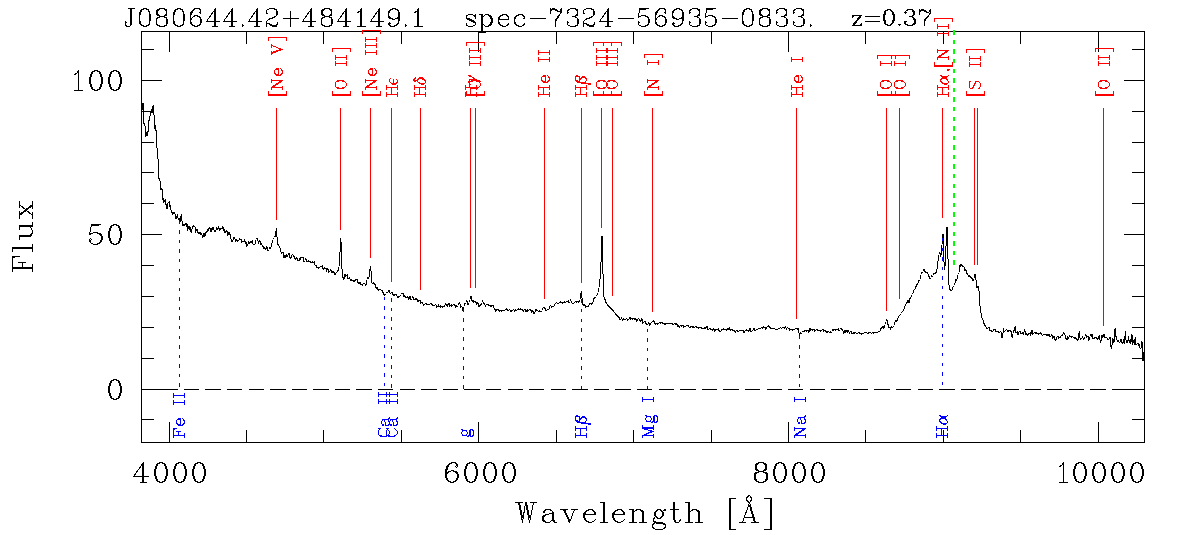} \\
\caption{Examples of QSOs with double-peaked or strongly lopsided broad Balmer lines. 
The unit of the flux and the meaning of the vertical lines are the same as in Fig.\ref{fig:LoBALs}.
}
\label{fig:dp_broad_Balmer}
\end{center}
\end{figure*}

\begin{figure*}[bhtp]
\begin{center}
\includegraphics[viewport= 0 -30 1185 535,width=9.1cm,angle=0]{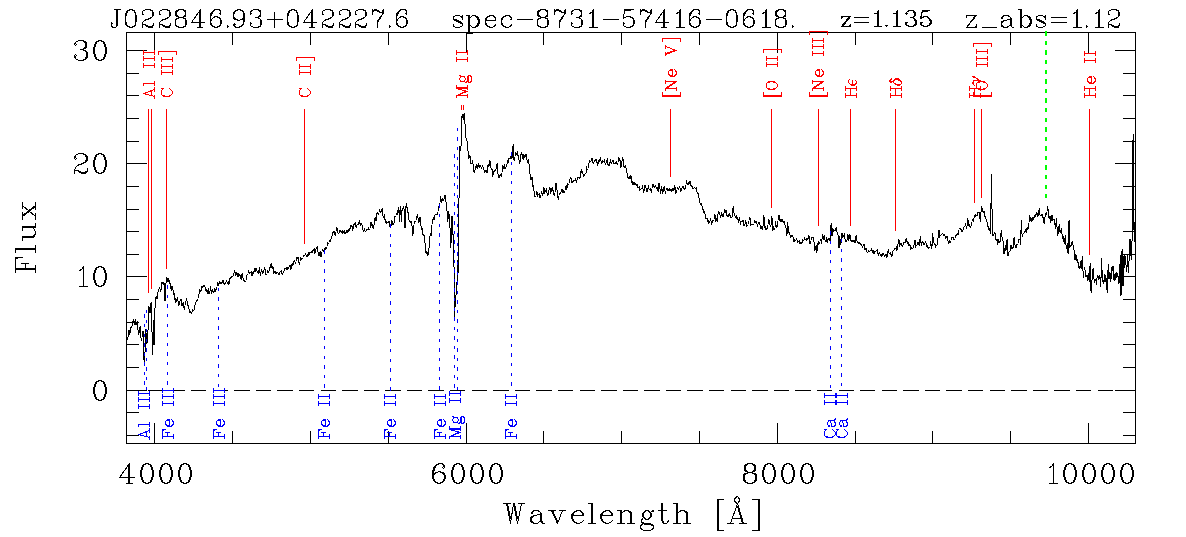} \
\includegraphics[viewport= 0 -30 1185 535,width=9.1cm,angle=0]{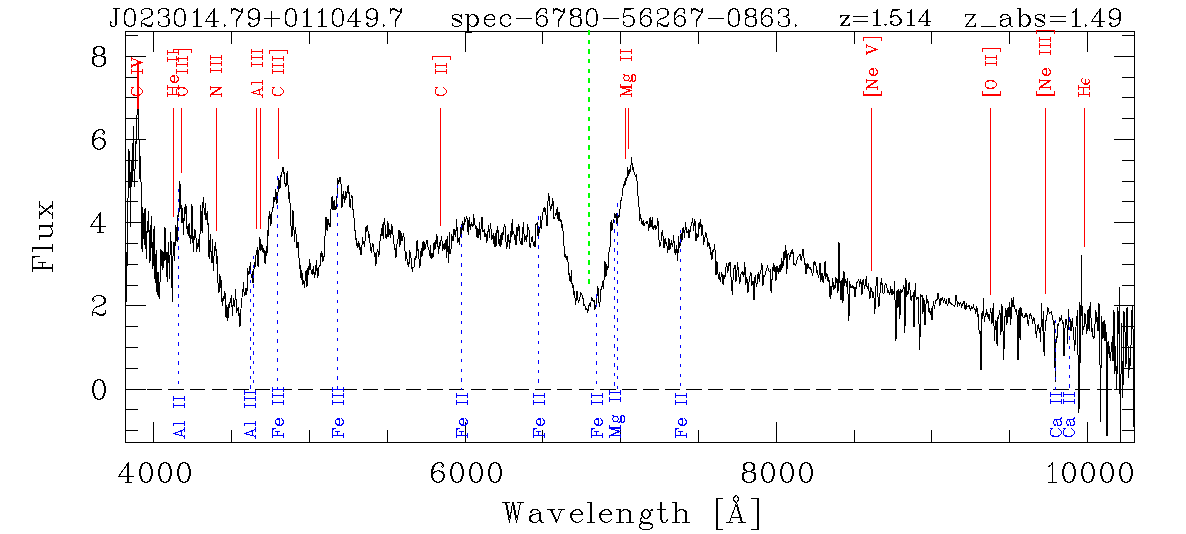} \\
\includegraphics[viewport= 0 -30 1185 535,width=9.1cm,angle=0]{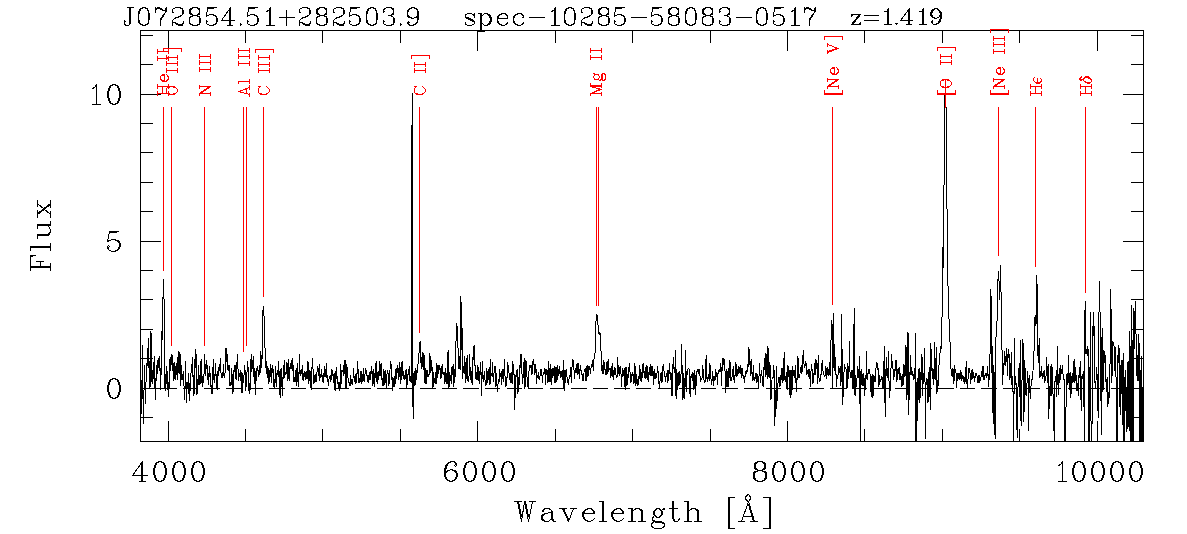} \
\includegraphics[viewport= 0 -30 1185 535,width=9.1cm,angle=0]{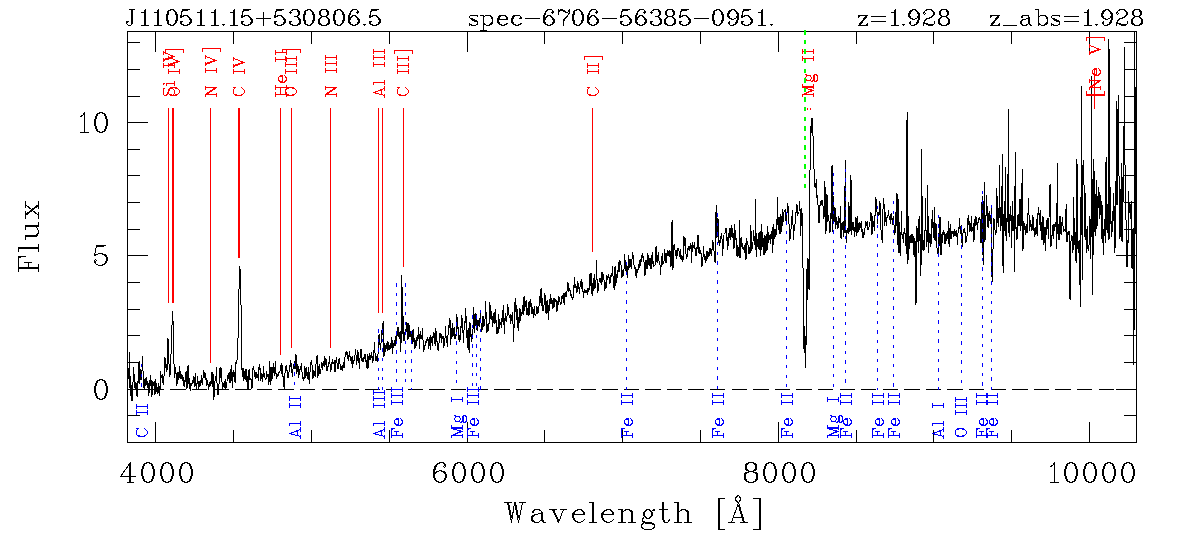} \\
\includegraphics[viewport= 0 -30 1185 535,width=9.1cm,angle=0]{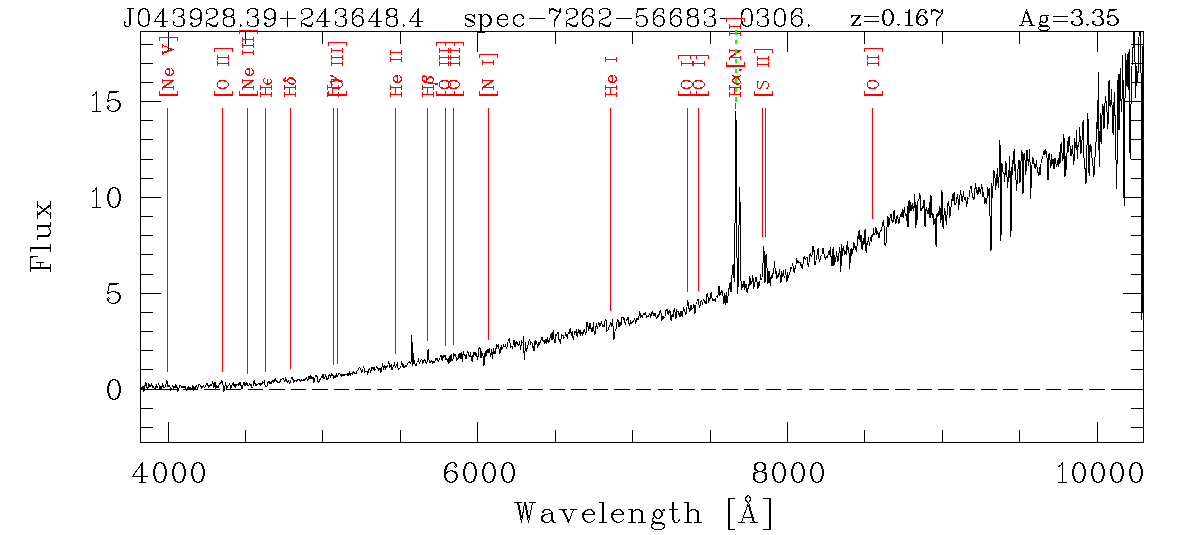} \
\includegraphics[viewport= 0 -30 1185 535,width=9.1cm,angle=0]{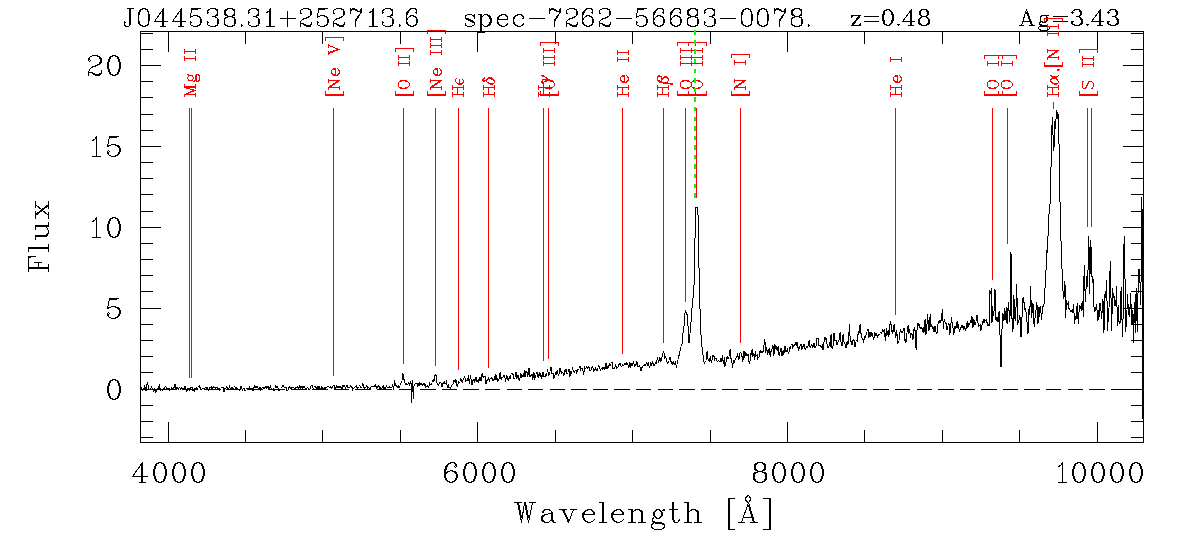} \\
\includegraphics[viewport= 0 -30 1185 535,width=9.1cm,angle=0]{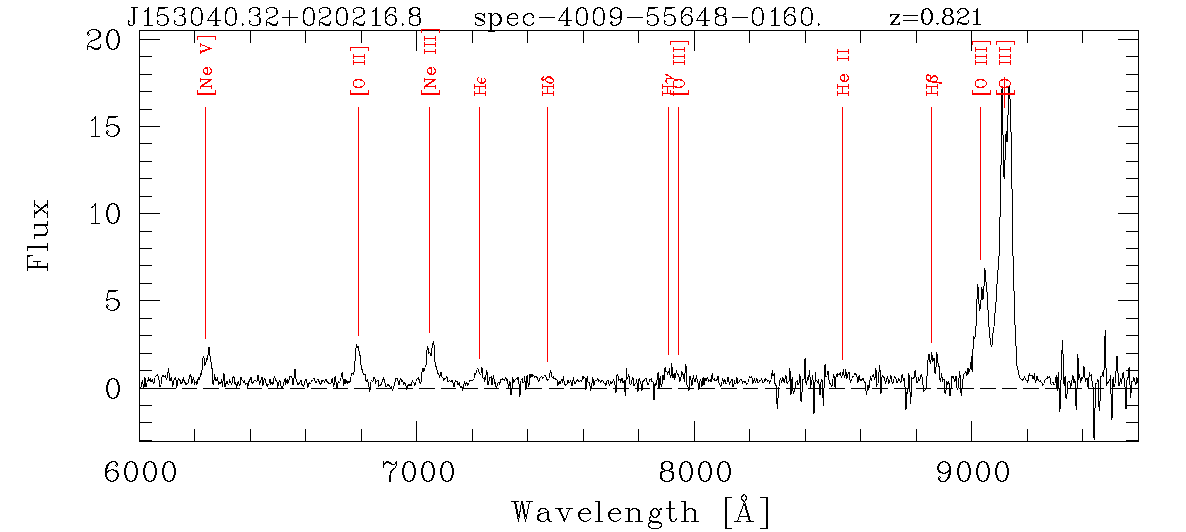} \
\includegraphics[viewport= 0 -30 1185 535,width=9.1cm,angle=0]{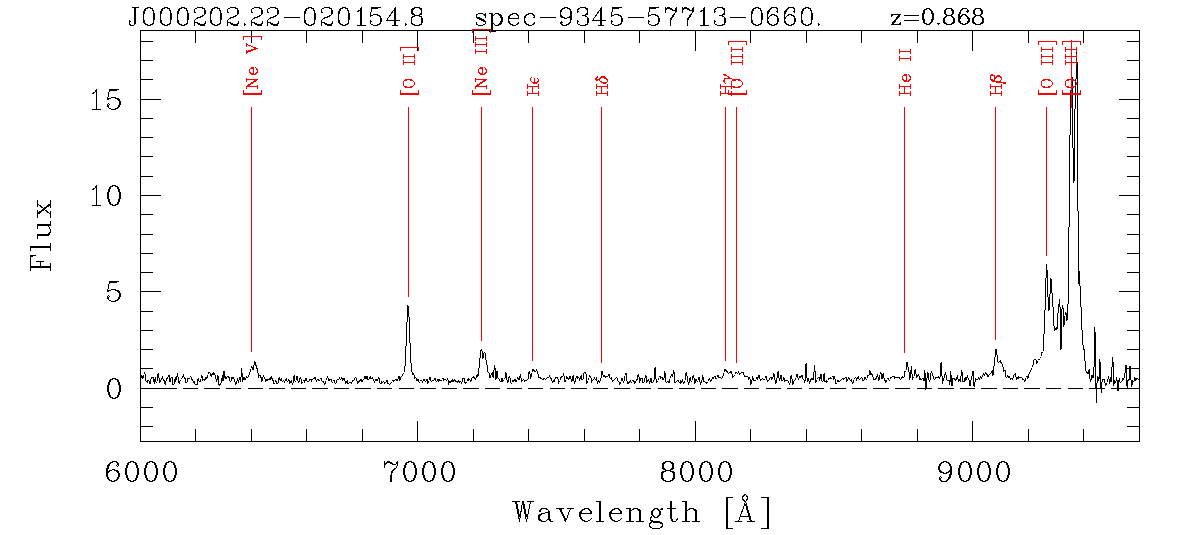} \\
\includegraphics[viewport= 0 -30 1185 535,width=9.1cm,angle=0]{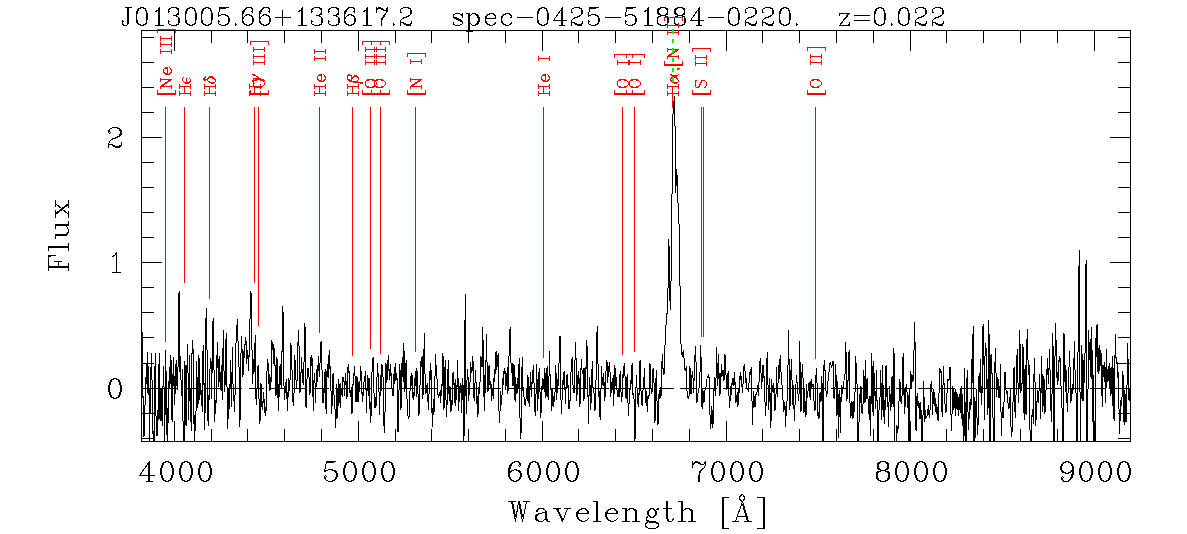} \
\includegraphics[viewport= 0 -30 1185 535,width=9.1cm,angle=0]{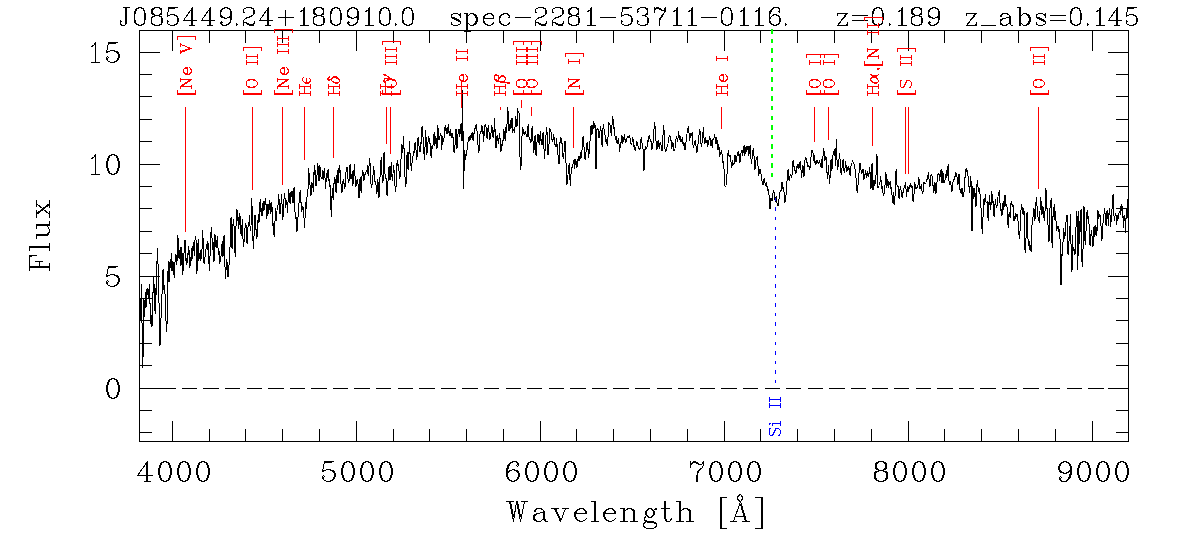} \\
\caption{Two examples each of five other types of rare spectra of QSOs or galaxies. 
Top to bottom: very strong Fe emission  features, type 2 AGNs, strongly reddened galaxies, double narrow emission lines, galaxies with supernovae. 
The unit of the flux and the meaning of the vertical lines are the same as in Fig.\ref{fig:LoBALs}.
}
\label{fig:mix_QSOs_gals}
\end{center}
\end{figure*}

\begin{figure*}[htbp]
\begin{center}
\includegraphics[viewport= 0 -30 1185 535,width=9.1cm,angle=0]{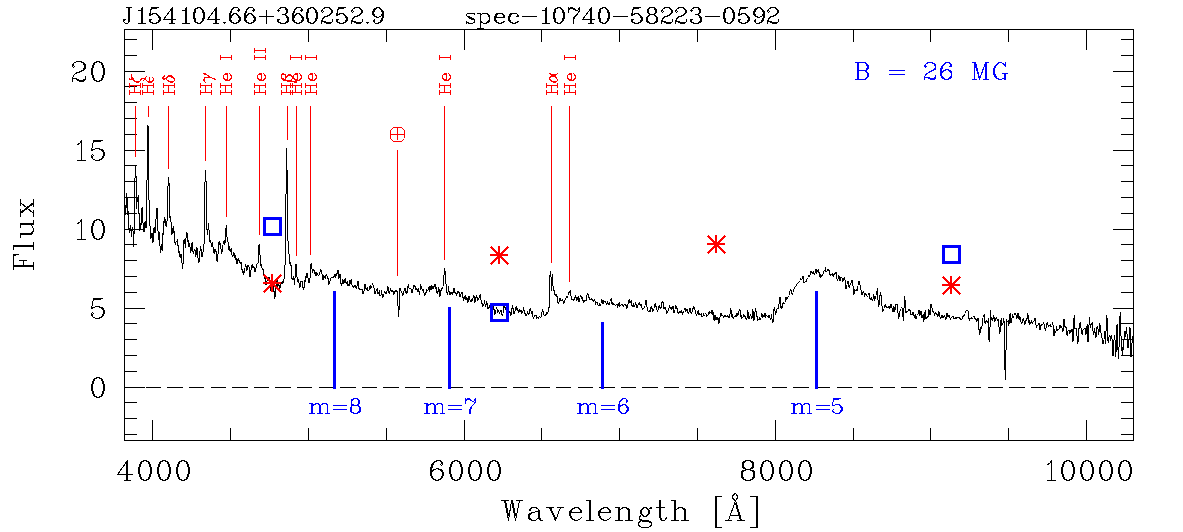} \ \ \ \ \
\includegraphics[viewport= 0 -30 256 256,width=3.8cm,angle=0]{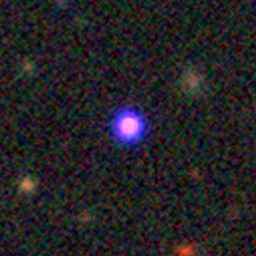} \\
\includegraphics[viewport= 0 -30 1185 535,width=9.1cm,angle=0]{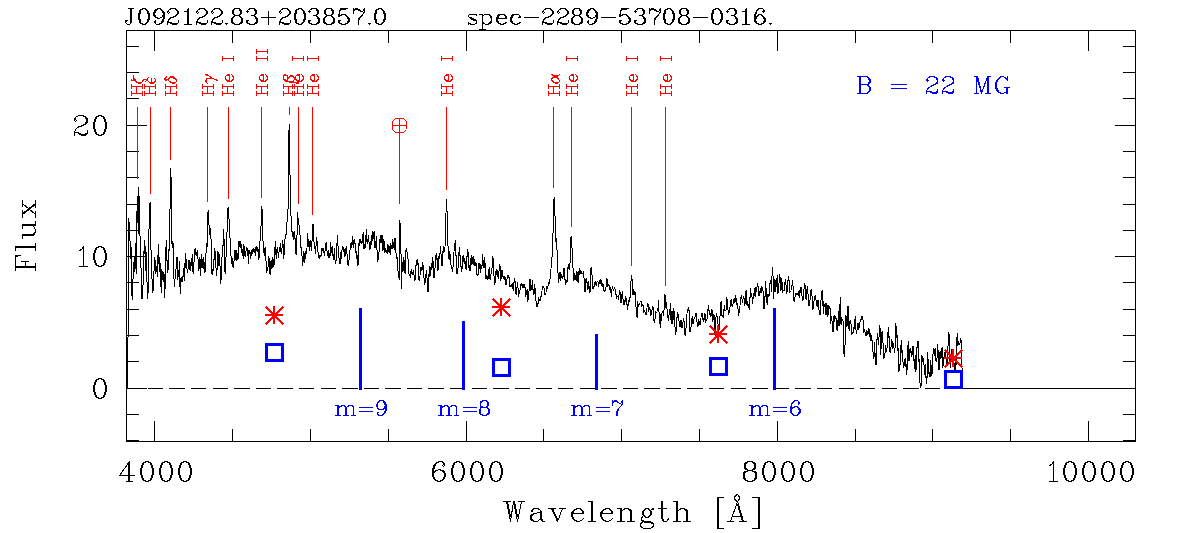} \ \ \ \ \
\includegraphics[viewport= 0 -30 256 256,width=3.8cm,angle=0]{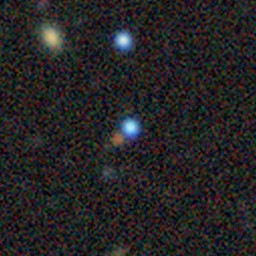} \\
\includegraphics[viewport= 0 -30 1185 535,width=9.1cm,angle=0]{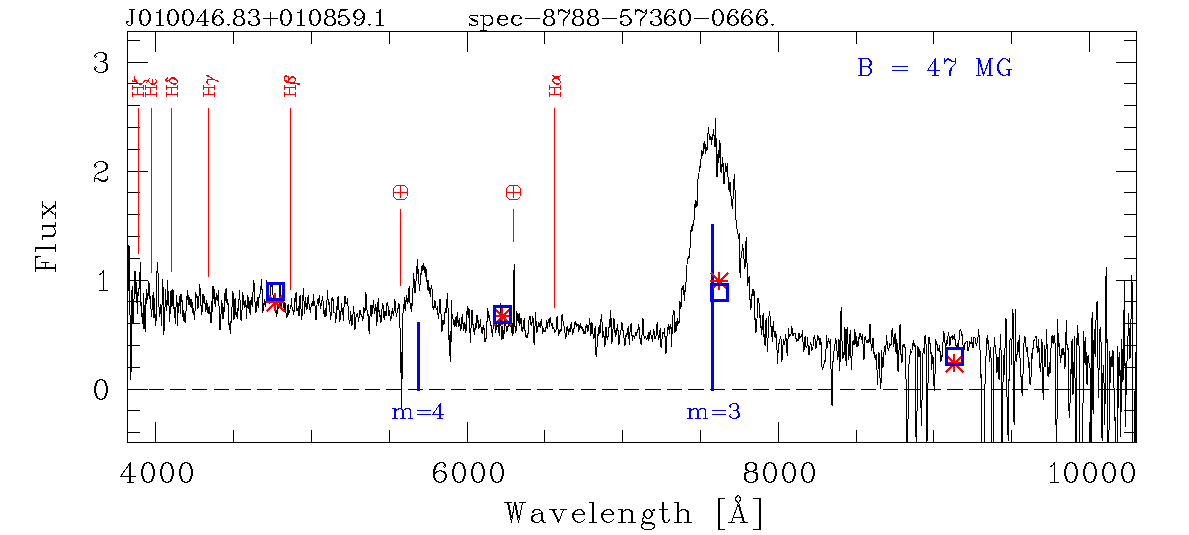} \ \ \ \ \
\includegraphics[viewport= 0 -30 256 256,width=3.8cm,angle=0]{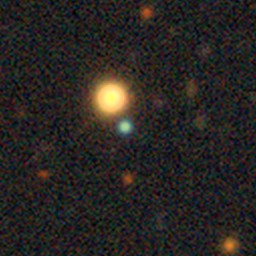} \\
\includegraphics[viewport= 0 -30 1185 535,width=9.1cm,angle=0]{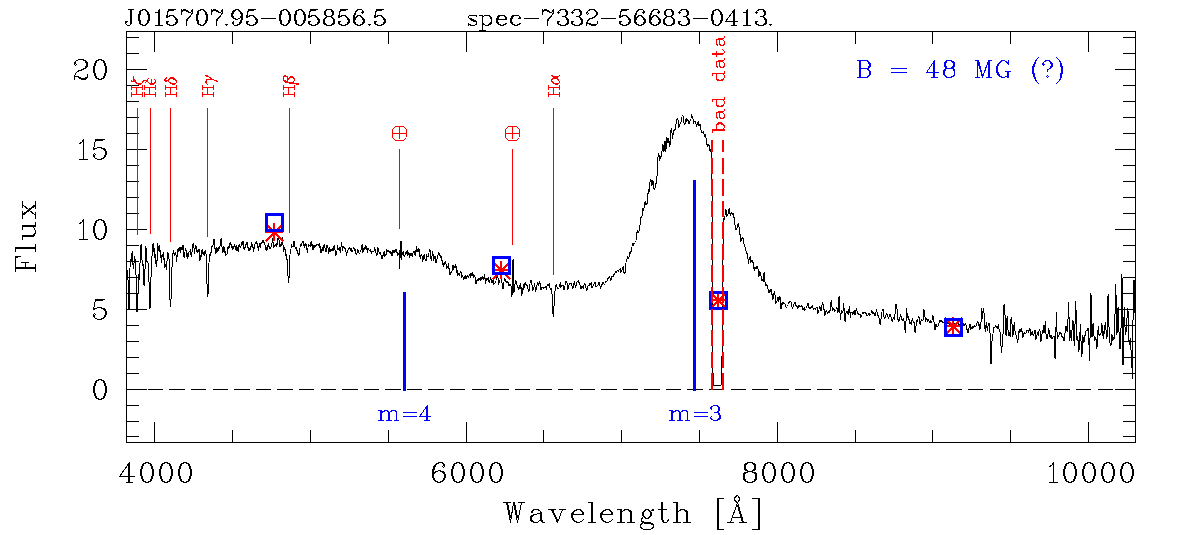} \ \ \ \ \
\includegraphics[viewport= 0 -30 256 256,width=3.8cm,angle=0]{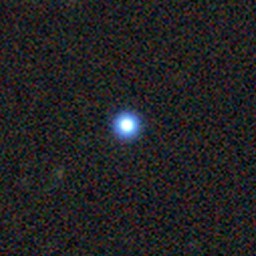} \\
\includegraphics[viewport= 0 -30 1185 535,width=9.1cm,angle=0]{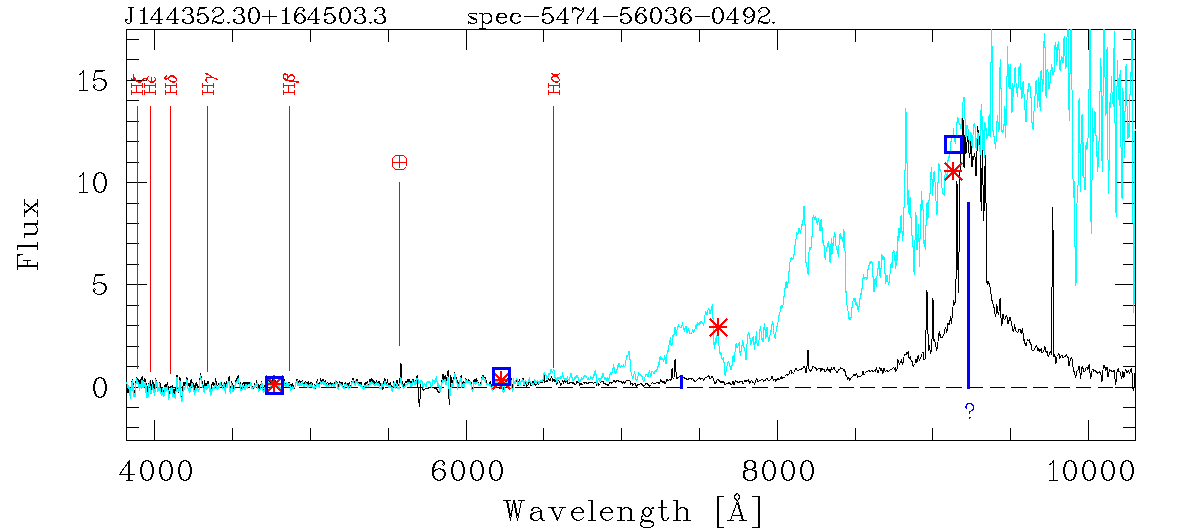} \ \ \ \ \
\includegraphics[viewport= 0 -30 256 256,width=3.8cm,angle=0]{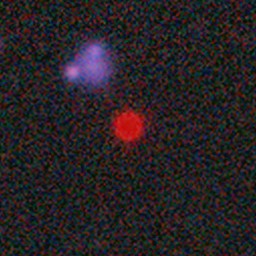} \\
\caption{Spectra with real or false emission humps that seem to indicate cyclotron radiation from polars. 
The flux is $F_\lambda$ in units of $10^{-17}$\,erg\,s$^{-1}$\,cm$^{-2}$\,\AA$^{-1}$. 
The best matches for the cyclotron harmonics $m$ are marked by blue vertical lines. 
The estimated magnetic flux density $B$ is given in the top right corner.  The positions of the Balmer lines and of identified He lines are displayed by red vertical lines, the $\oplus$ symbols indicate telluric lines. 
Red asterisks mark the flux densities derived from the SDSS magnitudes and open blue squares are the flux densities corresponding to the LS magnitudes in the griz bands. In the bottom panel  (SDSS\,J144352.30+164503.3), the black spectrum is  from DR16 whereas the cyan spectrum is the same spectrum from DR15 
(i.e. an earlier version of the pipeline). The RGB images on the right-hand side are cutouts from the LS and have side lengths of 20 arcsec.  The object in the image centre is the source of the spectrum.}
\label{fig:polars}
\end{center}
\end{figure*}

\end{appendix}

\end{document}